\pdfoutput=1 
\documentclass{JINST}

\title{Searching for a particle of unknown mass and lifetime in the presence of an unknown non-monotonic background}

\author{Mike Williams\\
Massachusetts Institute of Technology, Cambridge, MA, United States
}

\abstract{
Many extensions to the Standard Model of particle physics hypothesize the existence of new low-mass particles.  Typically there are few theoretical constraints on the mass or lifetime of such particles.  This requires the experimentalist to perform a search in which both the mass and lifetime of the particle are unknown.  Such searches for low-mass particles are complicated by the possible presence of resonances and other non-monotonic  backgrounds.  This paper presents a simple and fast approach to assigning significance and setting limits in such searches.   
}

\begin{document}

\section{Introduction}

Many extensions to the Standard Model (SM) of particle physics hypothesize the existence of new low-mass particles.  Typically there are few theoretical constraints on the mass or lifetime of such particles.  Therefore, the experimentalist is required to perform a search in which both the mass and lifetime of the particle are unknown.  
Such searches for low-mass particles are complicated by the possible presence of resonances and other non-monotonic backgrounds whose probability density functions (PDFs) are not well known or constrained.

The null results from numerous searches for new particles that decay into leptons has shifted the focus of many theorists towards {\em leptophobic} bosons.  
For example, Tulin~\cite{tulin} proposes a force that couples weakly to quarks at the QCD scale.  Experimentally, this requires a search for a MeV or GeV mass boson that decays to $\pi^+\pi^-\pi^0$.  
If such a boson decays promptly\footnote{Throughout this paper I refer to decays as either: (prompt) the separation of the decay point from the production point is too small to be resolved by the detector and (displaced) this separation is resolvable.}, then searching for it requires dealing with an irreducible resonance background.  It is preferable to perform this search with the data blinded; however, the reaction under study is governed by non-perturbative QCD and so the only {\em a priori} assumption about the SM PDF that can be made is: any resonance that couples to the same final state as the new boson may contribute to the SM background.  The number of known resonances is large and new resonances are still being discovered, even those that only contain light quarks~\cite{compass}.  Furthermore, parametrization of resonances is reaction-dependent and interference effects are often significant.  Therefore, the number of nuisance parameters in the background PDF is very large and there will likely always be some component to the PDF that is simply not modeled accurately to better than a few percent.   
 Even in dedicated amplitude analyses performed without blinding the data, it is often difficult to evaluate the systematic uncertainties associated with the fit PDF due to the large number of nuisance parameters.  It is also difficult to obtain a fit model which describes the data at the percent level for all masses.

\begin{figure}[]
\centering
\includegraphics[width=0.49\textwidth]{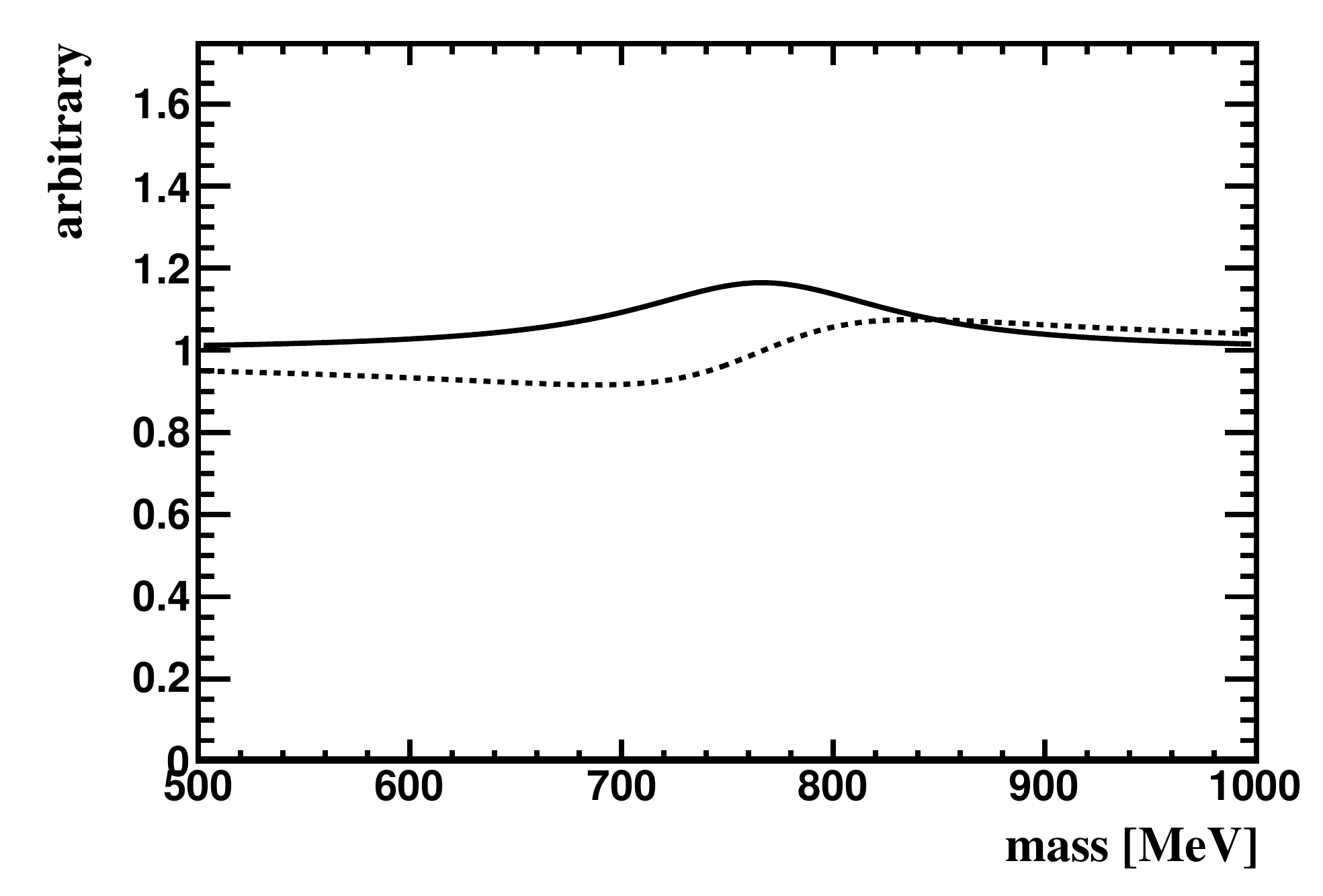}
\caption{\label{fig:rho}Simulated PDF including interference between $B\to K^*\mu\mu$ and $B\to K^*\rho(\mu\mu)$ for two arbitrary phase differences between the two amplitudes (shown as solid and dashed lines).  
}
\end{figure}

Seaches for new bosons decaying to leptons are not immune to the problem of resonances.  
Consider, {\em e.g.}, the rare decay $B\to K^*\mu\mu$ which is an excellent laboratory to search for a new low-mass boson that couples to mass~\cite{thaler}.
This decay can have a contribution from $B\to K^*\rho(\mu\mu)$.  
Using the published LHCb $B\to K^*\mu\mu$ results~\cite{kstrmm} and the ratio of branching fractions $\mathcal{B}(B\to K^*\rho)\mathcal{B}(\rho\to\mu\mu)/\mathcal{B}(B\to K^*\mu\mu)$ from the PDG~\cite{pdg} one obtains a predicted yield of $B\to K^*\rho(\mu\mu)$ that is less than one event; however, this calculation ignores interference.  
Figure~\ref{fig:rho} shows a toy prediction obtained using the measured branching fractions $\mathcal{B}(B\to K^*\mu\mu)$, $\mathcal{B}(B\to K^*\rho)$ and $\mathcal{B}(\rho\to\mu\mu)$, and the assumption that the full amplitudes for $B\to K^*\mu\mu$ and $B\to K^*\rho(\mu\mu)$ interfere\footnote{I ignore muon spin states here which produces an overestimate of the size of this interference effect, but for illustrative purposes this model is sufficient.}.  
 Naively using the ratio of branching fractions suggests $B\to K^*\rho(\mu\mu)$ is negligible, but the interference cross term could be large enough to generate a ``local'' 5\% effect. 
In principle the $\rho$ contribution could be parametrized (with some uncertainty); however, the branching fractions of many resonance decays to $\mu^+\mu^-$ and of $B\to K^*$ resonance have yet to be measured and so such contributions would be unconstrained.
Since the effect generated can also depend on interference, a full model allowing all resonances to interfere (which requires introducing unknown relative phases for each resonance) must be constructed.

For the case of displaced decays, the largest background contributions are often due to some mis-reconstructed candidates.  These include material interactions, photon conversions, {\em etc.}  Given that the searches considered in this paper must scan a large mass and lifetime range, predicting (blindly) the full list of displaced backgrounds and parametrizing them will often not be possible.  Furthermore, resonances may contribute to the displaced search via decays of beauty or charm particles. {\em E.g.}, in the aforementioned $B\to K^*\mu\mu$ search, it is possible to obtain the $K^*$ and $\mu^+\mu^-$ from two different $B$ decays at LHCb.  This could result in a candidate that contains what appears to be a long-lived charmonium resonance.  It is desirable to approach such a search by making as few assumptions as possible about the backgrounds.  

This paper presents a simple approach for performing a search for a new low-mass particle of unknown mass and lifetime.  The method is described in Sec.~2.  Obtaining global $p$-values, including the so-called trials factor or look elsewhere effect, is discussed in Sec.~3,  while setting upper limits and the coverage properties of such limits is discussed in Sec.~4.   Discussion and summarizing are provided in Sec.~5.

\section{Search Strategy}

The strategy that will be employed is to perform a scan in mass since the new particle mass is not known.  The step size of the scan will be $\sigma(m)/2$, where $\sigma(m)$ is the detector mass resolution.  For each test mass value a search will be conducted for evidence of a particle with no constraints on its lifetime.  For illustrative purposes, the following toy model is used:
\begin{itemize}
\item the searched region in mass is 1~GeV wide\footnote{I just take the mass region searched to be 0 to 1000~MeV since the absolute mass values do not matter; {\em i.e.}, shifting the mass window to a range that starts at some allowed kinematic limit has no affect on applying this method.  For this toy study I also do not specify what the decay products of the particle being searched for are since these also do not matter apart from identifying which resonances may contribute to the data.};
\item the mass resolution is taken to be $\sigma(m)=2$~MeV, giving a step size of 1~MeV;
\item the expected number of prompt events (in the absence of signal) is 1000;
\item the expected number of displaced events (in the absence of signal) is 100;
\item both prompt and displaced events are generated uniformly in mass.
\end{itemize}
This toy data set is sufficient for illustrating how the proposed method works.  Some variations are considered later in the text.

\subsection{Mass}

As an alternative to fitting the mass distribution in data with a PDF whose accuracy is {\em a priori} unknown and difficult to validate, I propose the following simple approach: For each test mass value in the scan, $m($test$)$, use the mass sidebands to estimate the expected background\footnote{Any candidate that does not come from the decay of a new particle is considered as a background in the search.} yield.  The signal region (where signal events would be observed if a new particle of mass close to $m($test$)$ exists) is defined as $|m({\rm test})-m| < 2\sigma(m)$, while the background region is the sidebands defined as $3\sigma(m) < |m({\rm test})-m| < (2x+3)\sigma(m)$, where $x$ is the ratio of the size of the signal to sideband regions (these regions do not need to be the same size).  The factors that contribute to choosing $x$ are discussed at the end of this subsection.

If the background PDF is approximately linear in the region $|m({\rm test})-m| < (2x+3)\sigma(m)$, then the observed yield in the background region provides an estimate of the expected yield in the signal region.  The presence of resonances and some types of background will violate this so-called local-linear approximation; this is dealt with below.   
Under the local-linear assumption, the likelihood is given by
\begin{equation}
  L(n_s,n_b|s,b) = \mathcal{P}(n_s,s+b) \cdot \mathcal{P}(n_b,xb),
\end{equation}
where $\mathcal{P}$ denotes the Poisson PDF, $n_{(s,b)}$ denote the yields in the (signal, background) regions and $s$ and $b$ are the signal and background rates in the signal region.
It is straightforward to account for uncertainty in the relationship between the sideband and search window background yields, or equivalently in deviations from local linearity in the background PDF, by augmenting the likelihood as follows:
\begin{equation}
  \label{eq:likegaus}
  L(n_s,n_b,x|s,b,y) = \mathcal{P}(n_s,s+b) \cdot \mathcal{P}(n_b,yb) \cdot \mathcal{G}(y,x,\sigma_y),
\end{equation}
where $\mathcal{G}$ denotes the Gaussian PDF and $\sigma_y$ is the uncertainty on the scaling factor between the background yield in the signal and background regions\footnote{This likelihood can be used for any value of $n_b$; however, Appendix~A shows that for the case where $n_b$ is large the background region Poisson term can be replaced with a Gaussian one which results in a much faster algorithm.}.  If one can estimate the size of possible deviations from local linearity, {\em e.g.}, due to resonance contributions, then these can be incorporated into the likelihood via $\sigma_y$.  
The profile likelihood can then be used to obtain the significance of any excess of events (see, {\em e.g.}, Ref.~\cite{kyle}) and/or to set limits~\cite{rolke}.  See Appendix~A for a more detailed discussion on the likelihood.

In the previous paragraph the nominal background PDF is taken to be linear but in principle any background PDF can be used.  For a low-statistics search the linear approximation will be sufficient; however, for cases where the sample sizes are large, then the uncertainty due to deviations from local-linearity may be large compared to the statistical uncertainties.  
All that is required to use this method is that based on the observed yields in the sidebands, an estimate for the expected yield in the signal region can be obtained.  If a non-linear background PDF is chosen as nominal, then $x$ represents the scaling factor between the expected background yields in the sidebands and signal region, which may no longer be simply the ratio of the size of the regions.  The $\sigma_y$ term is again the uncertainty in the scale factor (independent of how $x$ is defined).  See Sec.~5 for discussion on non-linear background PDFs.  For the remainder of the method overview, I will assume a nominal background PDF of local linear.  

Appendix~B provides a study of the deviation from local linearity due to resonance contributions.  These deviations are a function of the width of the resonance, $\Gamma$,  the fraction of the total PDF near $m($test$)$ due to the resonance, and of $\sigma(m)$ and $x$.  The conclusions are as follows: 
\begin{itemize}
\item for $\Gamma/\sigma(m) > 20$, even if the resonance makes up close to 100\% of the PDF the choice $x=1$ is still local-linear to about 10\% (wide resonances are {\em safe});
\item for $\Gamma/\sigma(m) < 5$, the deviation from local-linear is large unless the resonance contribution is small (narrow resonances must be vetoed);
\item for  $5 < \Gamma/\sigma(m) < 20$, the local-linear approximation is valid at the 10\% level for moderate resonance contributions, but not valid if the resonance is dominant.
\end{itemize}  
Assuming a $\mathcal{O}(10\%)$ value is chosen for $\sigma_y$, then wide resonances can effectively be ignored (including those that have yet to be discovered).  If nothing is known about the possible size of a contribution from a narrow resonance, then the region near the resonance mass should be vetoed.  If some limits can be placed on the size of a resonance contribution, then this veto may not be required.  Such limits must be determined in each analysis independently.
  
The key point is that narrow resonances are typically well known (including their branching fractions to many final states), whereas wide resonances are not well known.  The sideband approach allows the analyst to ignore wide resonances by accounting for their non-linear effects via the $\sigma_y$ term in the likelihood.  Narrow resonances likely must be vetoed, but there are few of these and their properties are well measured. 
Intermediate-width resonances are also accounted for automatically by $\sigma_y$ provided they do not dominate the local PDF.  Such cases should be rare and can be studied using alternative decays of the resonance or via the data directly using bins $\mathcal{O}(10\sigma(m))$ wide.

Other categories of background that have a broad peaking structure are also handled naturally in this approach.  The study in Appendix~B can be applied to non-resonant backgrounds with the same conclusions drawn: only a background whose peak is narrow relative to $\sigma(m)$ must be vetoed; all other backgrounds are accounted for via $\sigma_y$ (the analyst does not need to know what these are or account for them individually). {\em E.g.}, in the $B\to K^*\mu\mu$ search, partially reconstructed charm particle decays can be ignored, while any $J/\psi\to\mu^+\mu^-$ contribution must be vetoed.

\begin{figure}[]
\centering
\includegraphics[width=0.32\textwidth]{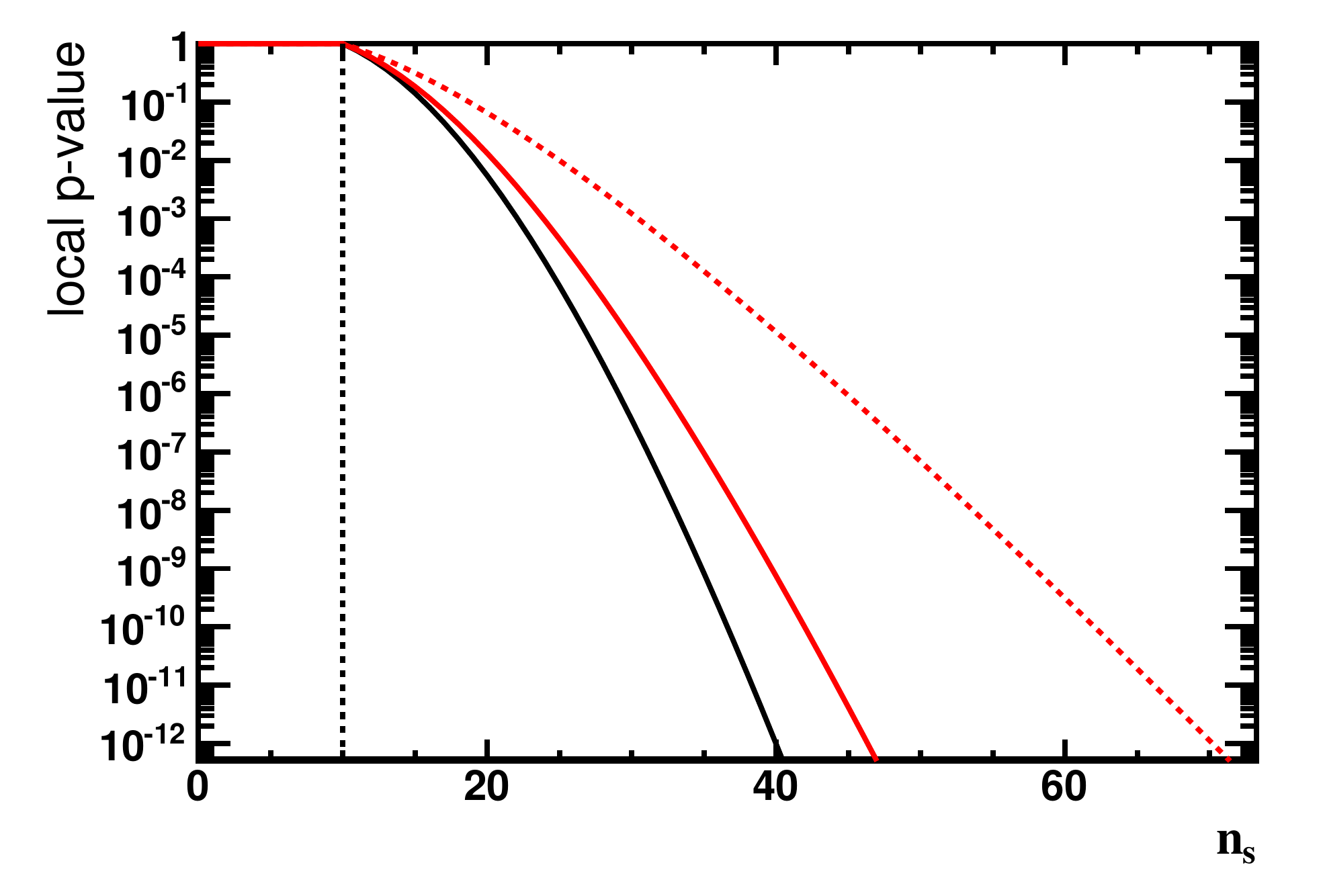}
\includegraphics[width=0.32\textwidth]{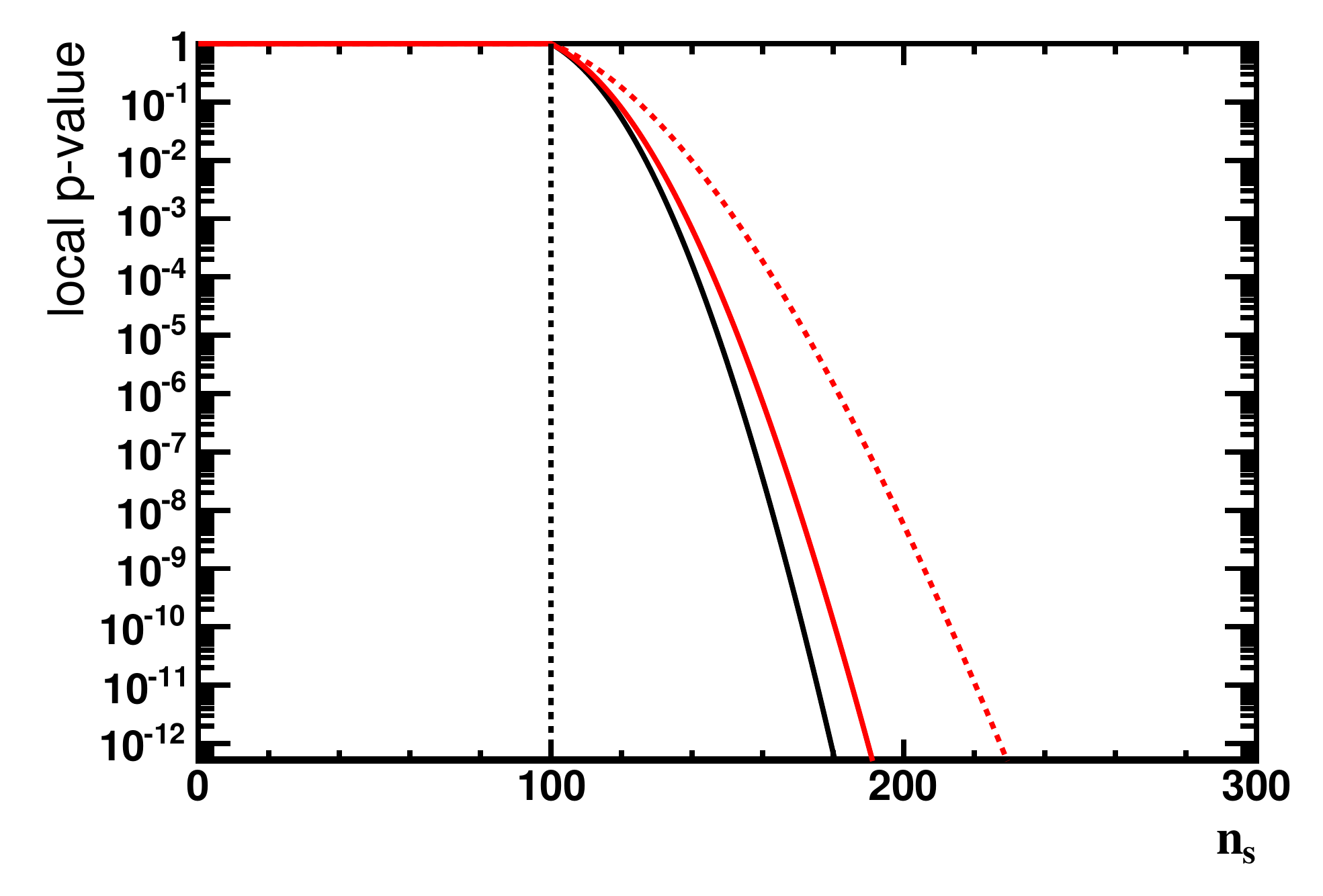}
\includegraphics[width=0.32\textwidth]{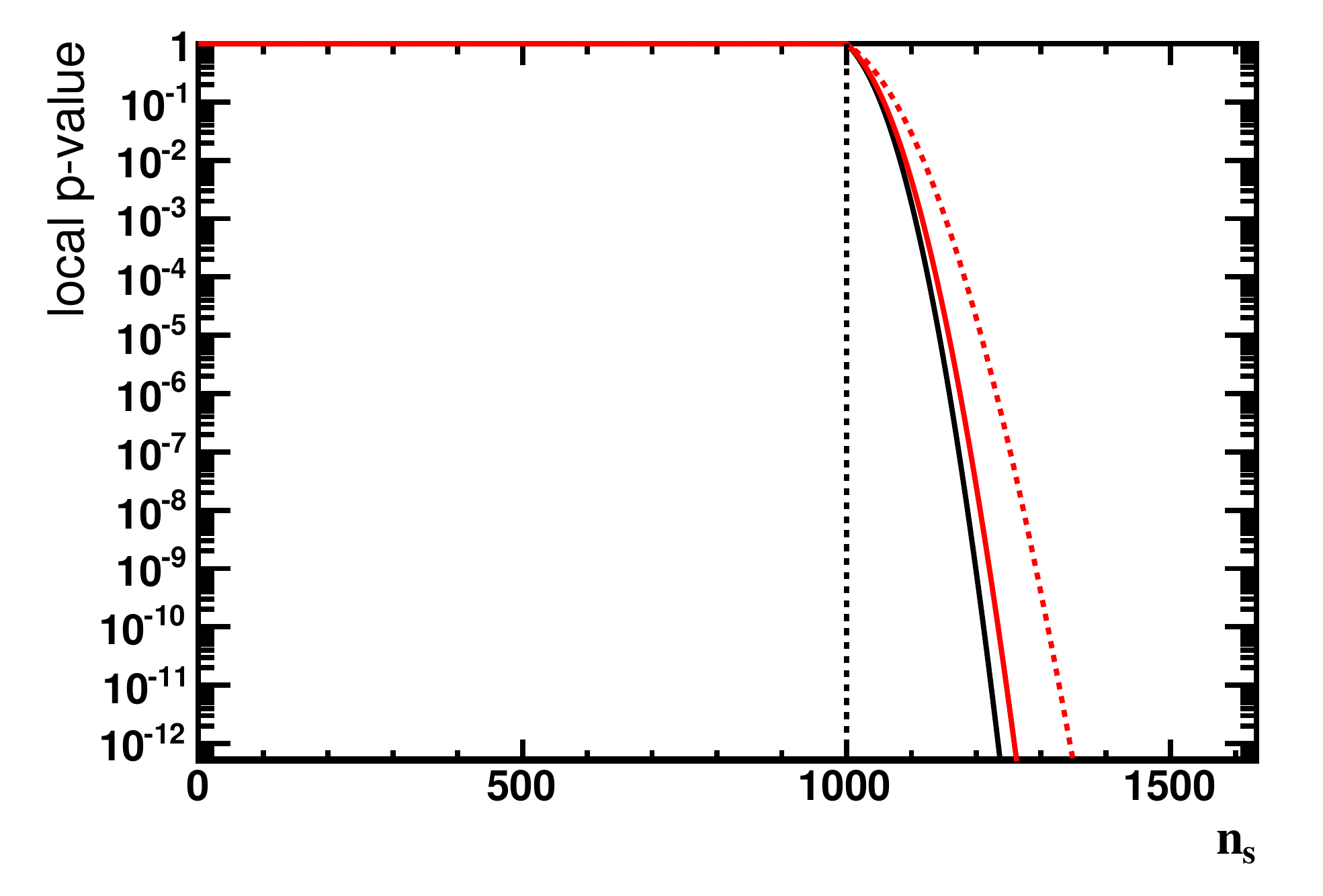}
\caption{\label{fig:x}
The $p$-value {\em vs} $n_s$ for (left) $b=10$, (middle) $b=100$, (right) $b=1000$ for (solid black) $b$ known with no uncertainty, (dotted red) $x=1$ and (solid red) $x=5$.  
}
\end{figure}

The parameter $x$ should be optimized for each analysis. The larger $x$ is chosen to be, the less statistical uncertainty there is on the background rate; however, this also increases the size of the region in which local-linearity is assumed.  Figure~\ref{fig:x} shows a comparison of using sidebands to estimate $b$ to when $b$ is known (which I assume here is not possible; this is added for illustrative purposes to show the best possible performance which could be obtained if $b$ could be known).  Setting $\sigma_y/x = 0.1$ typically loses very little power (at most 10-20\% sensitivity to the signal rate relative to the ideal situation of a known background PDF), so a good rule of thumb would be to choose $x$ as large as possible such that $\sigma_y/x = 0.1$ is still valid. 

\subsection{Lifetime}

I now proceed to considering lifetime, $\tau$, information and assume that the background can be factorized into two components: (1) a prompt background where the signal candidate's children all originate from the same spatial point as the rest of the final state and (2) a displaced background where they do not.  
The lifetime PDF for type (1) is given by the detector resolution.  The lifetime PDF for type (2) is assumed to be unknown.  In principle it may be obtainable from the mass sidebands; however, looking at these sidebands is forbidden in a blind analysis when the mass is unknown.  Furthermore, in many cases the displaced background data will be sparse making obtaining a reliable estimate of its PDF (which likely depends on $m$) impossible even if the data is not blinded. 

If the lifetime distribution of both prompt and displaced backgrounds is the same in the signal and background regions, then a non-parametric two-sample goodness-of-fit test can be used to test the null hypothesis at each $m($test$)$.  
Appendix~C discusses several such distribution-free tests, while Appendix~D shows the results of applying these to the toy data set described above.  
The conclusion of these studies is that such tests are far from optimal and require introducing the assumption that the displaced background $\tau$ PDF does not vary with $m$ (which may not be true and would be difficult to validate).

An alternative (and simple) approach is to define two $\tau$ regions at each $m($test$)$: (1) a prompt region ({\em e.g.}, $\tau < 3\sigma(\tau)$) and (2) a displaced region ($\tau > 3\sigma(\tau)$).  The mass sidebands in each region can be used to estimate the expected background rate.  The joint likelihood is defined as 
\begin{eqnarray}
L(n_s^{\rm prompt},n_b^{\rm prompt},n_s^{\rm displ},n_b^{\rm displ},x|s^{\rm prompt},b^{\rm prompt},y^{\rm prompt},s^{\rm displ}, b^{\rm displ},y^{\rm displ}) = \hspace{1.0in}\\ \nonumber
\hspace{0.0in}L(n_s^{\rm prompt},n_b^{\rm prompt},x|s^{\rm prompt},b^{\rm prompt},y^{\rm prompt})\cdot  L(n_s^{\rm displ},n_b^{\rm displ},x|s^{\rm displ},b^{\rm displ},y^{\rm displ}),
\end{eqnarray}
{\em i.e.}, the likelihood is the product of the likelihoods from each region individually (each obtained in the same manner as discussed in the previous section).  The profile likelihood $(\Lambda)$, which is the ratio of the maximum likelihood with the signal rate fixed to the maximum likelihood with the signal rate free to vary (see Appendix~A), for this two-region test is the product of the profile likelihoods from each region. 
This is true because no assumption is made about the lifetime of the new particle and so there is no assumed relationship about the signal yields in each region.
The asymptotic distribution of $-2\log{\Lambda}$ is a $\chi^2$ with two DOFs.  
Appendix~D shows that this simple approach, which only uses the lifetime information to determine which region each candidate falls in, is nearly optimal except when $\tau \sim \sigma(\tau)$.  

The approach presented in this paper only assumes that variation in the lifetime distribution {\em vs} mass is slow enough that the linear approximation holds in both the prompt and displaced regions.  The possible deviations from linearity are accounted for by $\sigma_y$.  In this paper I use a constant $\sigma_y$ in the notation but $\sigma_y$ is allowed to depend on $m$ and different values of $\sigma_y$ can be used in the prompt and displaced regions.  This is also true of $x$: it can be chosen to be different values for different test masses and in the prompt and displaced regions. Using variable $\sigma_y$ and $x$ does not introduce any additional complexity.  

Finally, I note that one can run both the two-region profile likelihood test and a two-sample test.  Appendix~D shows that this approach provides a small increase in performance in the region near $\tau \sim \sigma(\tau)$; however, there is an important assumption required to use the two-sample test.   
This approach requires that all lifetime PDFs are the same in the signal and background regions, {\em i.e.}, that locally the $\tau$ PDF is independent of $m$.  This may not be true for displaced backgrounds and would be difficult to validate unless the number of displaced-background candidates is large.  
Given that the gain in performance is small and the additional complexity introduced into the analysis is non-negligible,  I conclude that unless one has a reason to expect $\tau \sim \sigma(\tau)$ and a method for validating the $\tau$ PDF $m$ dependence, that the two-region profile likelihood test is the best option.

While it may be surprising to the reader that such a simple approach performs so well, it is often the case that analysis-specific information provides great power.  In this case, factorizing the background into two categories using the known detector resolution is key to this search.
Note that this procedure allows the analyst to completely ignore the $m$ and $\tau$ distributions for any background that does not form a narrow peak relative to $\sigma(m)$.  One instead relies on the fact that such backgrounds will populate the signal and sideband regions and that locally their PDFs are approximately linear.  The possible deviation from linearity is incorporated into the likelihood via $\sigma_y$.  
Narrow peaking backgrounds, {\em e.g.}, $J/\psi\to\mu\mu$ in the $B\to K^*\mu\mu$ search, must be vetoed in each lifetime region unless it can be shown (or known) that they can only contribute to one region. Such backgrounds should be few and easy to identify.  

\section{$p$-Values}

The full procedure involves first determining the local $p$-values at each $m($test$)$, then obtaining the global $p$-value of the most significant excess observed in the full mass range.  
An outline of the procedure is as follows:
\begin{itemize}
\item The full mass range is scanned in steps of $\sigma(m)/2$.  
\item An independent test is performed for each $m($test$)$ where the signal and background regions are defined as $|m-m({\rm test})| < 2\sigma(m)$ and $3\sigma(m) < |m({\rm test})-m| < (2x+3)\sigma(m)$, respectively ($x$ should be optimized for each analysis).
\item The signal and background regions are divided into prompt and displaced sub-regions.  The quantities $n_s^{\rm prompt},n_b^{\rm prompt},n_s^{\rm displ},n_b^{\rm displ},x$ are the inputs to the local two-region likelihood.  
\item The profile likelihood provides the local test statistic (see Appendix~A).  Since the lifetime of the new particle is unknown, there is no assumed relationship between the signal rate in the prompt and displaced regions. 
\item For each $m($test$)$ an approximate local $p$-value is obtained using the asymptotic formula for the profile likelihood test statistic.   As discussed below, the accuracy of the asymptotic formula for this test only needs to be good enough to properly select the most significant local excess.  
\item The minimum $p$-value from the full mass scan is selected as the test statistic to which a significance is to be assigned.  This approach is motivated by the assumption that there is at most one new particle contributing to the data sample.  
\end{itemize}
The global $p$-value is not the minimum local $p$-value as this would ignore the so-called trials factor, or look elsewhere effect, induced by the fact that a large number of tests have been performed.  

Appendix~E discusses the fact that this test has been designed such that it can be run on 10M data sets for $\mathcal{O}(1000)$ $m($test$)$ values in about 2 hours on a single CPU core.  This permits determining the global $p$-value without the need for using asymptotic formulae.  The only reliance on asymptotic formulae is in selecting the minimum local $p$-value; therefore, the accuracy of the asymptotic formula only needs to be sufficient to properly select the most significant local excess.  There is no need to interpret the local $p$-values as probabilities under the null hypothesis.

\begin{figure}[]
\centering
\includegraphics[width=0.49\textwidth]{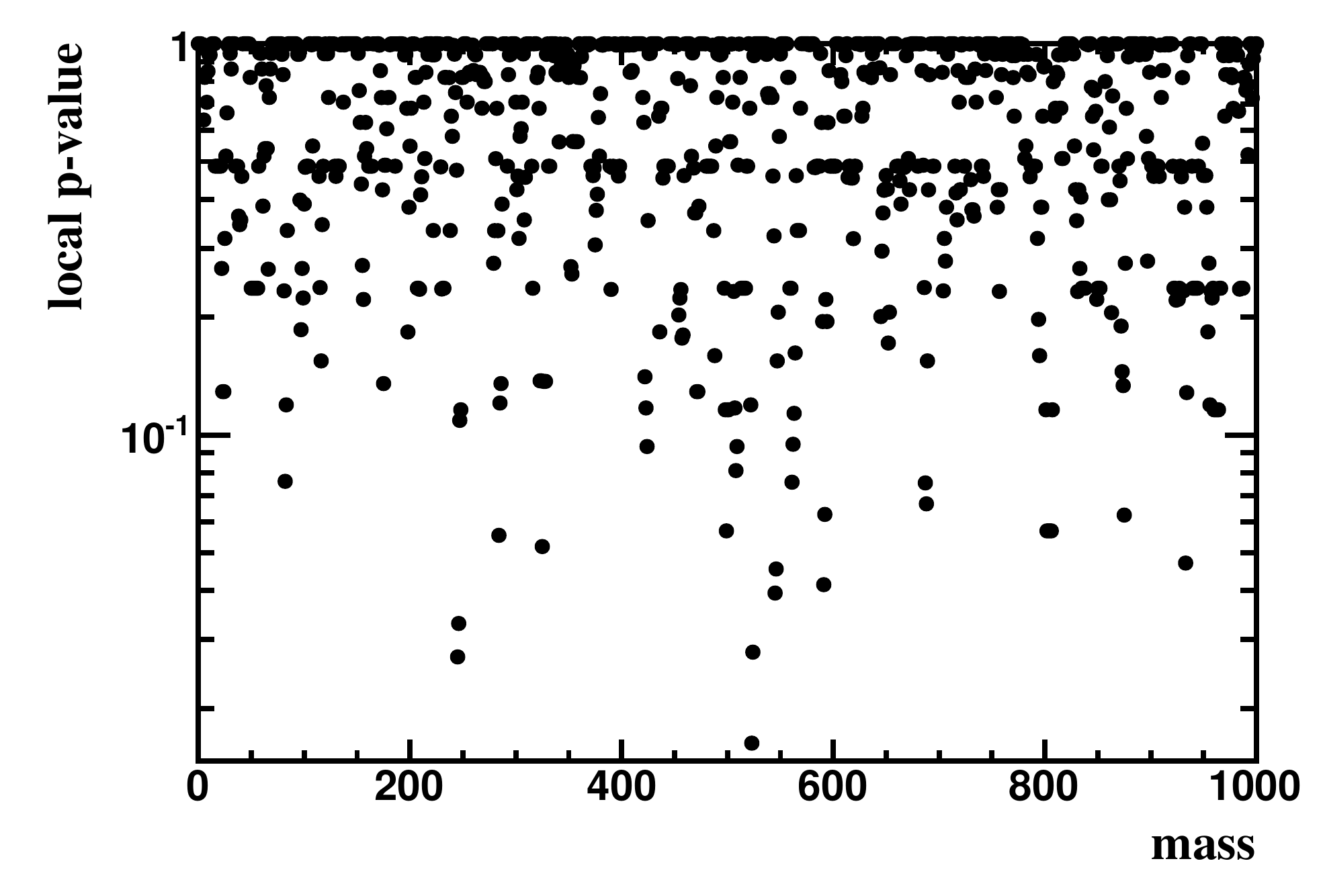}
\caption{\label{fig:local_p} 
Local $p$-values obtained from a single simulated data set ($x=1$).
}
\end{figure}

\begin{figure}[]
\centering
\includegraphics[width=0.49\textwidth]{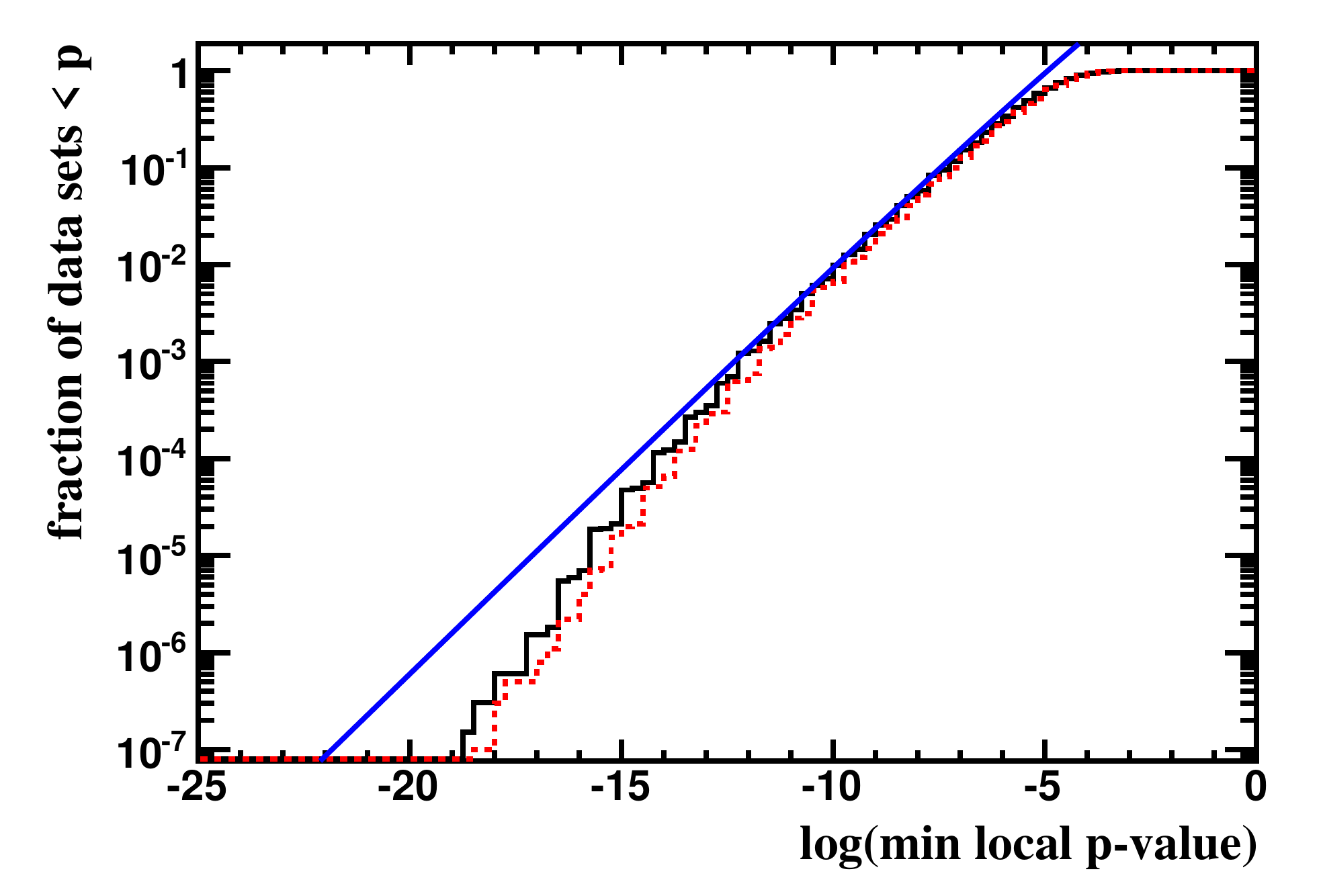}
\caption{\label{fig:global_p} 
Cumulative distribution of minimum local $p$-values obtained using simulated (black) toy-model events compared to the (blue) asymptotic expectation\cite{lee}.  Two variations of the test are shown (both use $x=1$): (solid) no uncertainty in the relationship between the signal and background regions and  (dashed red) $\sigma_y/y = 0.1$.  The discrepancy at very small $p$-values between the asymptotic and solid distributions is due to low statistics in each local test region.  The data sets were generated using a PDF that is local linear so it is expected that using $\sigma_y/y = 0.1$ is an overestimate of the scaling uncertainty which results in an underestimate of the significance. 
}
\end{figure}

Figure~\ref{fig:local_p} shows the local $p$-values obtained from a single simulated toy data set.  Since the step size in $m($test$)$ is smaller than the signal and sideband regions, neighboring tests are correlated.  This produces a jagged-looking distribution.  The test mass near 520 has the minimum $p$-value so would be selected to be assigned a significance in this data set.  
To obtain the significance the procedure is as follows:
\begin{itemize}
\item Obtain the minimum local $p$-value from the data as described above.
\item Get an approximate null PDF from the data by ignoring the region near the most significant excess and obtaining a smooth PDF from the remaining data (interpolating into the removed region).  Below I show that the details of how this is done are not important.
\item Generate an ensemble of simulated data sets from the PDF from the previous bullet point.  The global $p$-value is the fraction of simulated data sets that have a minimum local $p$-value less than that observed in the data. 
\item The number of generated data sets determines the statistical uncertainty on the global $p$-vaule.  The most likely outcome, no evidence for a new particle, will require only $\mathcal{O}(100)$ data sets.  To confirm $>3\sigma$ requires $\mathcal{O}(1000)$ while $>5\sigma$ requires $\mathcal{O}(10^7)$.  
\item Confirming $>5\sigma$ can be done on a single CPU core on a laptop.  In the unlikely (and exciting) event that not a single simulated data set in $10^8$ has a minimum $p$-value less that that of the data, the asymptotic formula from Ref.~\cite{lee} can be used to obtain an approximate significance (if one is desired).
\end{itemize}
Figure~\ref{fig:global_p} shows the cumulative distribution of minimum local $p$-values obtained from 10M toy data sets.  One can see that to obtain a global $3\sigma$ in this example requires a local $p$-value of about $e^{-13}$.  The trials factor then is $\mathcal{O}(400)$ which is roughly the width of the full mass range divided by $\sigma(m)$ (that is 500 in this example).    
The asymptotic distribution provides an underestimate of the significance in this example for small $p$-values due to the small sample sizes used in each local test.  Note that the true PDF used in the toy model is local-linear and so setting $\sigma_y/y = 0.1$ is an overestimate of the local non-linearity.  In this example, such an overestimate produces only a minor shift (towards lower significance; it produces a conservative estimate).  

\begin{figure}[]
\centering
\includegraphics[width=0.49\textwidth]{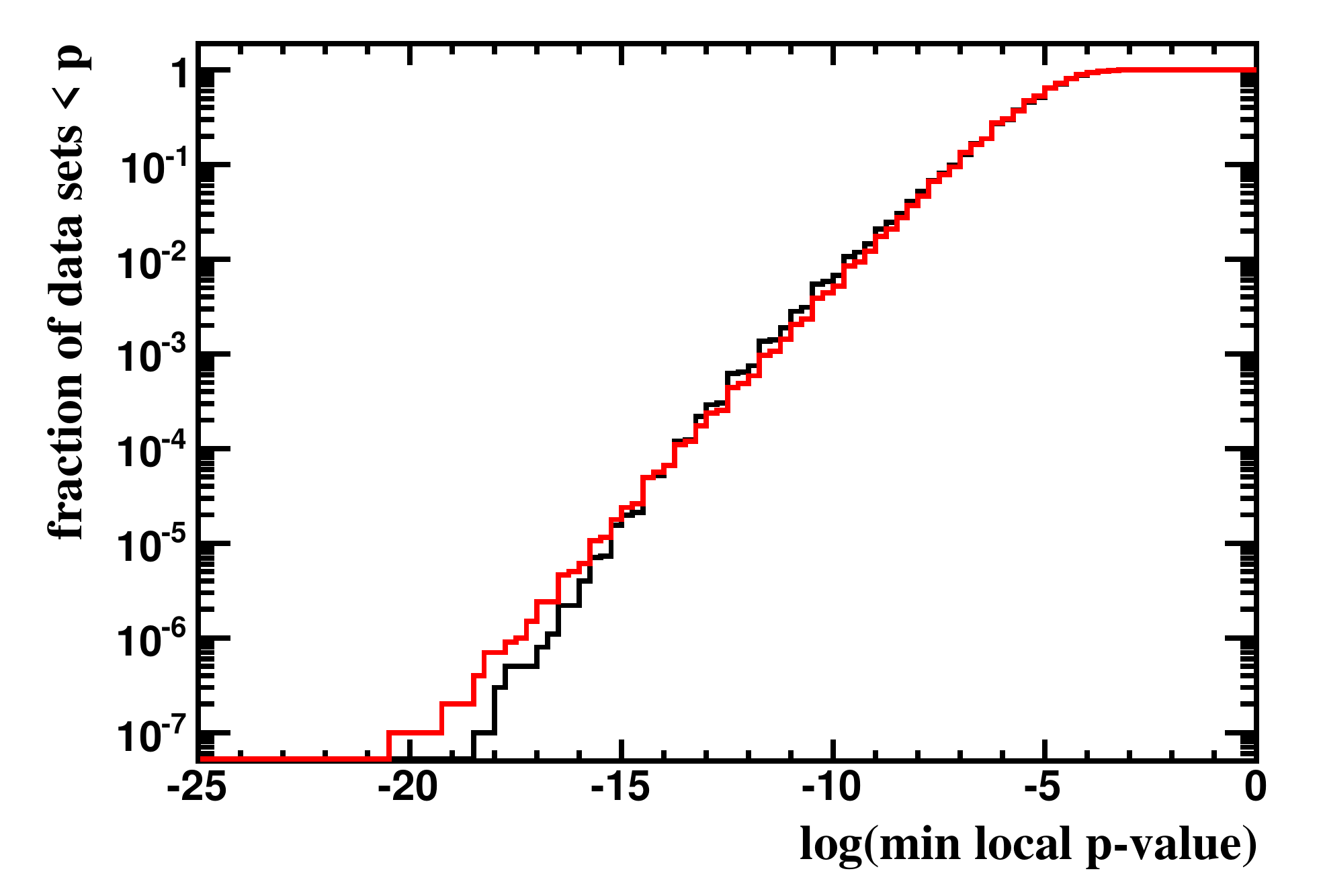}
\includegraphics[width=0.49\textwidth]{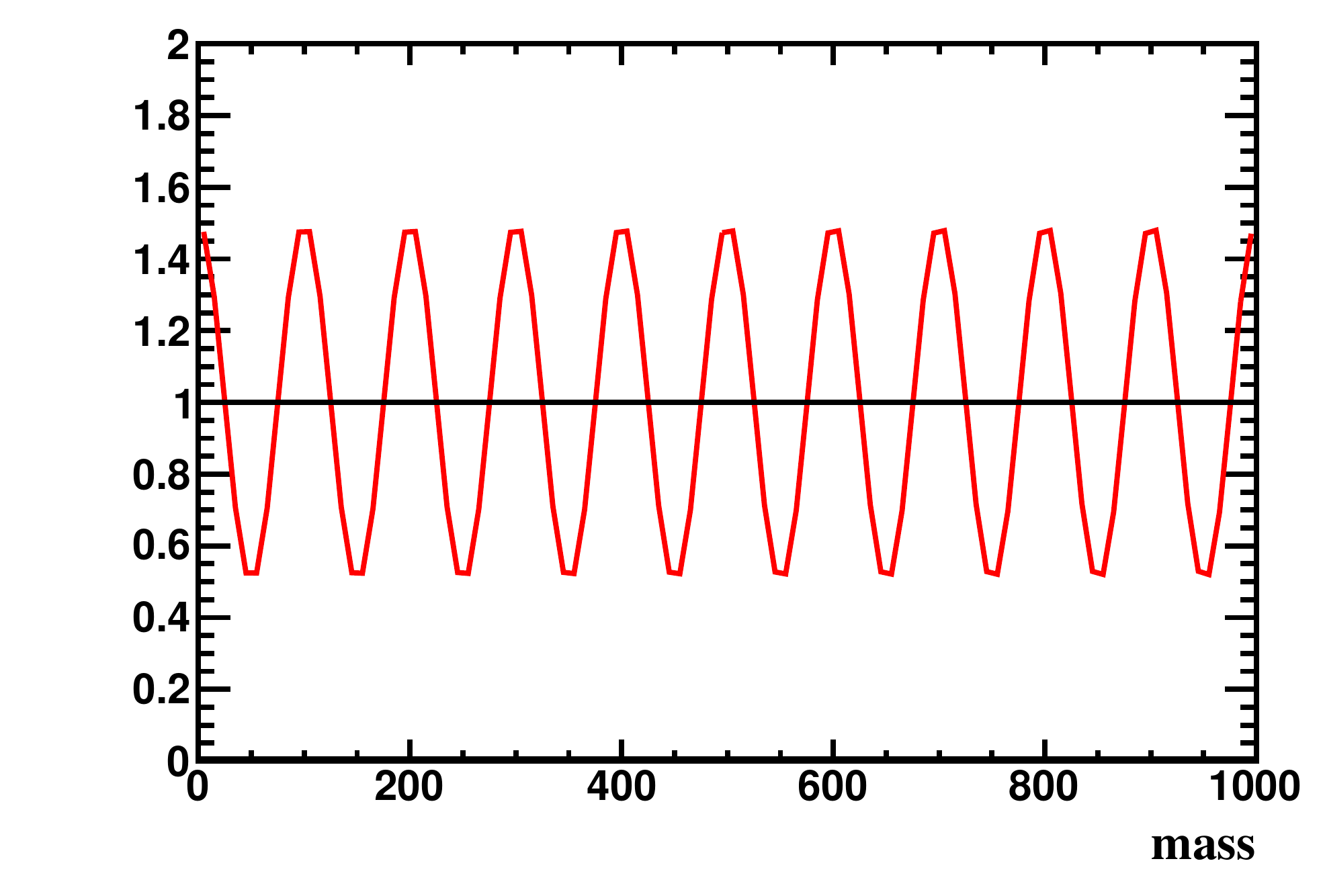}
\caption{\label{fig:global_p_dist} 
Cumulative distribution of minimum local $p$-values obtained using simulated toy-model events distributed according to the PDFs shown at right (black is the nominal linear model, while red is a non-monotonic alternative).  The distribution of local $p$-values has little dependence on how the data is distributed as expected.  This distribution is driven by the size of the mass range being searched and $\sigma(m)$.
Both results here are shown for $\sigma_y/y=0.1$ and $x=1$.
}
\end{figure}

Figure~\ref{fig:global_p_dist} shows that the cumulative $p$-value distribution obtained using an alternate (highly non-monotonic) PDF but the same sample size and detector resolution (and, thus, test mass values).  There is very little dependence on the data PDF.  
This means that the exact PDF used to generate the pseudo data sets is not important.
For example, one could simply bin the data in a histogram with wide bins (relative to $\sigma(m)$), remove the most-significant excess region, and use spline interpolation to obtain a background PDF (which interpolates into the removed most-significant region).
This will be accurate enough to produce a reliable cumulative $p$-value distribution.

To summarize: One can confirm up to $>5\sigma$ without using asymptotic formulae in about an hour on a laptop.   Assignment of a significance beyond this level can be done using the asymptotic formula as an estimate (if this is desired).  Figure~\ref{fig:global_p_dist} demonstrates that even an oscillatory PDF is handled naturally (I did not input any knowledge of this PDF to the method except that $\sigma_y/y=0.1$) provided the features of the PDF are wide relative to $\sigma(m)$.

\section{Limits}

Upper limits are to be set as a function of $m$ and $\tau$ for all $m$.  In the event that a globally significant excess is observed, the analyst can additionally perform PDF-based fits to determine estimators for the new particle mass and lifetime. 
The two-region profile likelihood can be used to set the limits after making the following modifications to the likelihood function:
\begin{itemize}
\item for each value of $\tau$, there is a relationship between the number of signal events expected in the prompt and displaced regions;
\item a Gaussian term is added to the likelihood to account for uncertainty (due to detector efficiency) in the fraction of signal expected in the prompt and displaced regions;
\item another Gaussian term is added to the likelihood to account for uncertainty in the absolute detector efficiency scale of the signal (most likely relative to a normalization decay mode).
\end{itemize}
The likelihood for each $(m({\rm test}),\tau)$ is then given by
\begin{eqnarray}
\label{eq:likeul}
L(n_s^{\rm prompt},n_b^{\rm prompt},n_s^{\rm displ},n_b^{\rm displ},x,\tau|\ldots) &=&  L(n_s^{\rm prompt},n_b^{\rm prompt},x|\epsilon \cdot s \cdot f,b^{\rm prompt},y^{\rm prompt}) \nonumber \\
          &\times&   L(n_s^{\rm displ},n_b^{\rm displ},x| \epsilon \cdot s \cdot (1-f),b^{\rm displ},y^{\rm displ}) \nonumber \\
          &\times& \mathcal{G}(f,f_{\rm MC}(\tau),\sigma(f)) \times  \mathcal{G}(\epsilon,\epsilon_{\rm MC}(\tau),\sigma(\epsilon)),
\end{eqnarray}
where $f$ is the fraction of signal events in the prompt region with expected value from simulation $f_{\rm MC}(\tau)$ and uncertainty $\sigma(f)$, and $\epsilon$ is the efficiency (typically relative to some normalization reaction) with expected value from simulation $\epsilon_{\rm MC}(\tau)$ and uncertainty $\sigma(\epsilon)$.
The limits are then set by scanning the profile likelihood using the same method, including the handling of special circumstances, discussed in detail in Ref.~\cite{rolke} but using the likelihood given in Eq.~\ref{eq:likeul}
\footnote{This method actually returns a confidence interval whose lower limit may be $>0$.  Since the significance is discussed above, here I only study upper limits but the method will produce a lower limit as well.}.
For setting limits there is no need to generate 10M toy data sets; thus, I do not provide analytic solutions and, instead, use {\sc Minuit} to numerically scan the profile likelihood.

In the toy analysis I choose to normalize the new particle yield to the observed prompt yield in the full mass region.  This emulates the situation where the prompt background is dominantly a well-known SM process.  I take the ratio of the efficiency for detecting the new particle to the normalization process to be one.  In reality there will be some dependence of $\epsilon$ on $m$ and $\tau$ and there will be some candidates in the prompt region that do not come from the normalization mode, but these just scale the limits (and contribute to $\sigma(\epsilon)$) so I will not discuss them here.  

\begin{figure}[]
\centering
\includegraphics[width=0.49\textwidth]{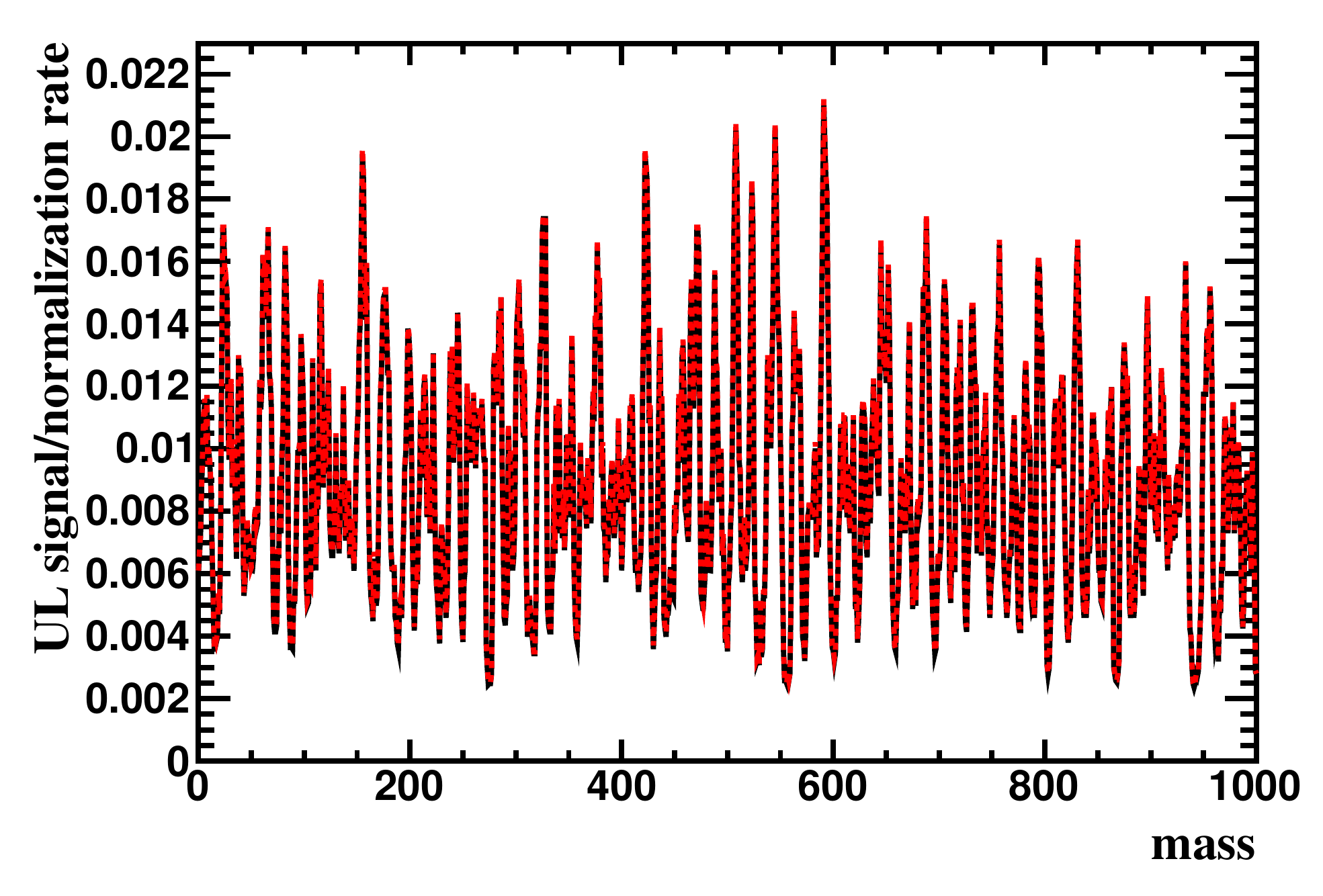}
\includegraphics[width=0.49\textwidth]{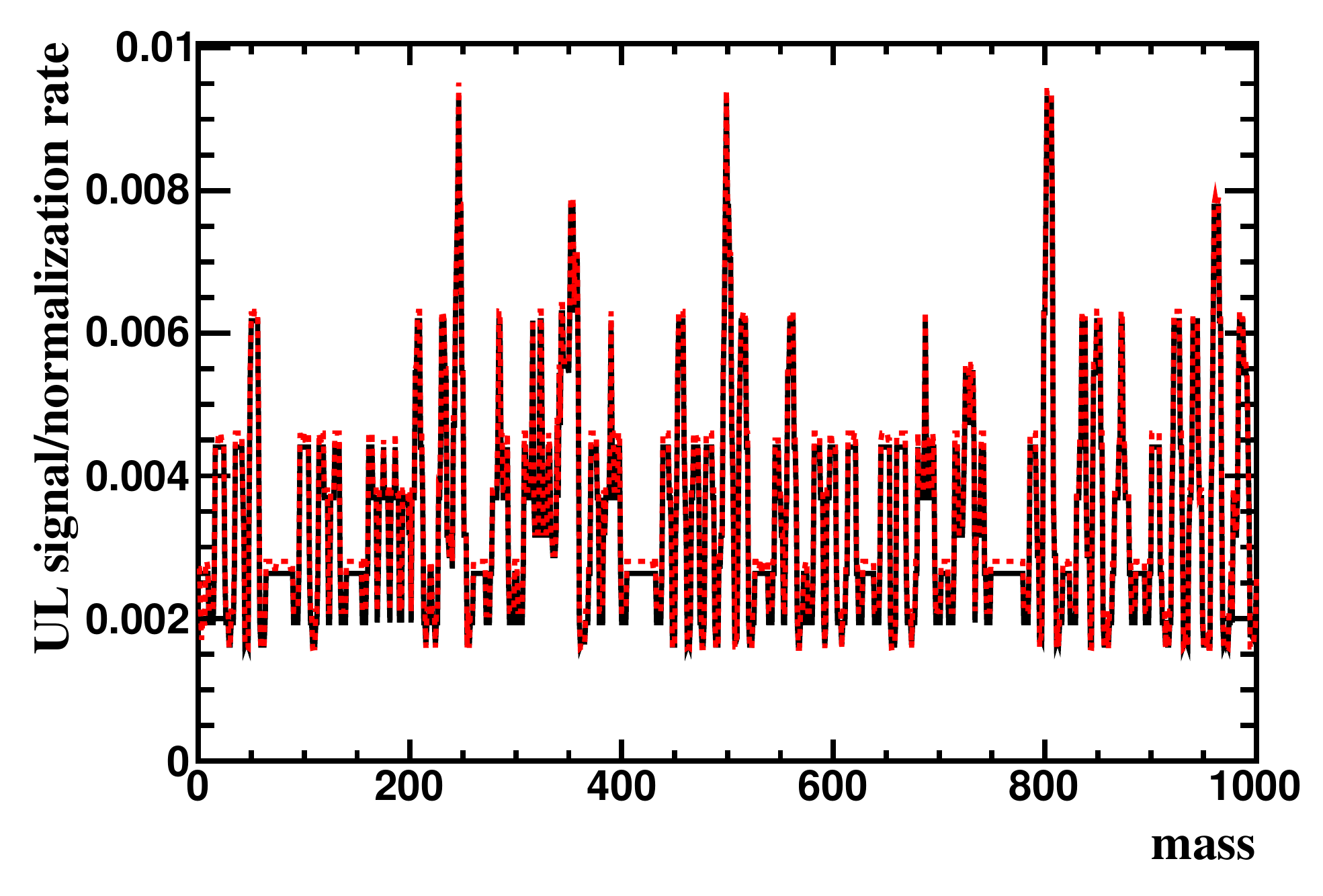}
\caption{\label{fig:comp2rolke} 
Comparison between limits obtained for a single simulated data set from (black) the TRolke class (only Poisson uncertainties) and (dashed red) the method discussed in this work (with $\sigma_y/y = 0.1, x=1$) for (left) test $\tau = 0$ and (right) $\tau = 1000\sigma(\tau)$.  In the left plot only the prompt region is used with TRolke, while in the right plot only the displaced region is used (TRolke only handles one region).  For very small and large $\tau$, using only one region is an excellent approximation to the full method presented here; thus, the agreement between TRolke and this method for these $\tau$ values is expected. 
}
\end{figure}

For the cases where the test $\tau$ is $\ll \sigma(\tau)$ or $\gg \sigma(\tau)$, to a good approximation only the prompt or displaced region matters.  Therefore, in such cases the one-region results obtained using the {\tt TRolke} class in {\sc ROOT}~\cite{ROOT} should be close to those produced here provided $\sigma_y/y$ is small.  
Figure~\ref{fig:comp2rolke} shows that this is the case.  With $\sigma_y/y = 0.1$, the limits returned by this method are slightly larger than {\tt TRolke} (which here takes the uncertainty on $b$ to be purely Poisson).  If $\sigma_y = 0$ then for small or large $\tau$ the limits produced by this method are the same as {\tt TRolke}.  

\begin{figure}[]
\centering
\includegraphics[width=0.49\textwidth]{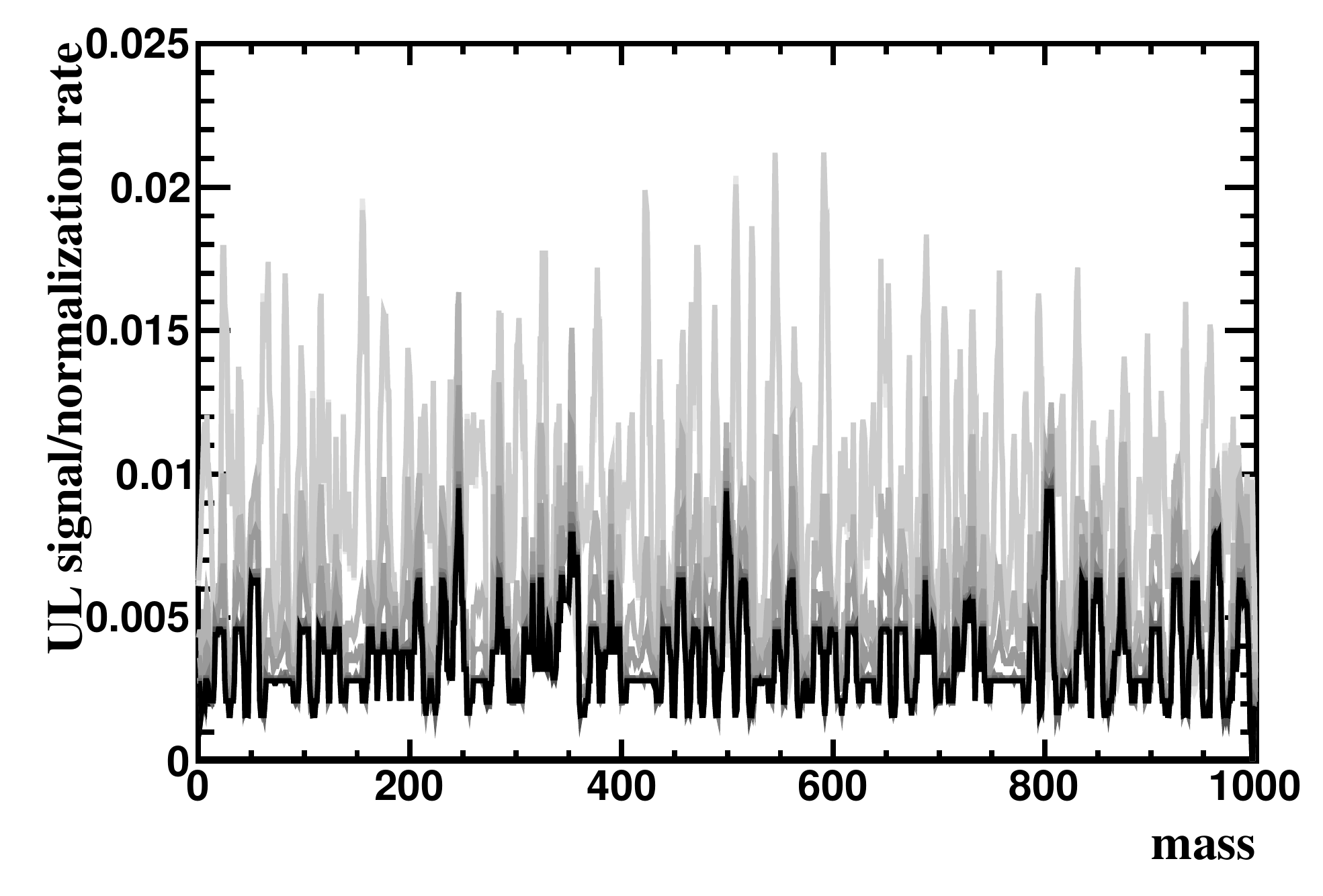}
\caption{\label{fig:tau_lims}  Upper limits from a single data set on the ratio of new particle production to the SM reaction for $\tau$ values (lighter to darker) $\tau = (0,1,5,10,50,100,500,1000)\sigma(\tau)$.
}
\end{figure}

\begin{figure}[]
\centering
\includegraphics[width=0.49\textwidth]{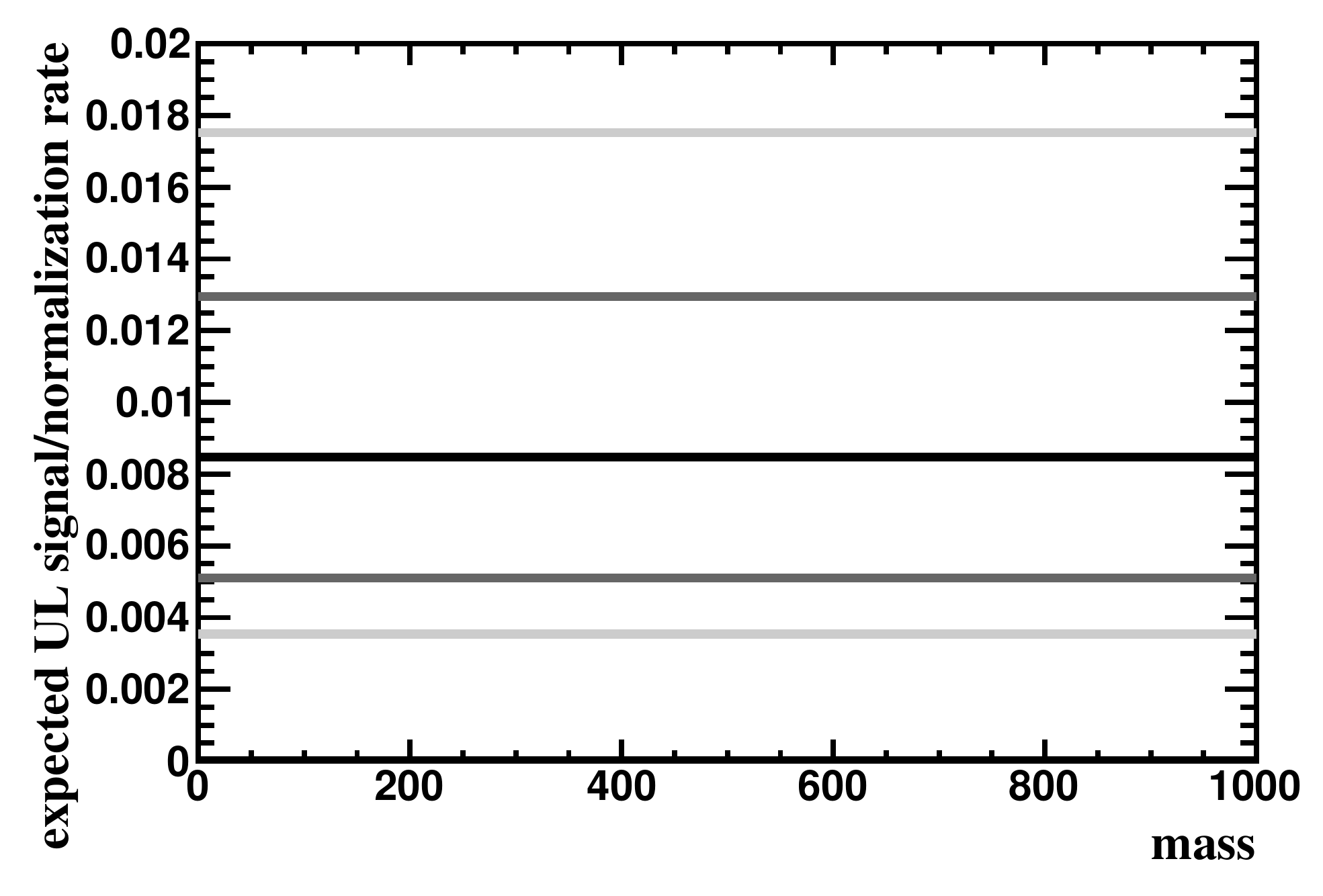}
\includegraphics[width=0.49\textwidth]{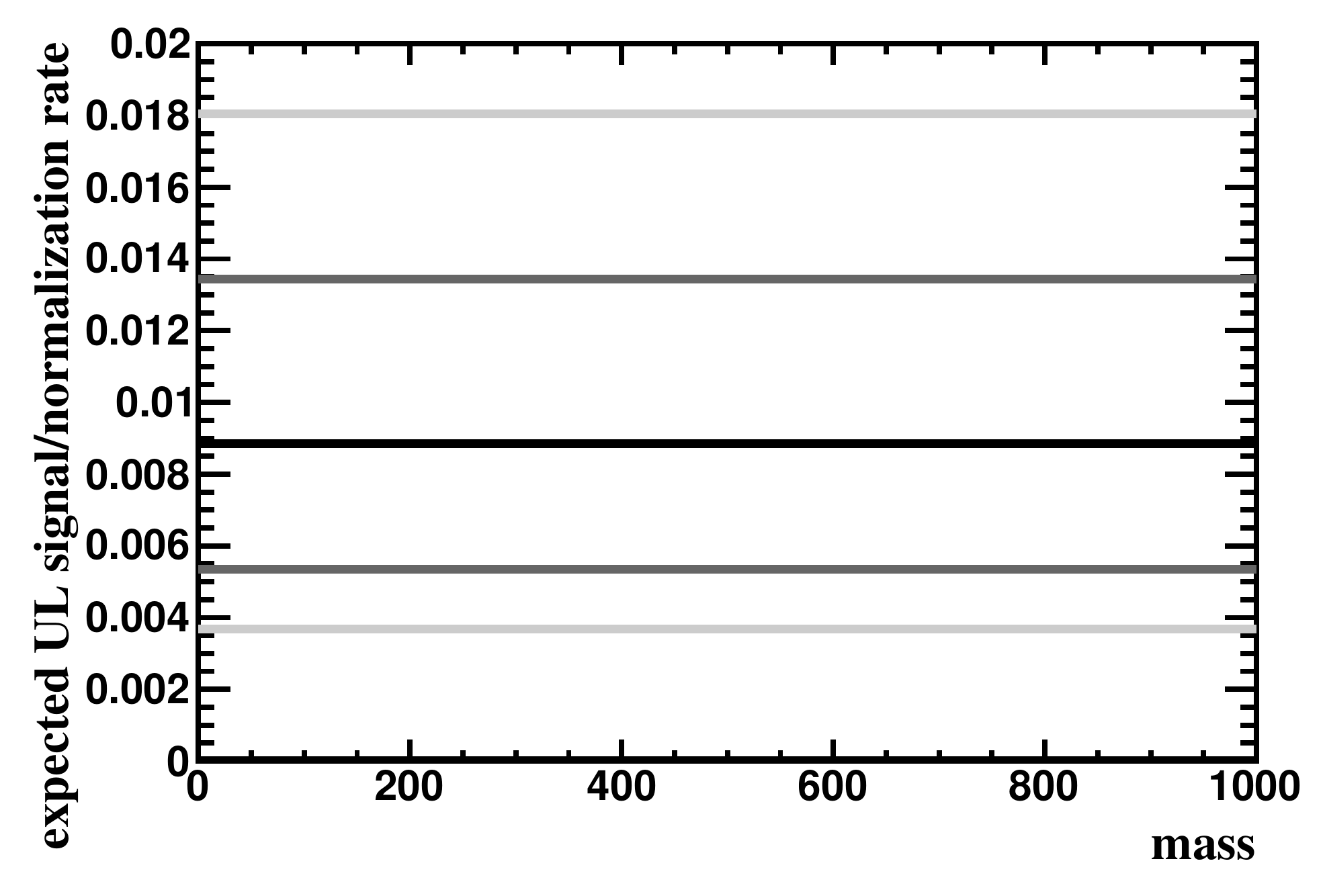}\\
\includegraphics[width=0.49\textwidth]{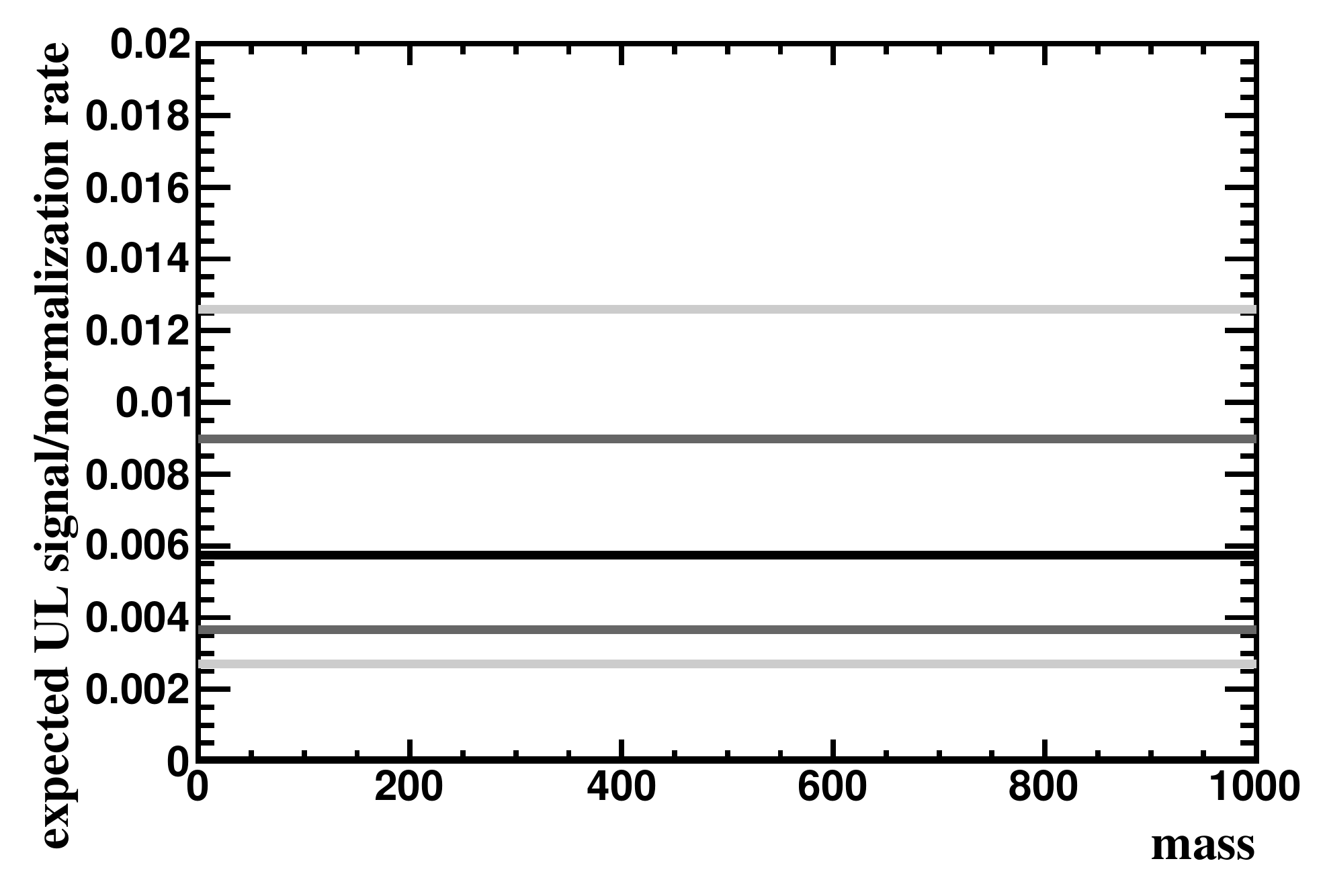}
\includegraphics[width=0.49\textwidth]{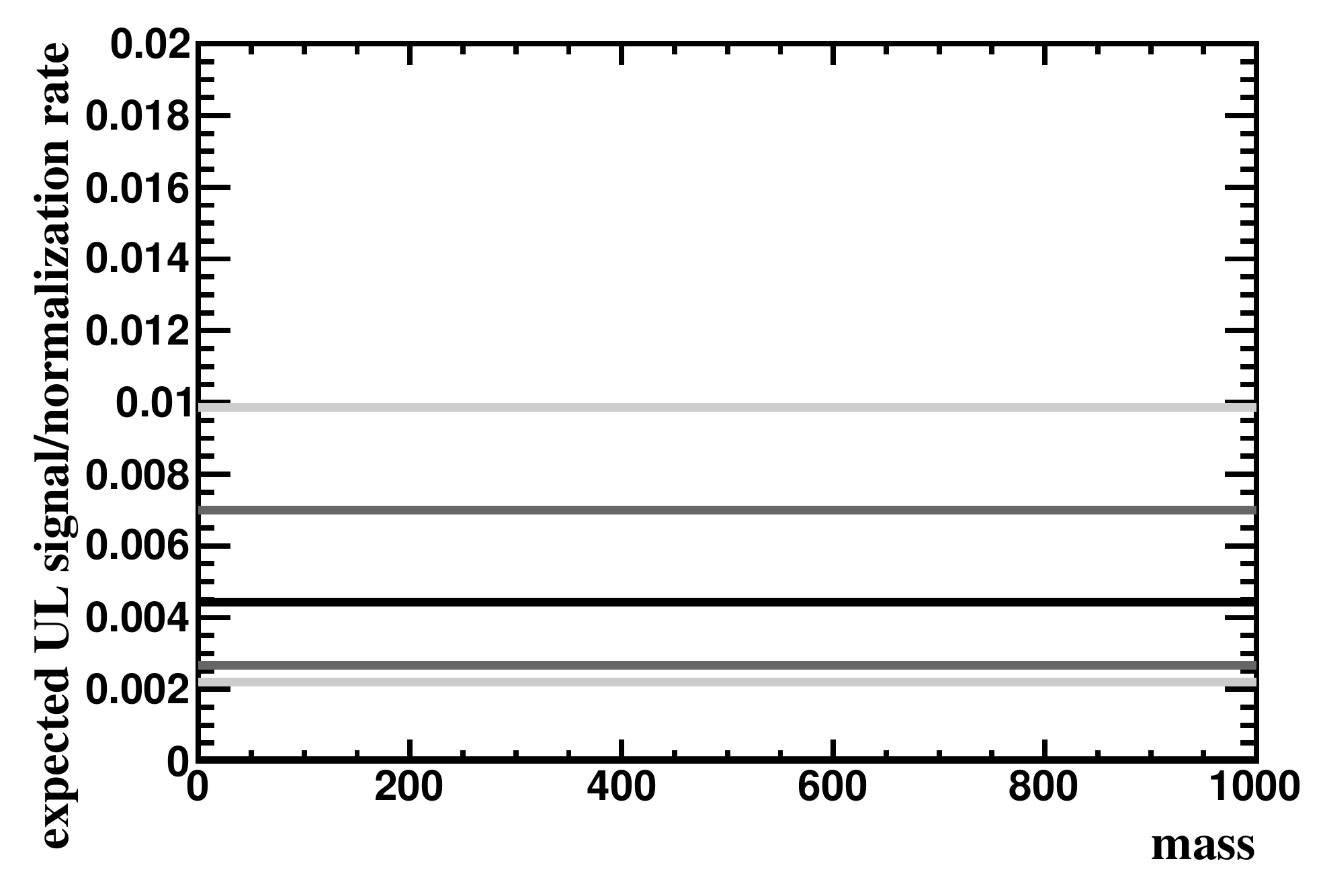}\\
\includegraphics[width=0.49\textwidth]{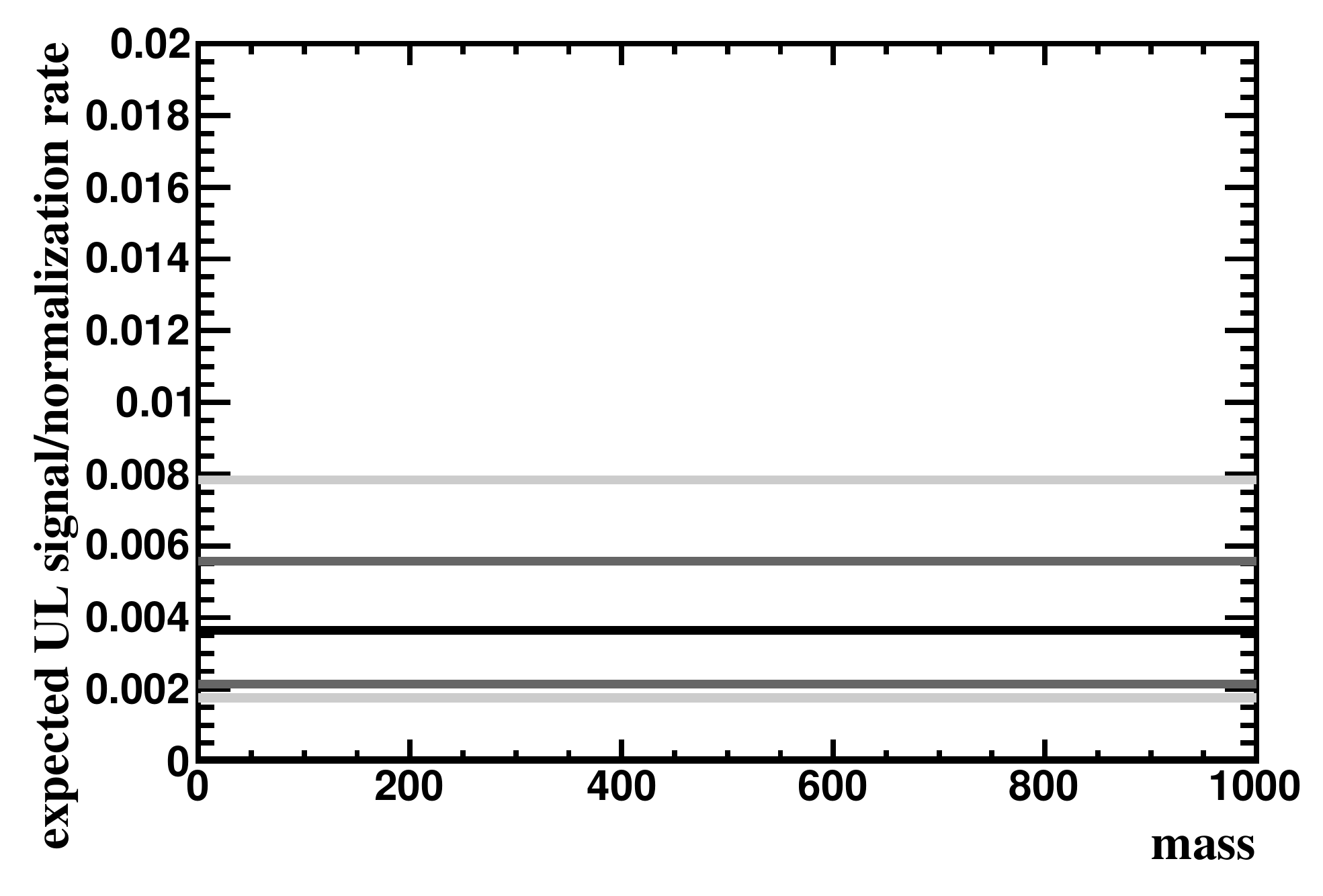}
\includegraphics[width=0.49\textwidth]{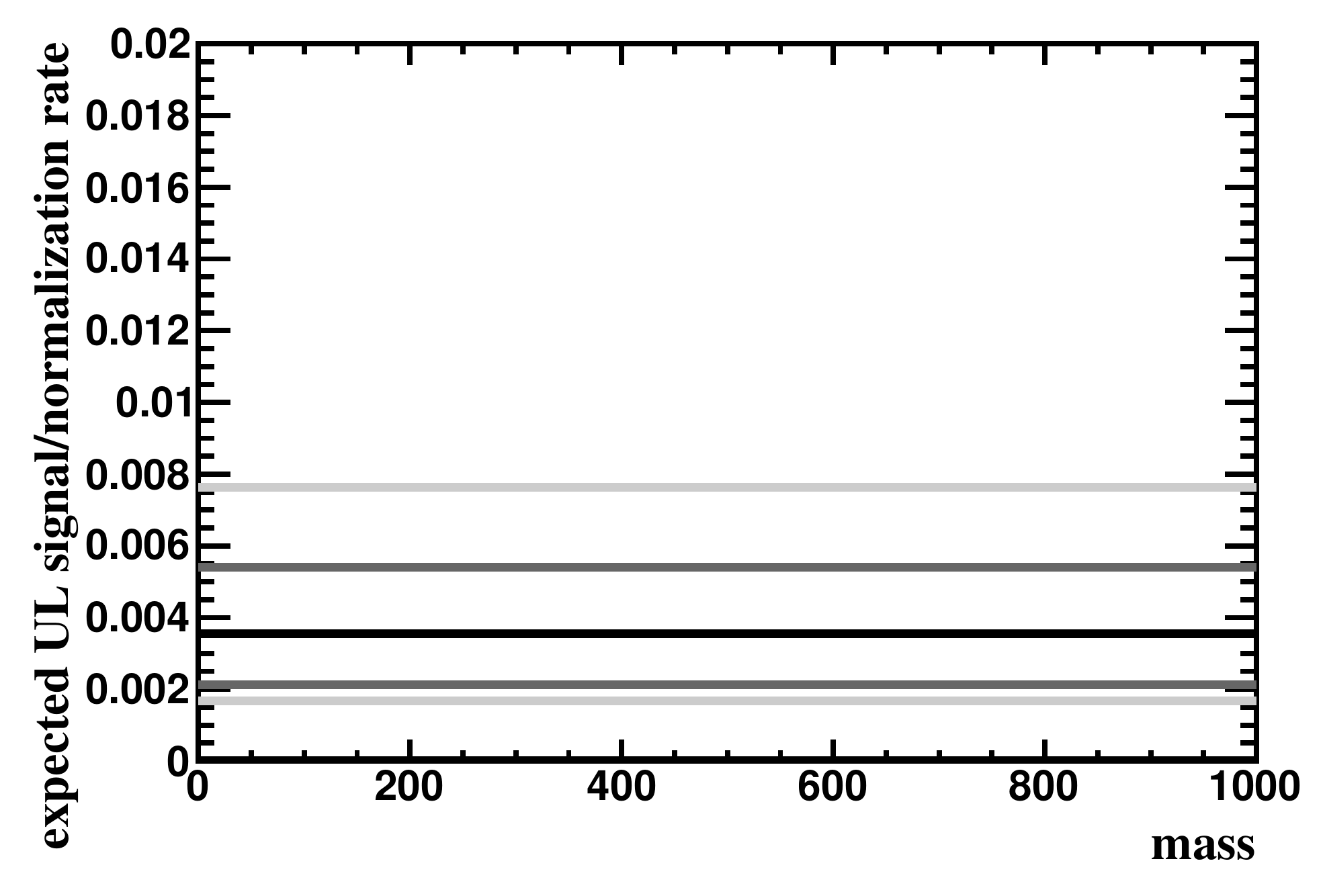}\\
\includegraphics[width=0.49\textwidth]{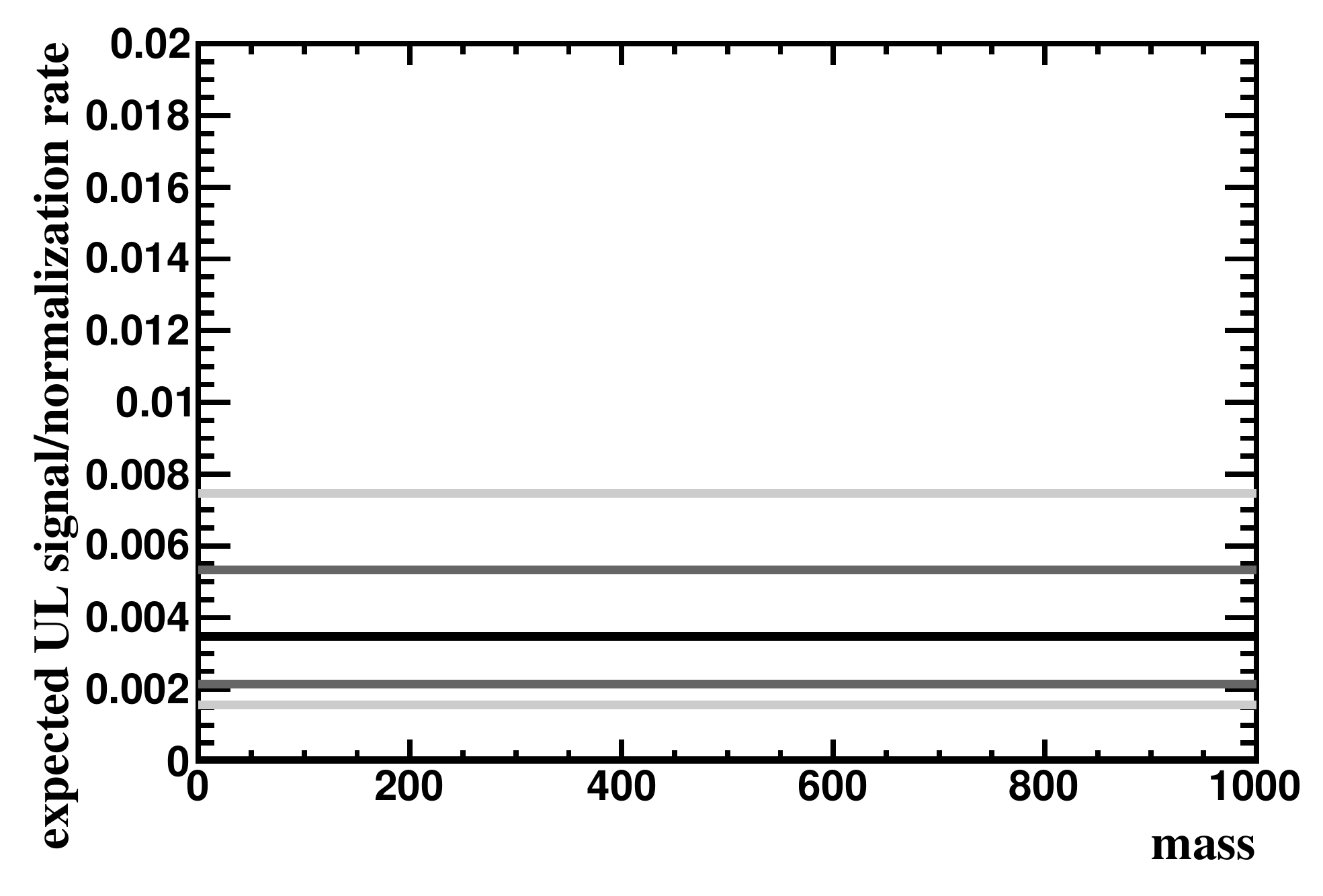}
\includegraphics[width=0.49\textwidth]{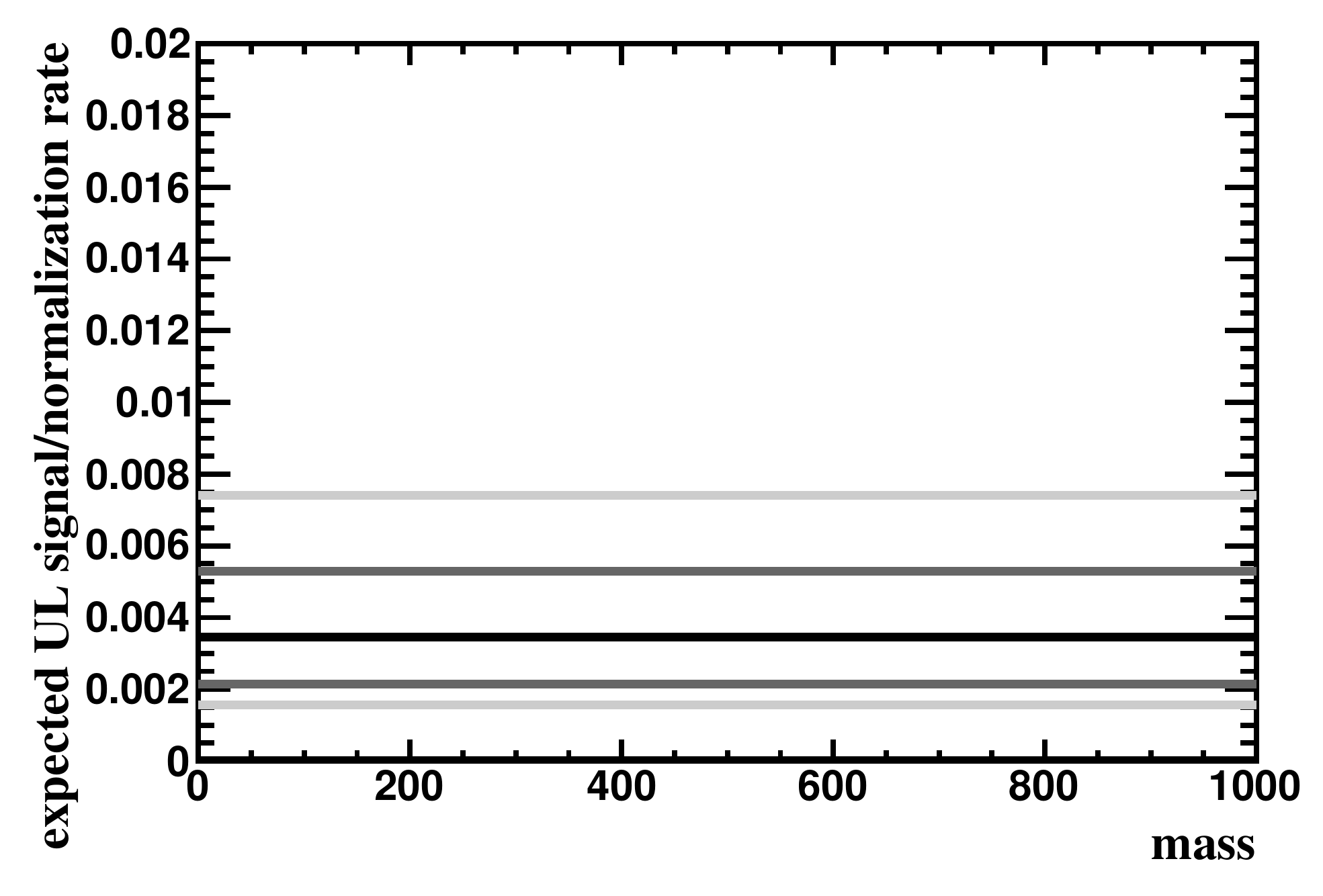}
\caption{\label{fig:brazil} 
Expected upper limits on the signal rate relative to the normalization mode {\em vs} $m$ for simulated toy-model data ($\sigma_y/y = 0.1, x=1$).  The lines are (black) mean, (dark gray) enclose the $1\sigma$ and (light gray) $2\sigma$ intervals.  From left to right, top to bottom $\tau = (0,1,5,10,50,100,500,1000)\sigma(\tau)$. 
}
\end{figure}

Figure~\ref{fig:tau_lims} shows how the limits depend on $\tau$ for a single toy-model data set.  Since in this example $n_b^{\rm prompt} > n_b^{\rm displ}$, the limits decrease with increasing $\tau$.  
Figure~\ref{fig:brazil} shows the dependence of the limits on $\tau$ expected (obtained from the 10M toy data sets generated).  For the toy-model data, whose PDF is uniform in $m$, the expected limits have no $m$ dependence.  In general this will not be the case.

\begin{figure}[]
\centering
\includegraphics[width=0.32\textwidth]{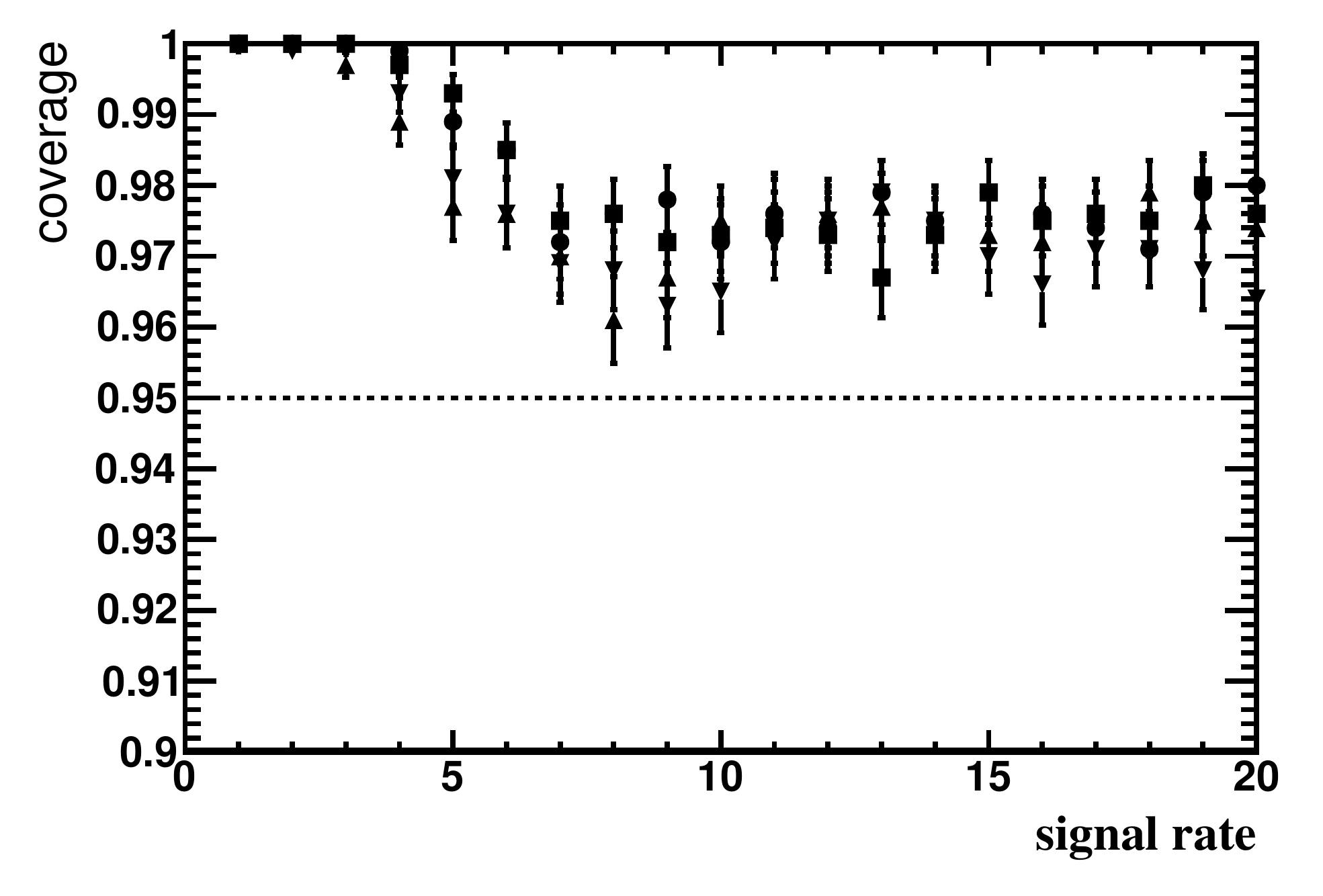}
\includegraphics[width=0.32\textwidth]{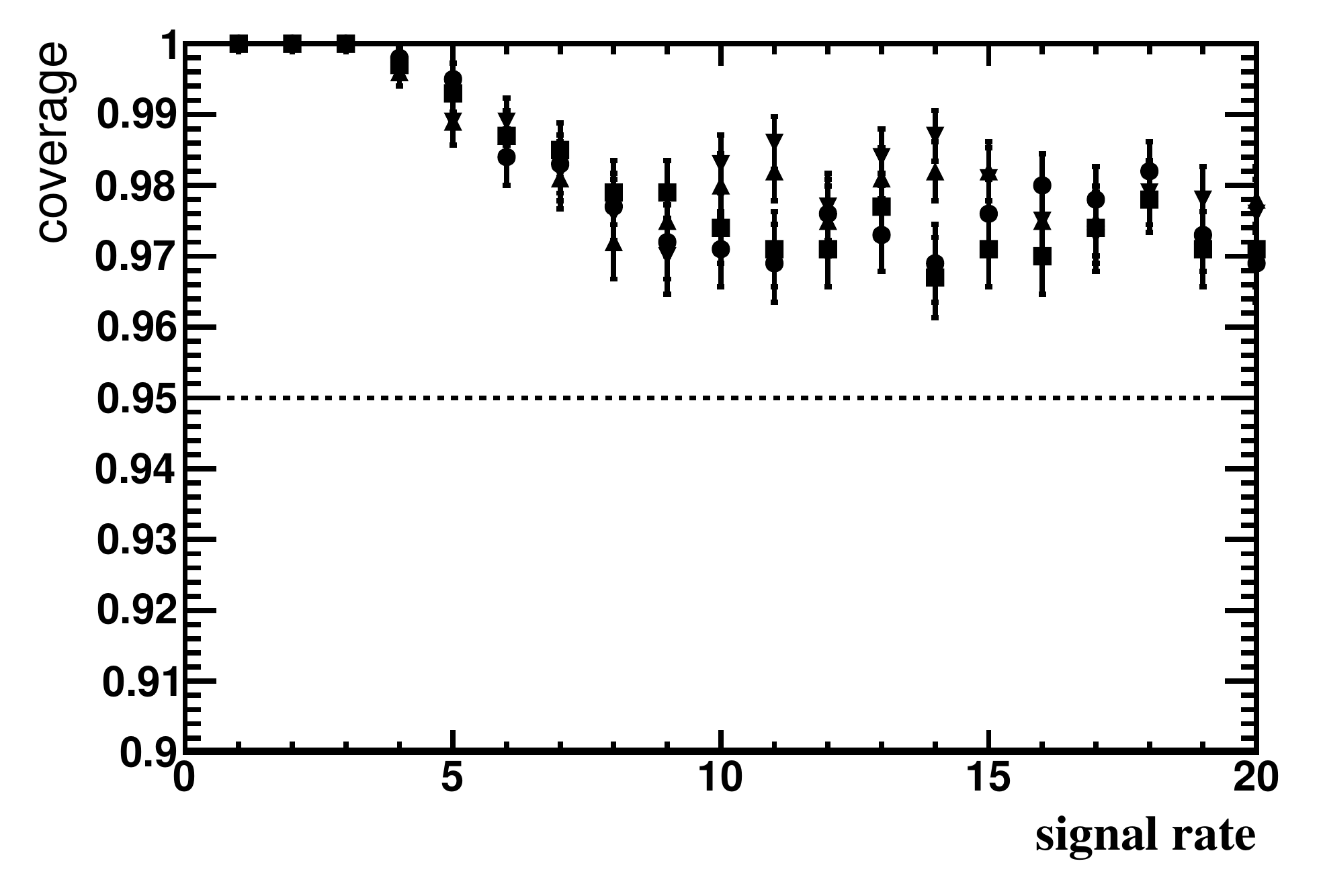}
\includegraphics[width=0.32\textwidth]{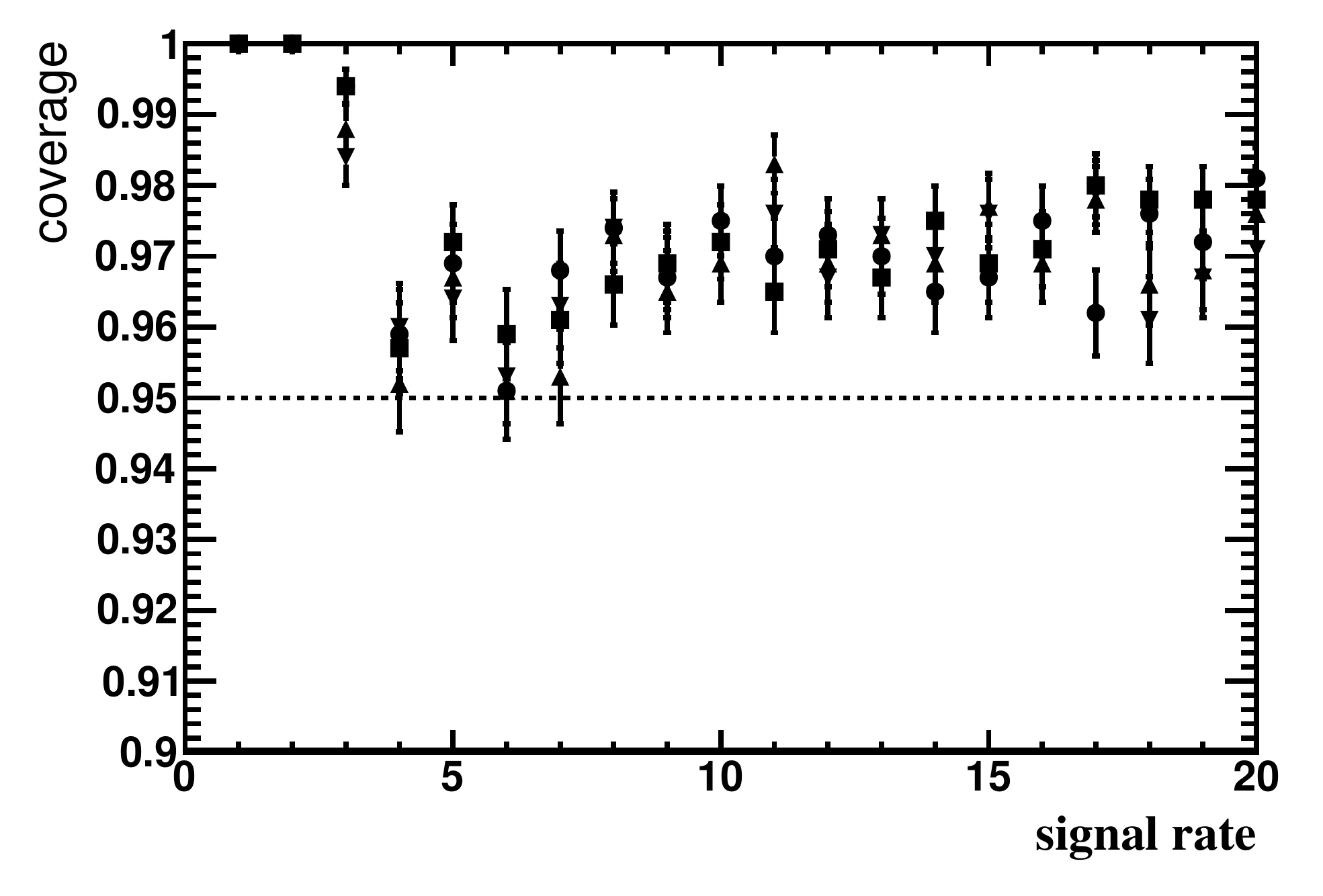}\\
\includegraphics[width=0.32\textwidth]{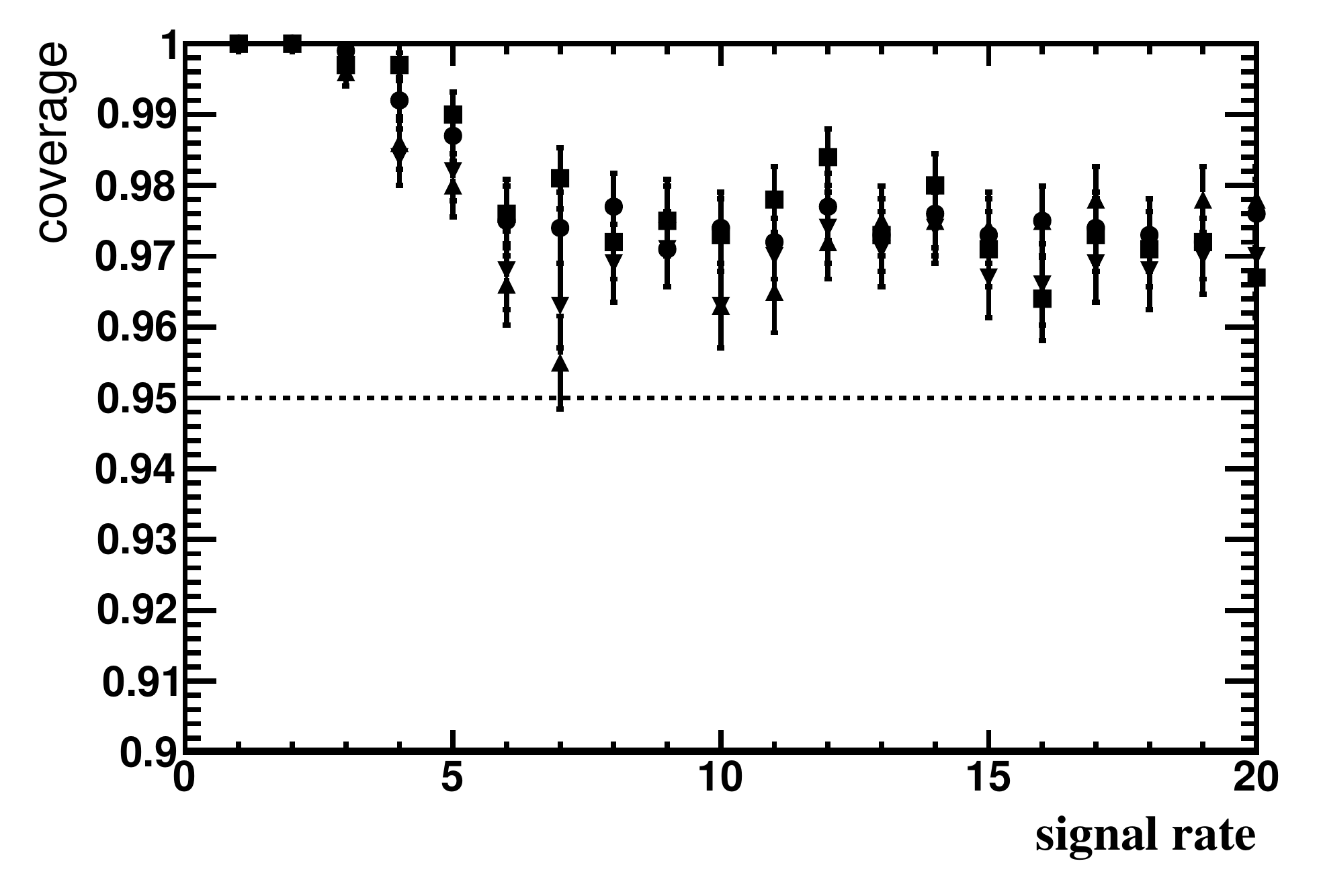}
\includegraphics[width=0.32\textwidth]{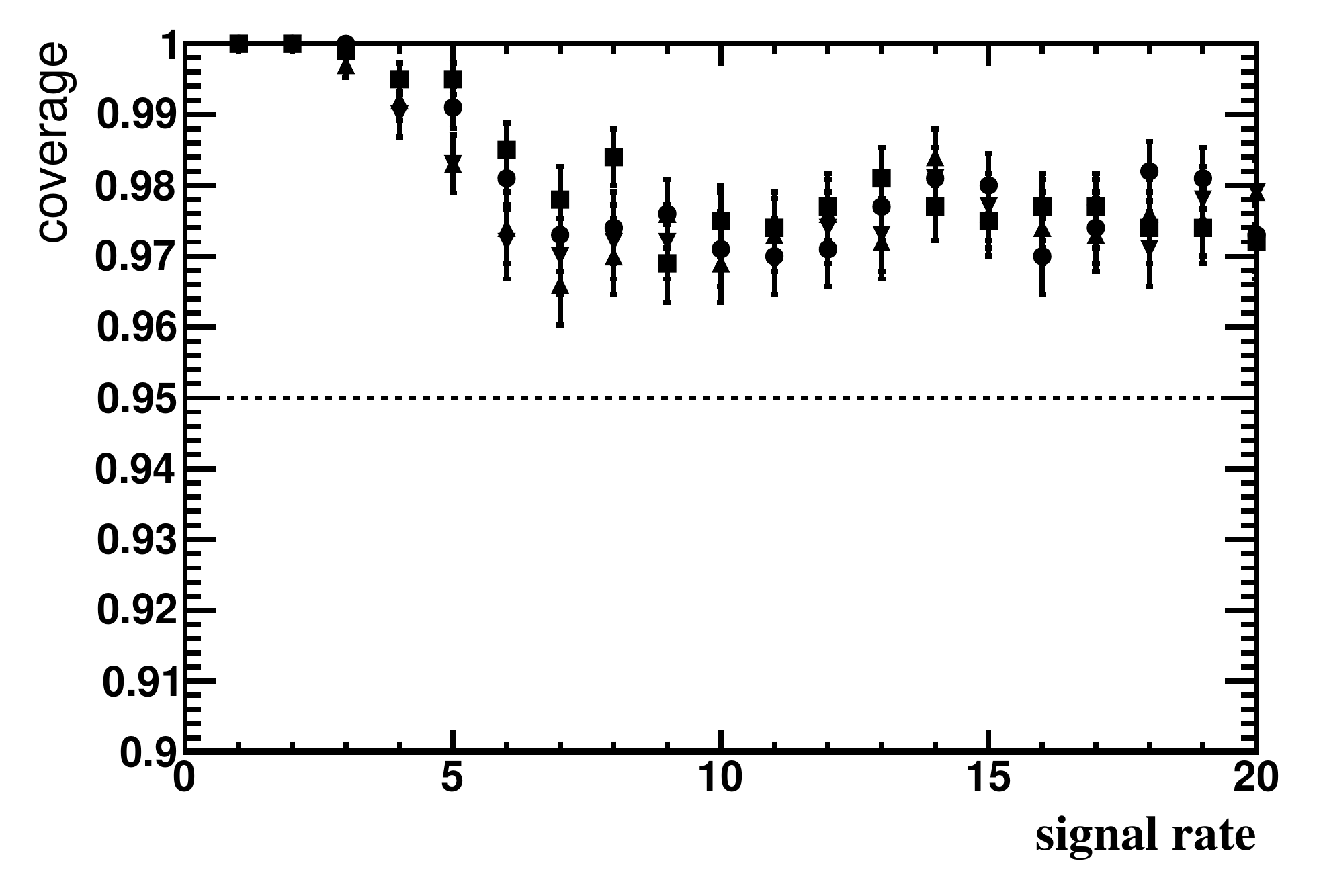}
\includegraphics[width=0.32\textwidth]{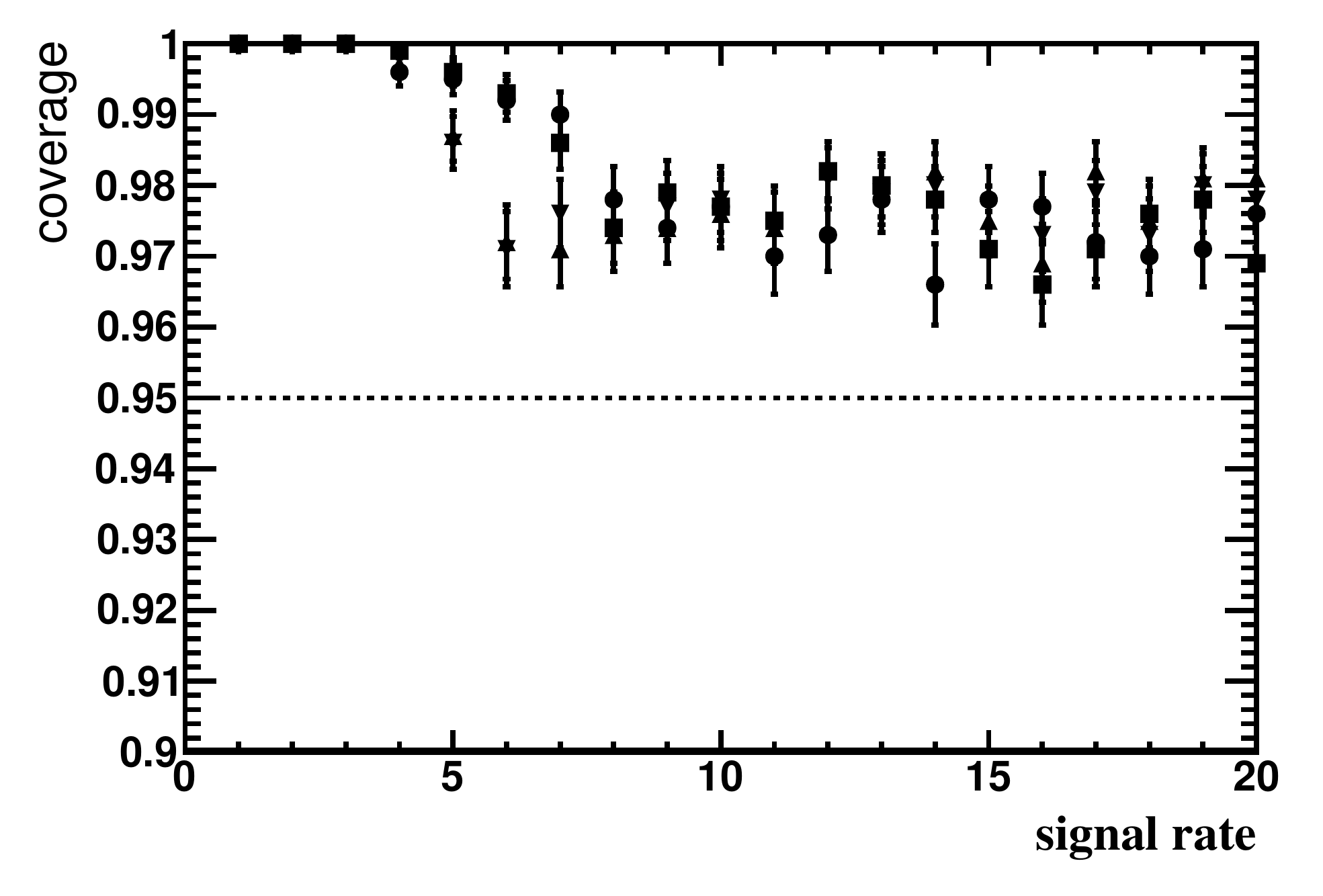}
\caption{\label{fig:cover} 
Coverage obtained using
(circles) $x=1,\sigma_y=0$, (squares)  $x=1,\sigma_y=0.1$, (up triangles)  $x=10,\sigma_y=0$, (down triangles)  $x=10,\sigma_y=0.1$ for
(top) $b^{\rm prompt}=10,b^{\rm displ}=0.1$ and (bottom) $b^{\rm prompt}=10,b^{\rm displ}=10$ and
(left) $\tau=0$, (middle) $\tau=\sigma(\tau)$ and (right) $\tau=10\sigma(\tau)$.
The dashed line shows the desired 95\% coverage. This method typically over covers by a few percent for the low-statistics cases shown here. 
}
\end{figure}

Figure~\ref{fig:cover} shows the coverage obtained for various configurations of this method and various background rates expected in the prompt and displaced regions.  The coverage properties are good except for at small $s$.  At small $s$ the method over covers but this is expected and unavoidable.  Otherwise, the method tends to over cover by only a few percent.  
Note that the small over coverage shown in Fig.~\ref{fig:cover} is due to the very low statistics of the samples studied.  The possible observations are discrete, and with such low statistics it is not possible to obtain perfect coverage.

\section{Extensions}

The method is not restricted for use where the local-linear approximation is applicable.  For example, in high-statistics searches, one could consider using multiple sideband regions for each $m($test$)$ and using spline interpolation (of whatever order is sufficiently constrained) instead of the local-linear approximation.  
Another approach could be to unblind some small fraction of the data and obtain estimates for the background PDF there.  
However the background estimates are obtained for each $m($test$)$, they will have some uncertainty $\sigma_y$ and so this method can still be applied using Eq.~A.10 in Appendix~A.   
There is no restriction to any particular local background shape.

\section{Summary \& Discussion}

This paper presents a simple likelihood-based approach for searching for a particle of unknown mass and lifetime in the presence of unknown non-monotonic backgrounds.  
Instead of exhaustively fitting the data with background PDFs containing hundreds of nuisance parameters, I propose to use the local-linear (or alternative local PDF) assumption and simple sideband counting.  Deviations from the nominal local PDF are parametrized via a single parameter $\sigma_y$ in the likelihood.  If $\sigma_y$ is overestimated, then only minor underestimation of the $p$-values is found.  This allows the analyst to concentrate on a small number of possible narrow peaking backgrounds and effectively ignore all other contributions.

The lifetime information is used in a simple two-region approach which is nearly optimal except when $\tau \sim \sigma(\tau)$. This permits the avoidance of attempting to parametrize displaced background PDFs from a few sparse observed data.
This method is fast enough to verify significance $>5\sigma$ in an hour on a laptop; thus, reliance on asymptotic formulae is not required.  
Furthermore, when setting limits only the integrated detector efficiency in each lifetime region is required to be determined for each mass.  A detailed determination of the uncertainty on the detector efficiency {\em vs} lifetime is not required since the $\tau$ information is only used to classify candidates as prompt or displaced.  For limit setting, uncertainties on the integrated detector efficiency and on the fraction of events that fall in each region are included in the likelihood and the coverage is shown to be good.

\begin{figure}[]
\centering
\includegraphics[width=0.49\textwidth]{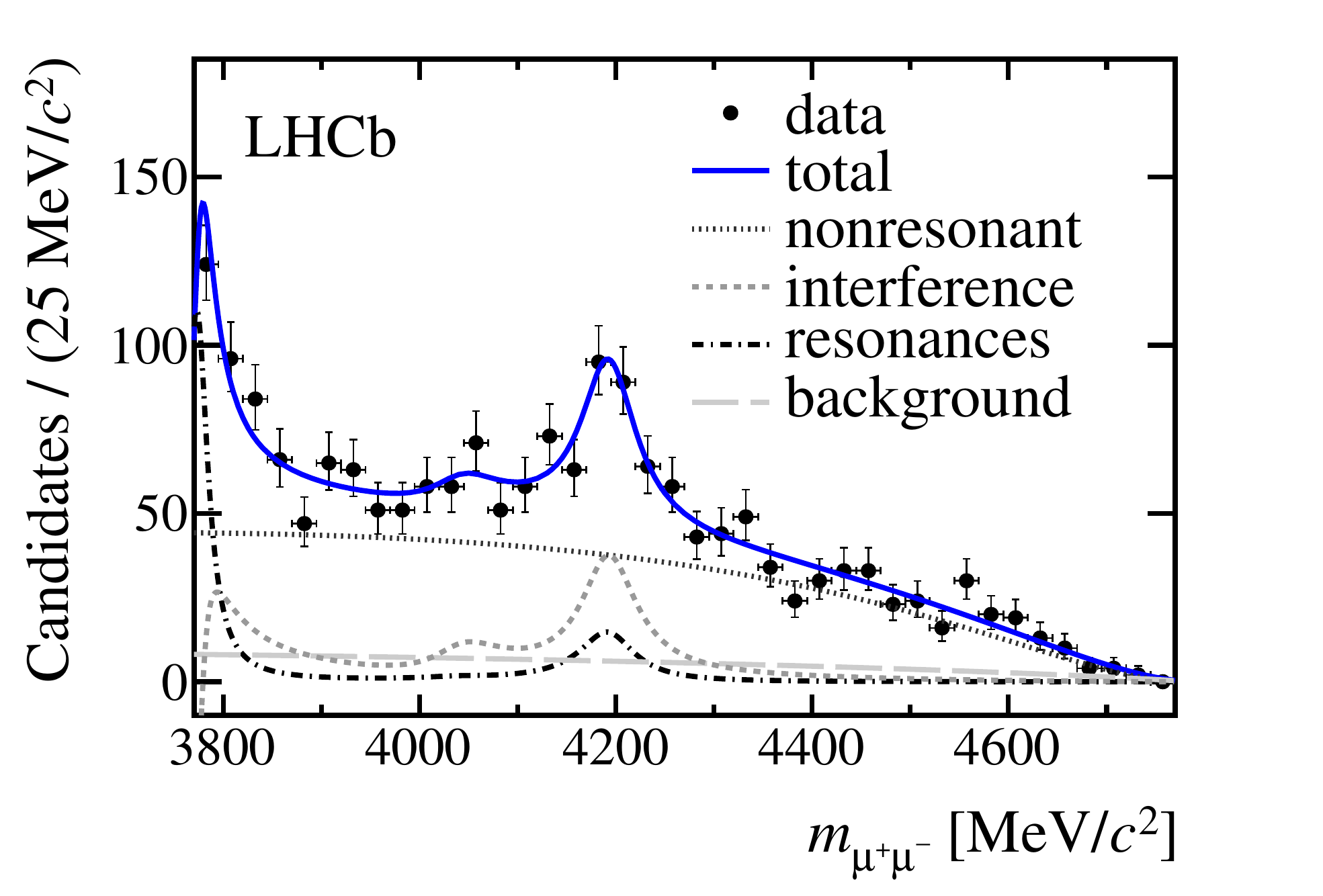}
\caption{\label{fig:kmm}
LHCb $B\to K\mu\mu$ data observed at low recoil (image taken from Ref.~\cite{kmm}). The peaking structure is due to the $\psi(4160)$.
}
\end{figure}

Finally, Fig.~\ref{fig:kmm} shows the low-recoil (high $M(\mu\mu)$) data observed by LHCb in the decay $B\to K\mu\mu$\cite{kmm}.
There is a clear and sizable contribution from the $\psi(4160)$ resonance.  The size of this contribution was unexpected.  
What would have happened if a blind analysis of this data had been performed to search for a new prompt particle that decays to $\mu\mu$?
If a fit using a monotonic background PDF and no $\psi(4160)$ term had been used\footnote{This is unlikely to have happened since prior to observing this data one would still have expected some charmonium contributions.  The more likely scenario would be that a fit would have been performed that contained every $\psi$ state with masses and widths free to vary within their nominal values; various other resonant shape parameters free to vary; and the relative phase of each amplitude free.  This would then have required a serious systematic study to determine the $p$-values.  Here I am simply using this as an example of how an unexpectedly large wide resonance contribution is handled naturally in the method presented in this paper.}, then I estimate that the local $p$-value near 4200 would have given a local significance of $4-5\sigma$.  
Using the method presented in this paper (in only the prompt region), I estimate that the $p$-value is about 0.3(0.4) for $\sigma_y=0(0.1)$ for $x=1$.  This is due to the fact that $\Gamma(\psi(4160)) \sim 10\sigma(m)$ and so locally the resonant peak only deviates from local-linearity by about 15\%.  
No false claim of a discovery would have been made.  
Furthermore, 
given the large number of nuisance parameters and sample size, a fit-based approach would not provide much (possibly no) additional sensitivity to new particles (no matter how much effort was put into developing the fit model).  

\acknowledgments

I thank Tim Gershon for useful comments that helped improve this paper.  This work was supported by US NSF grant PHY-1306550.

\clearpage

\appendix

\section{Profile Likelihood}

The Poisson likelihood for observing $n_s$ events in the signal region and $n_b$ events in the background region is
\begin{equation}
  L(n_s,n_b|s,b) = \mathcal{P}(n_s,s+b) \cdot \mathcal{P}(n_b,xb),
\end{equation}
where the Poisson PDF is defined as
\begin{equation}
  \mathcal{P}(n,\lambda) = \frac{\lambda^{n}}{n!}e^{-\lambda},
\end{equation}
and $s$ is the signal rate, $b$ is the background rate and $x$ is the ratio of the size of the signal and background regions.   
The profile likelihood is then defined as
\begin{equation}
  \Lambda(s|n_s,n_b) = \frac{L(s,\hat{b}(s)|n_s,n_b)}{L(\hat{s},\hat{b}|n_s,n_b)},
\end{equation}
where $\hat{b}(s)$ denotes the value of $b$ that maximizes the likelihood for fixed $s$, and $\hat{s},\hat{b}$ maximize $L$ in general.
These parameters can be obtained analytically\footnote{This requires differentiating $\log{L}$ with respect to $s$ and $b$, setting these to zero, then solving the system of equations.} and are $(\hat{s},\hat{b})=(n_s-n_b/x,n_b/x)$ and $\hat{b}(0) = (n_s+n_b)/(1+x)$.  In general, the estimator for $b$ for any value of $s$ is 
\begin{equation}
\label{eq:bhatpois}
\hat{b}(s) = \left(n_s + n_b - (1+x)s + \sqrt{(n_s+n_b-(1+x)s)^2 + 4(1+x)sn_b}\right)/2(1+x).
\end{equation}
Asymptotically, $-2\log{\Lambda}$ behaves as a $\chi^2$ with 1 DOF and so an approximate $p$-value can be obtained from $\Lambda$.  
{\em N.b.}, one may worry about the possibility that when $s<0$ $\hat{b}(s)$ could become imaginary.  See discussion at the end of this appendix on how the $s < 0$ case is dealt with.

\begin{figure}[]
\centering
\includegraphics[width=0.49\textwidth]{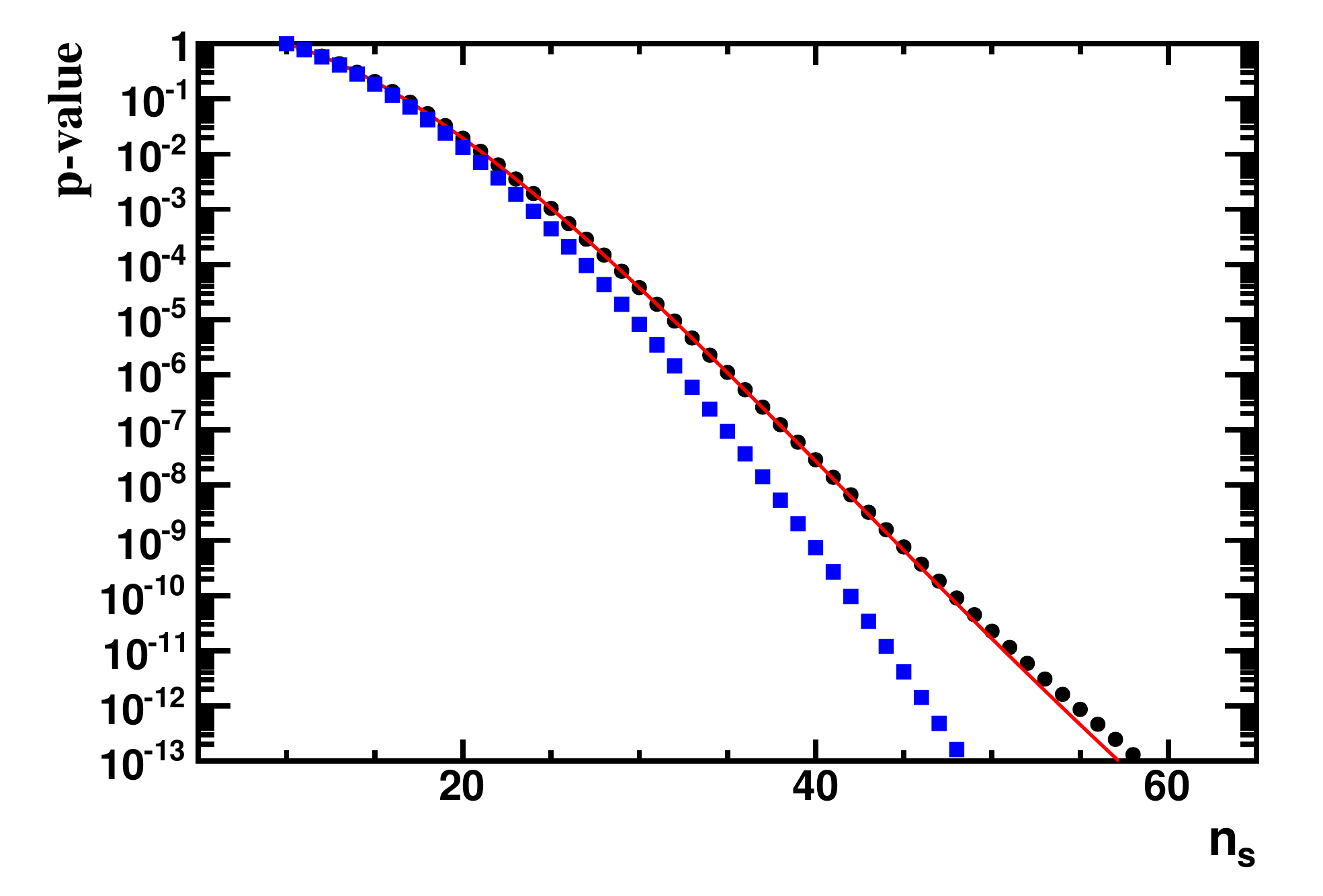}
\caption{\label{fig:pvalcheck} 
$p$-values obtained from the asymptotic formula for the profile likelihood 
as maximized (black points) numerically using MINUIT and (red line) using the analytic approximation. 
This example used $n_b=50$, $x=5$ and $\sigma_y/y=0.1$.  The (blue squares) show the $p$-values for $\sigma_y=0$ (no uncertainty in the scaling between regions).  
}
\end{figure}

It is straightforward to account for uncertainty in the relationship between the sideband and search window yields by augmenting the likelihood as follows:
\begin{equation}
  \label{eq:likegaus}
  L(n_s,n_b,x|s,b,y) = \mathcal{P}(n_s,s+b) \cdot \mathcal{P}(n_b,yb) \cdot \mathcal{G}(y,x,\sigma_y),
\end{equation}
where the Gaussian PDF is defined as 
\begin{equation}
\mathcal{G}(z,\mu,\sigma) =  \frac{1}{\sqrt{2\pi}\sigma}e^{-(z-\mu)^2/2\sigma^2}.
\end{equation}
Now $y$ is the scale factor that relates the yields in the sideband and search regions whose PDF is taken to be a Gaussian with mean $x$ and uncertainty of $\sigma_y$.  Following the same approach to maximizing $L$ produces an algebraically intractable set of three equations.  Making the approximation that $\sigma_y/y < 1/\sqrt{n_b}$, then to leading order in $\sigma_y/y$
\begin{eqnarray}
\label{eq:gausapprox}
\hat{s} &\approx& n_s - n_b/x \\
\hat{y} &\approx& x + \sigma^2_y(n_b/x-\hat{b}(\hat{s},\sigma_y=0)) \\
\hat{b} &\approx& \hat{b}(\hat{s},\sigma_y=0,x\to\hat{y}),
\end{eqnarray}
where $\hat{b}(\hat{s},\sigma_y=0)$ is the result given in Eq.~\ref{eq:bhatpois} and $\hat{b}(\hat{s},\sigma_y=0,x\to\hat{y})$ uses the same equation but replaces $x$ with $\hat{y}$ everywhere.  Figure~\ref{fig:pvalcheck} shows that this approximation is accurate out to $p$-values of about $\mathcal{O}(10^{-12})$ when the relative uncertainty on the scale factor is smaller than the relative statistical uncertainty in the background rate.  In my tests I find that for $\sigma_y/y = 0.1$ this holds for $n_b$ up to about 200.  

For the case where $\sigma_y/y > 1/\sqrt{n_b}$, the analytic approximation given is not valid so the values $\hat{s},\hat{y},\hat{b}$ must be found numerically, {\em e.g.}, using {\sc Minuit}.  This is a valid approach but increases the CPU time required by a factor $\mathcal{O}(100)$.  
In such cases, however, $n_b$ is large enough that the Poisson term for $n_b$ can be replaced by a Gaussian term.  The likelihood is then given by
\begin{equation}
  \label{eq:likegaus}
  L(n_s,n_b,x|s,b) = \mathcal{P}(n_s,s+b) \cdot \mathcal{G}(b,n_b/x,\sigma_b),
\end{equation}
where the statistical and scale factor uncertainties on the background rate are now included in a single term $\sigma_b^2 = (n_b/x^2)(1+n_b\sigma_y^2/x^2)$.  For large $n_b$ this reduces to $\sigma_b = (n_b/x^2)\sigma_y$.  The likelihood in this case can again be maximized analytically giving the following results:
$(\hat{s},\hat{b}) = (n_s-n_b/x,n_b/x)$ and for $s=0$
\begin{equation}
\hat{b}(0) = \frac{1}{2}\left(n_b/x-\sigma_b^2 + \sqrt{(\sigma_b^2-n_b/x)^2 + 4\sigma_b^2n_s}\right).
\end{equation}
Thus, it is possible to provide analytical solutions for all cases.

When searching for a new particle the physical region is $s \ge 0$ and the test statistic used for discovery is $\Lambda(s=0)$ if $\hat{s} \ge 0$.  If $\hat{s} < 0$, I take $\hat{s}=0$ which gives $\Lambda = 1$.   One would expect this to happen at half of all $m({\rm test})$ values considered which is handled naturally by the pseudo-experiment method when obtaining global $p$-values.
When setting limits, I use the {\em bounded} method from Ref.~\cite{rolke} where the increase in the likelihood is taken from $s=0$ instead of $\hat{s}$ for the case where $\hat{s}<0$.  This produces limits that are slightly more conservative but also never produces unphysical limits.

\clearpage

\section{Resonances}

\begin{figure}[]
\centering
\includegraphics[width=0.49\textwidth]{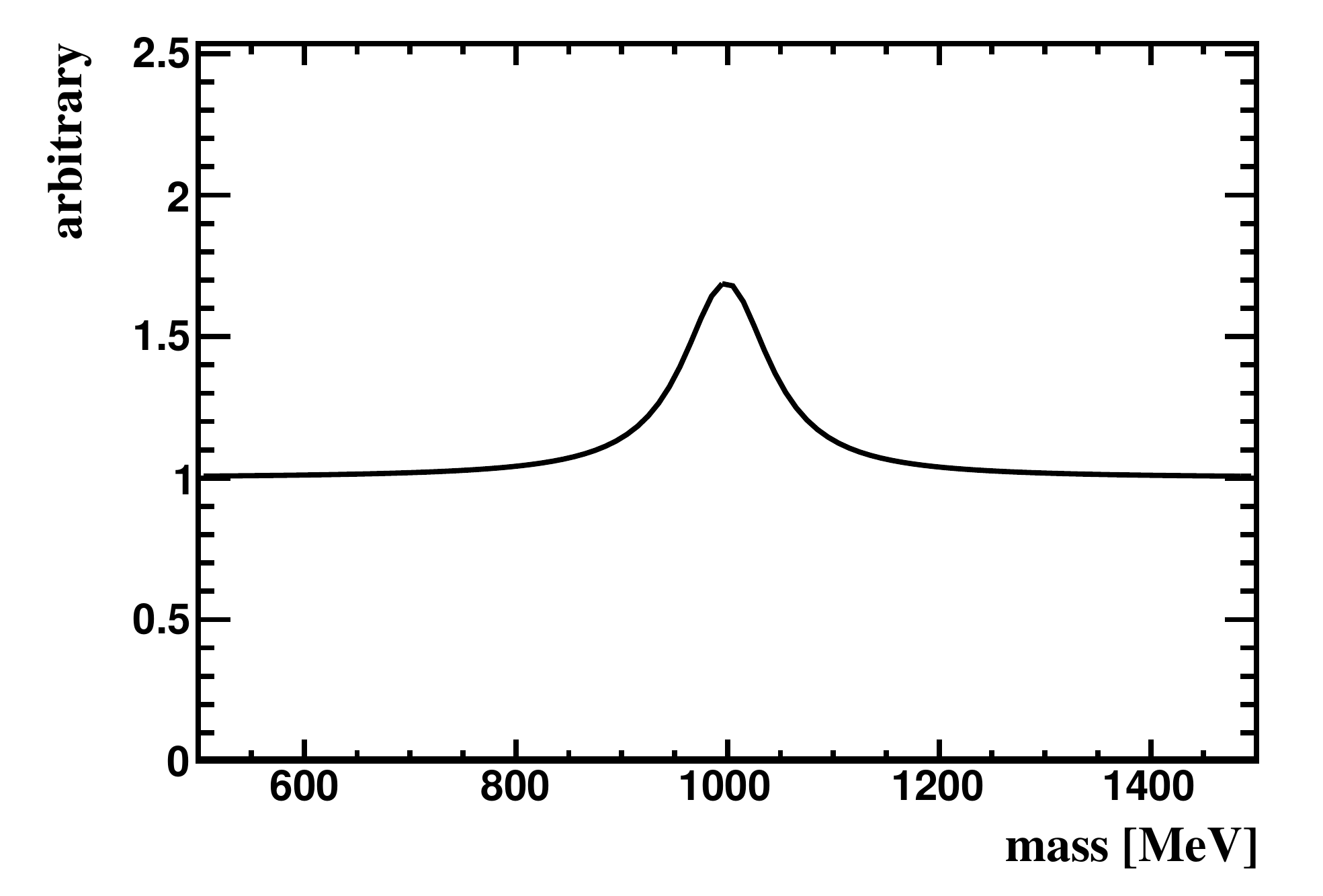}
\includegraphics[width=0.49\textwidth]{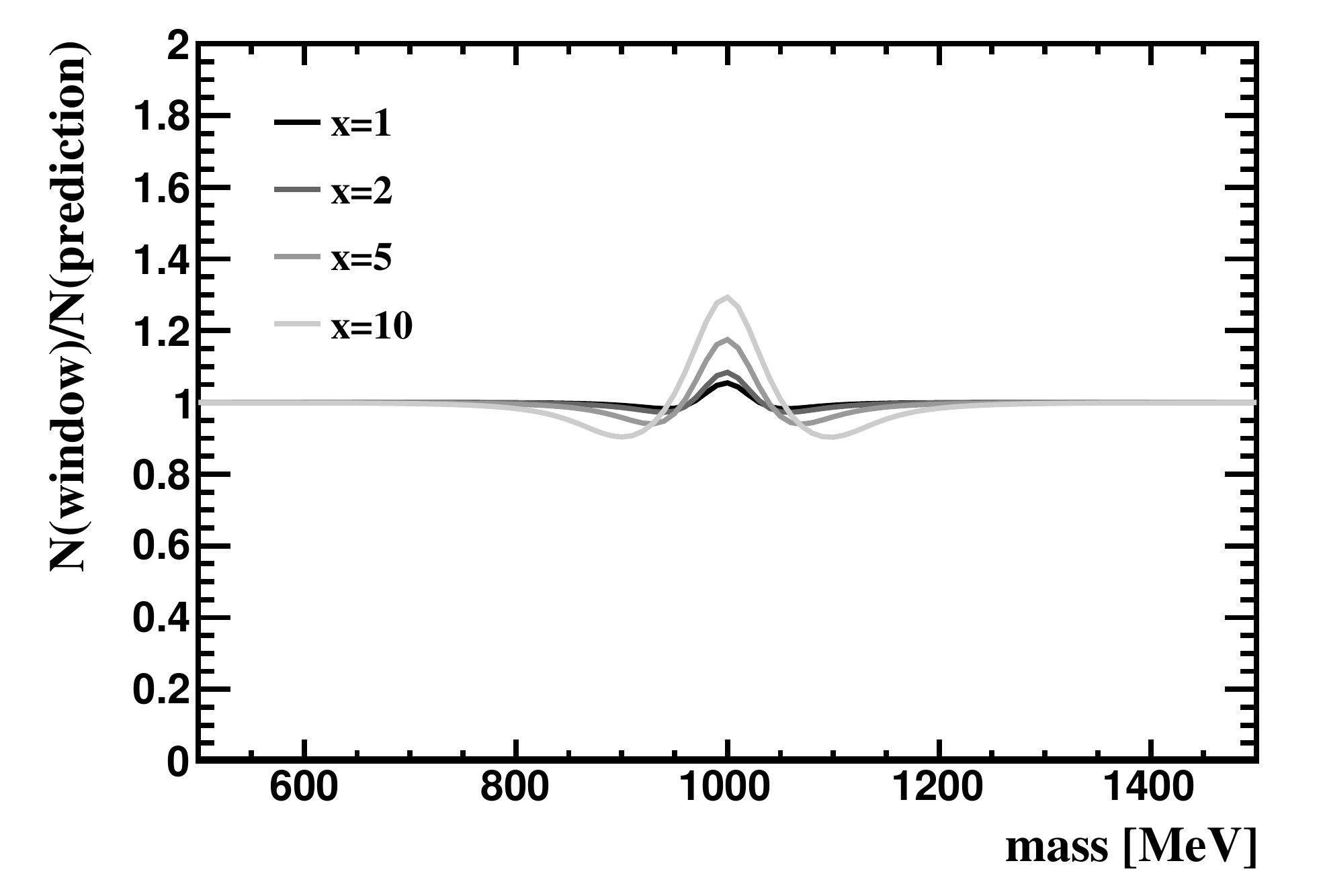}\\
\includegraphics[width=0.49\textwidth]{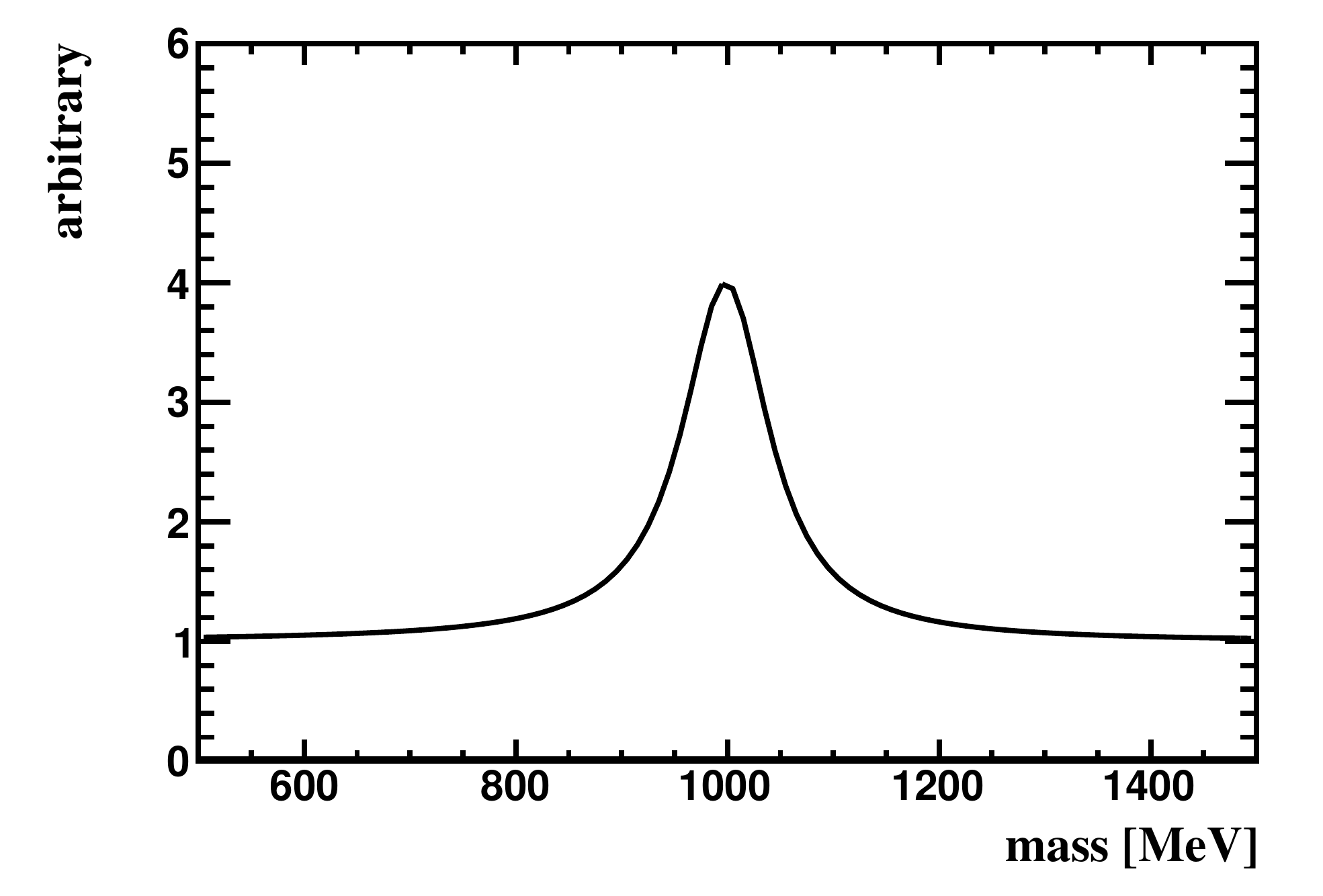}
\includegraphics[width=0.49\textwidth]{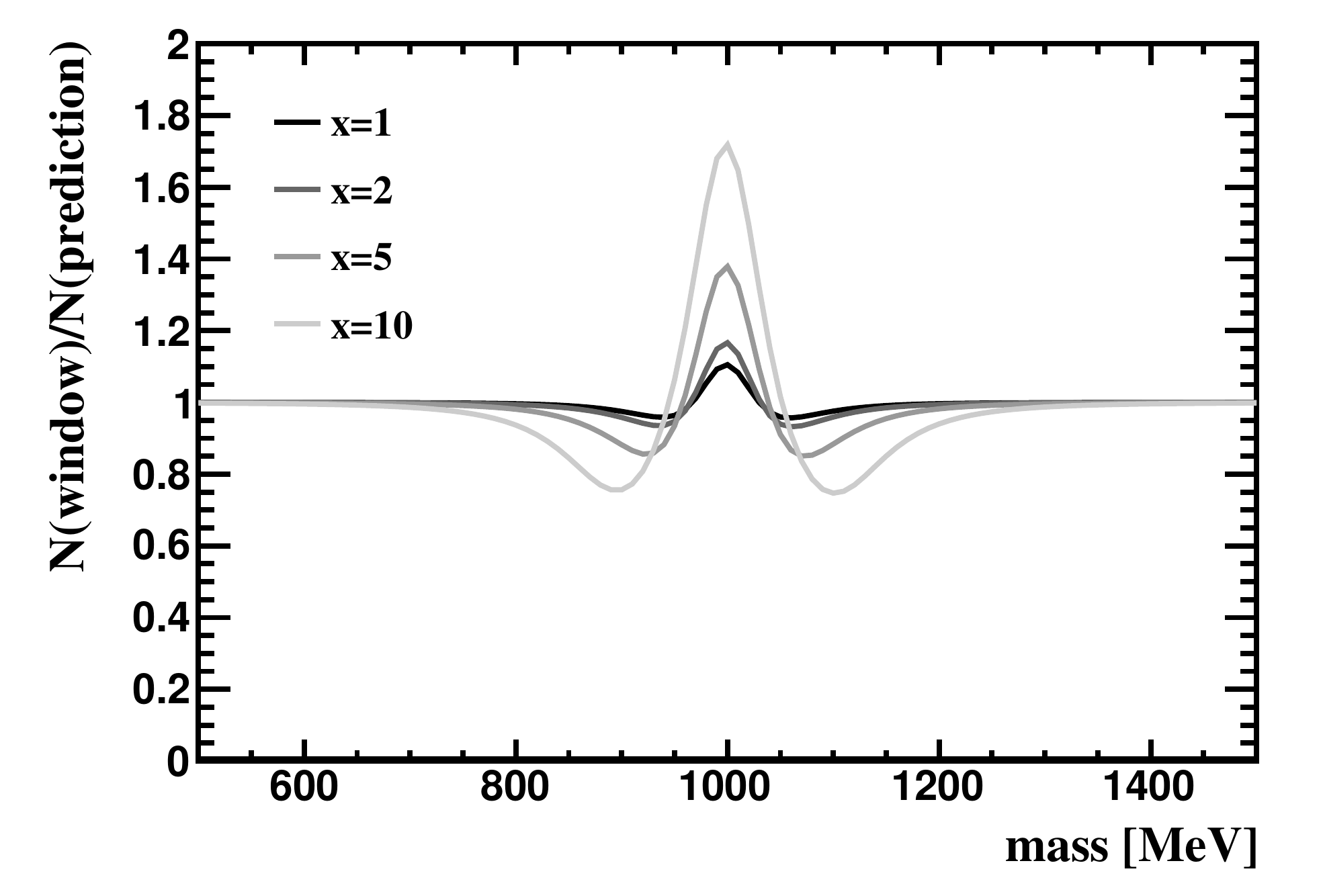}\\
\includegraphics[width=0.49\textwidth]{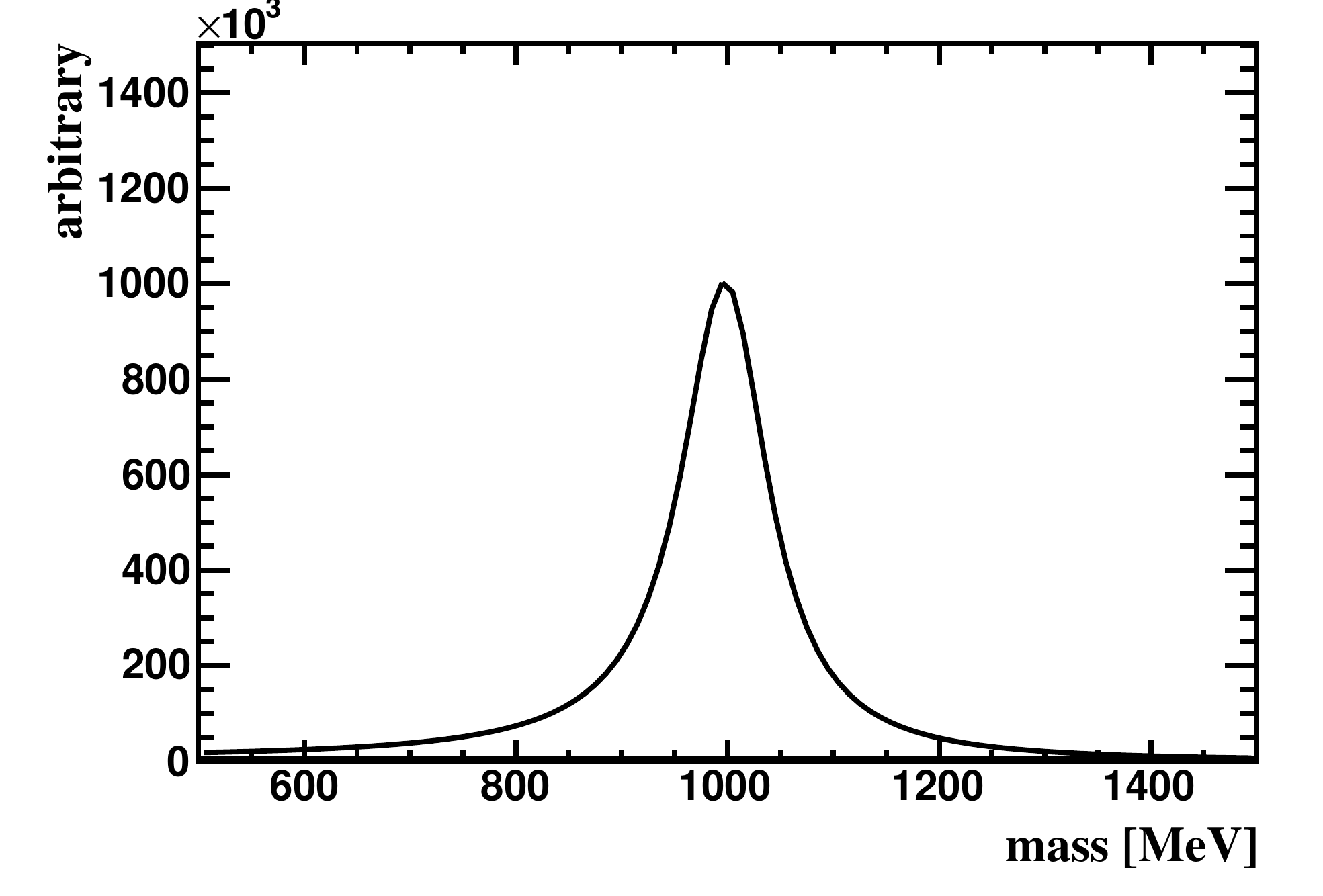}
\includegraphics[width=0.49\textwidth]{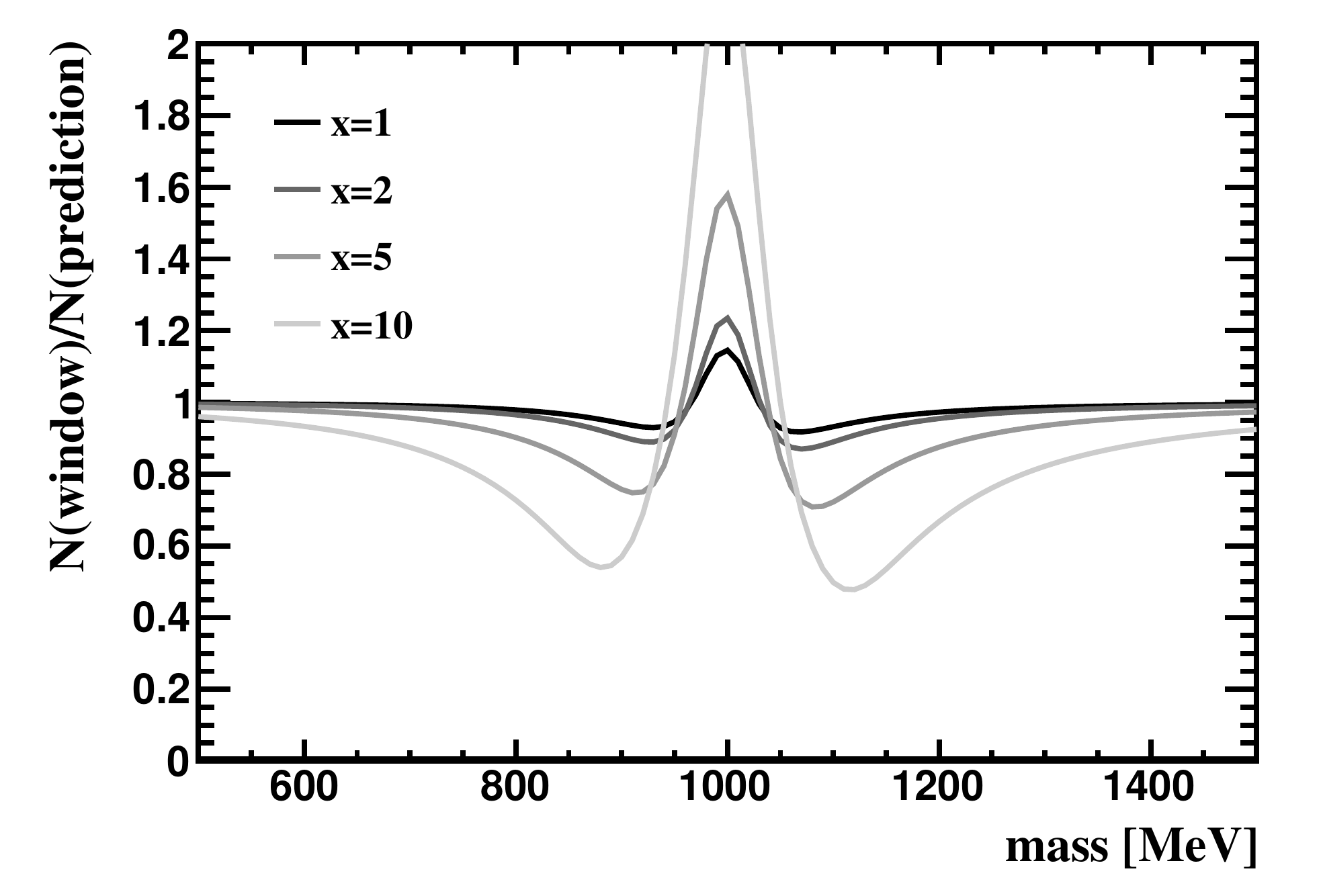}
\caption{\label{fig:res100} Deviations from local-linear for a resonance with $m=1000$~MeV, $\Gamma = 100$~Mev, where $\sigma(m) = 5$~MeV.  The left plots show the yield distribution (in arbitrary units) for various choices of resonance contribution fraction.  The right plots show the ratio of events expected in a signal region to the prediction from the sidebands for various test masses ($x$-axis) and for various sideband to signal region size ratios $x$.}
\end{figure}

\begin{figure}[]
\centering
\includegraphics[width=0.49\textwidth]{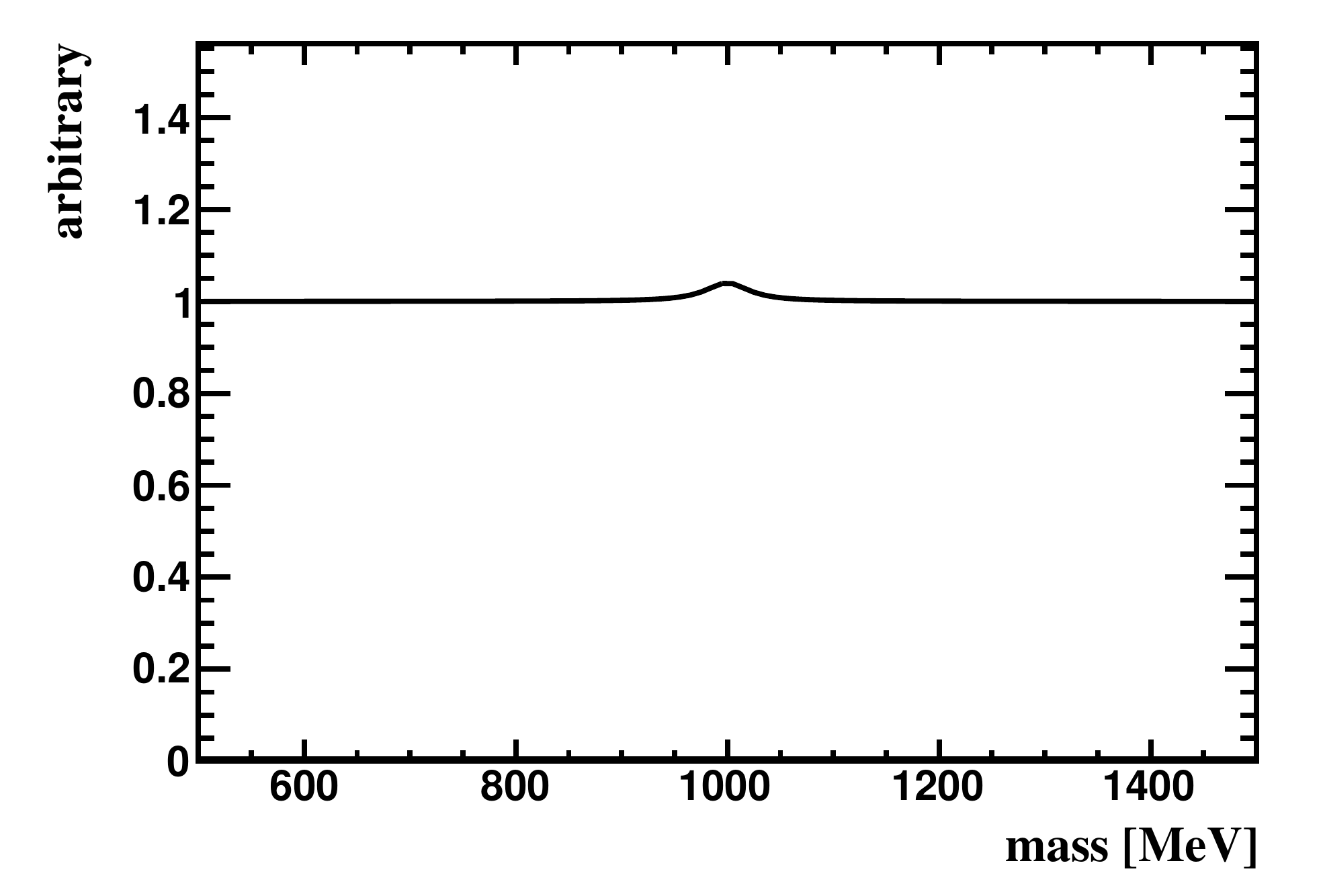}
\includegraphics[width=0.49\textwidth]{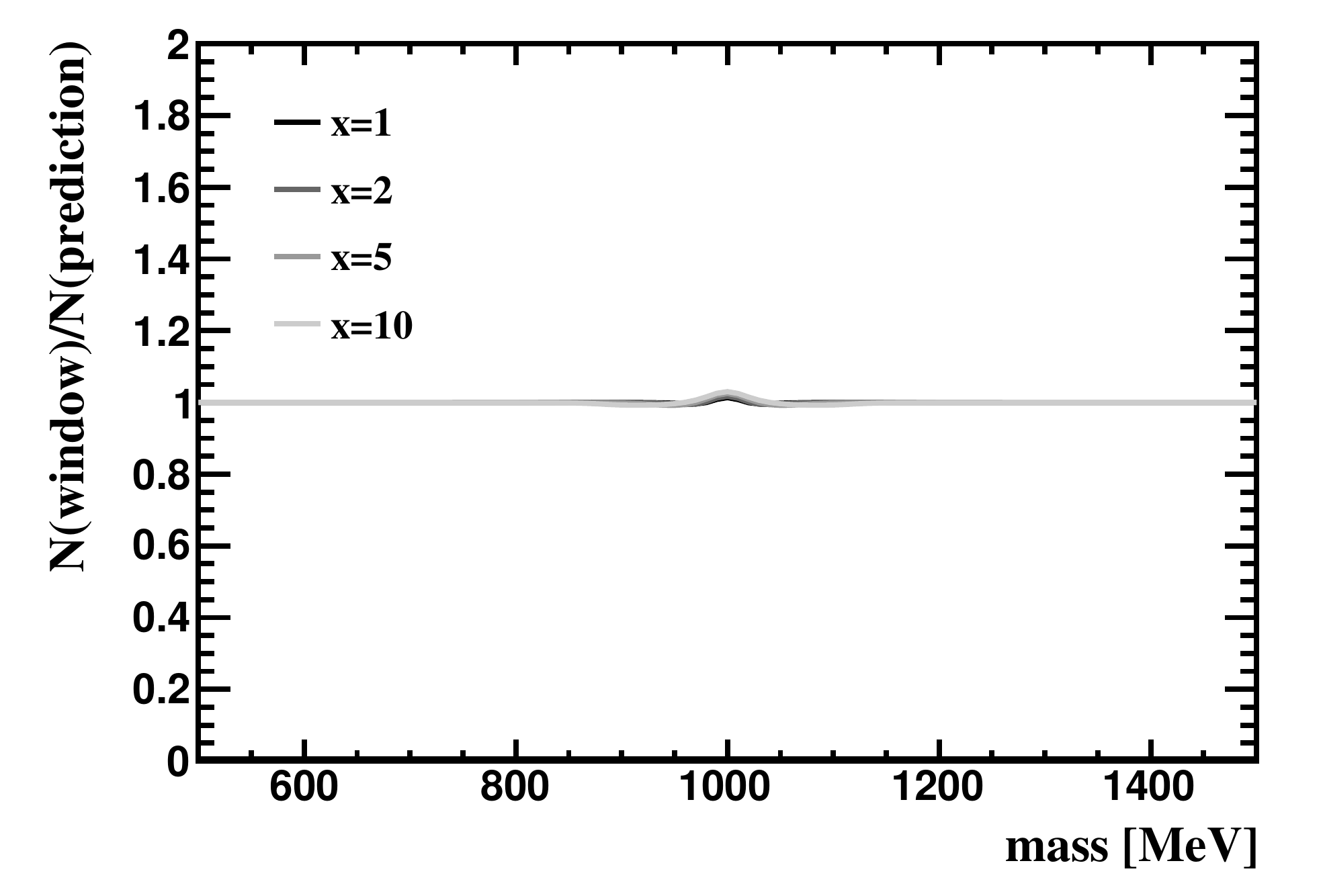}\\
\includegraphics[width=0.49\textwidth]{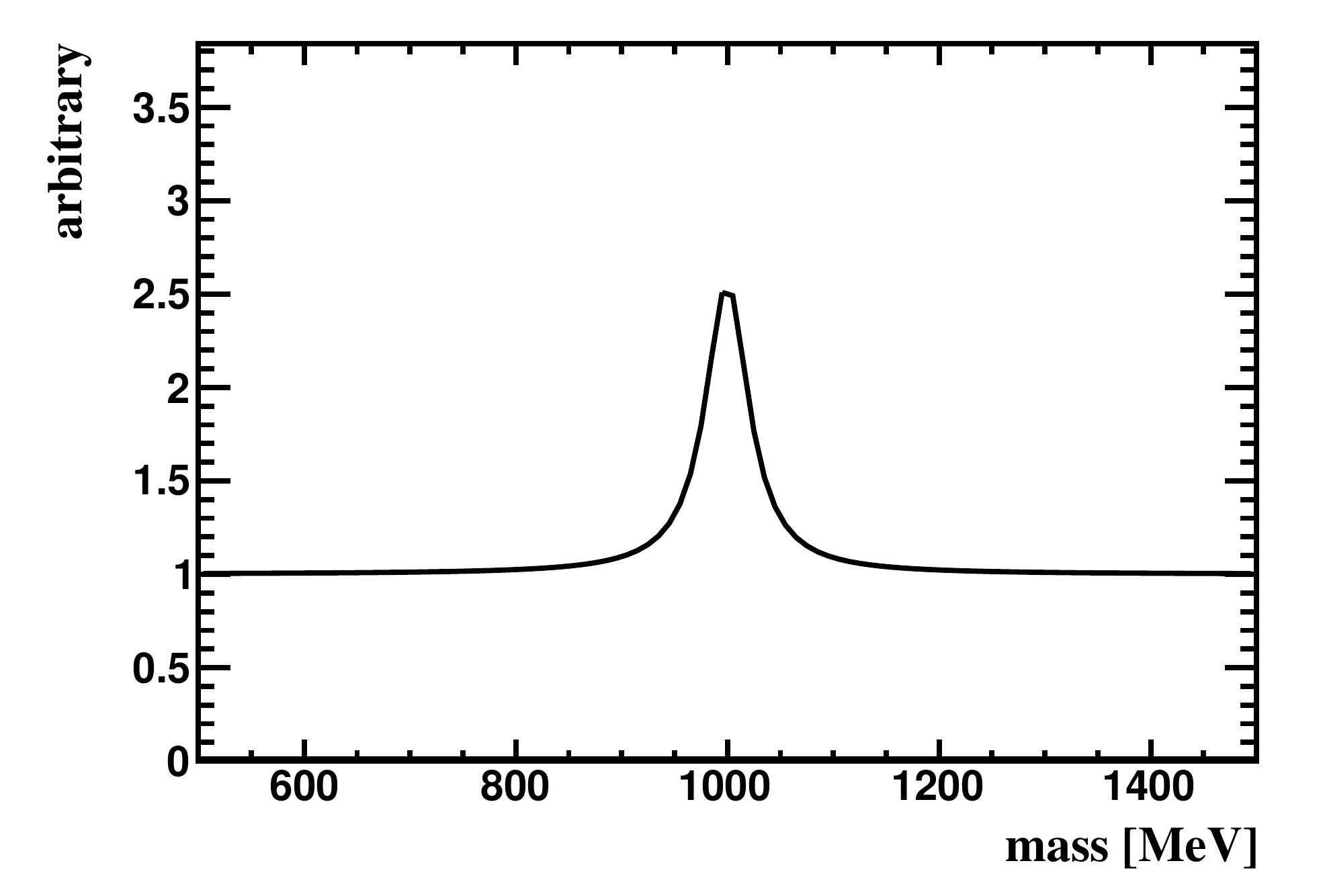}
\includegraphics[width=0.49\textwidth]{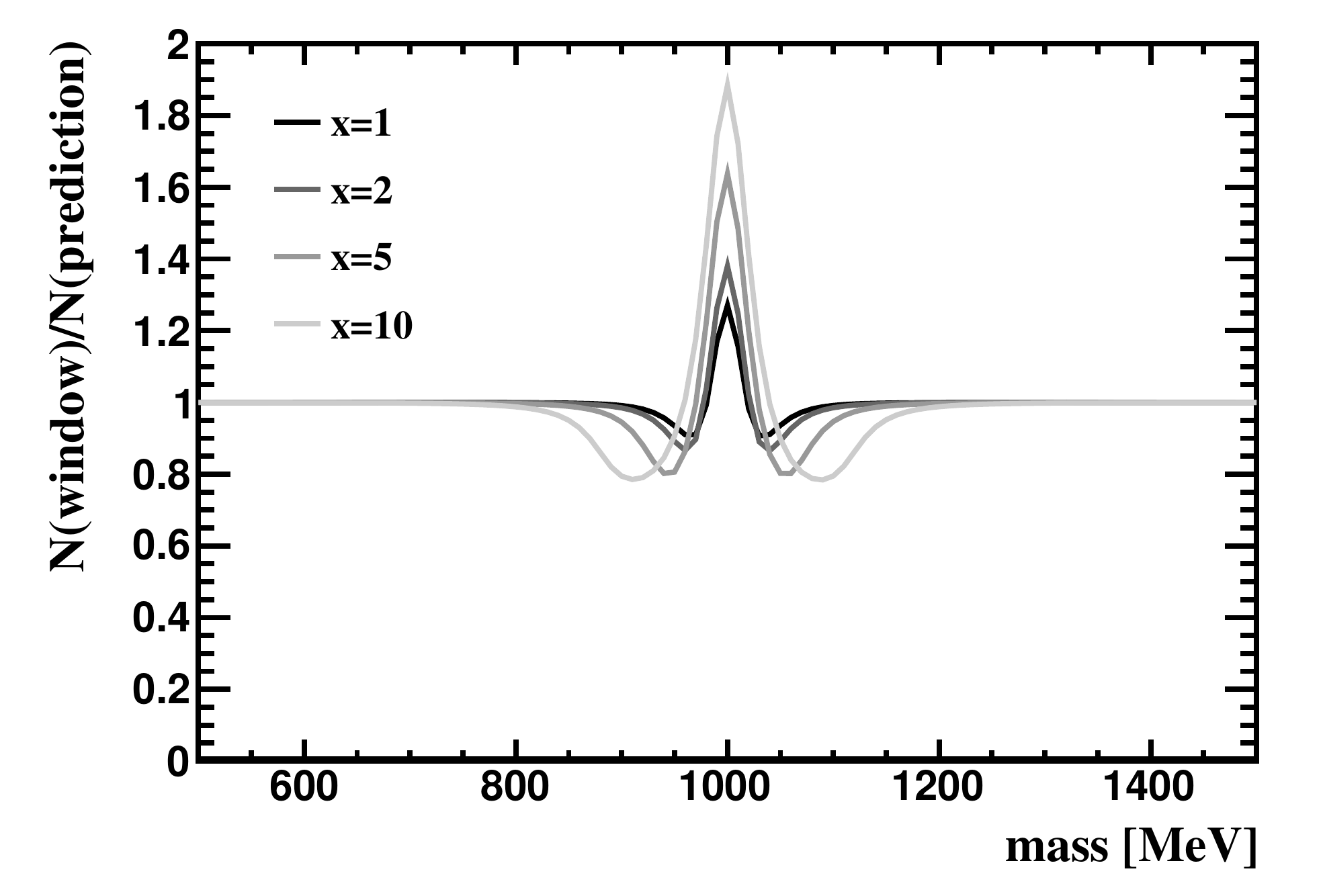}\\
\includegraphics[width=0.49\textwidth]{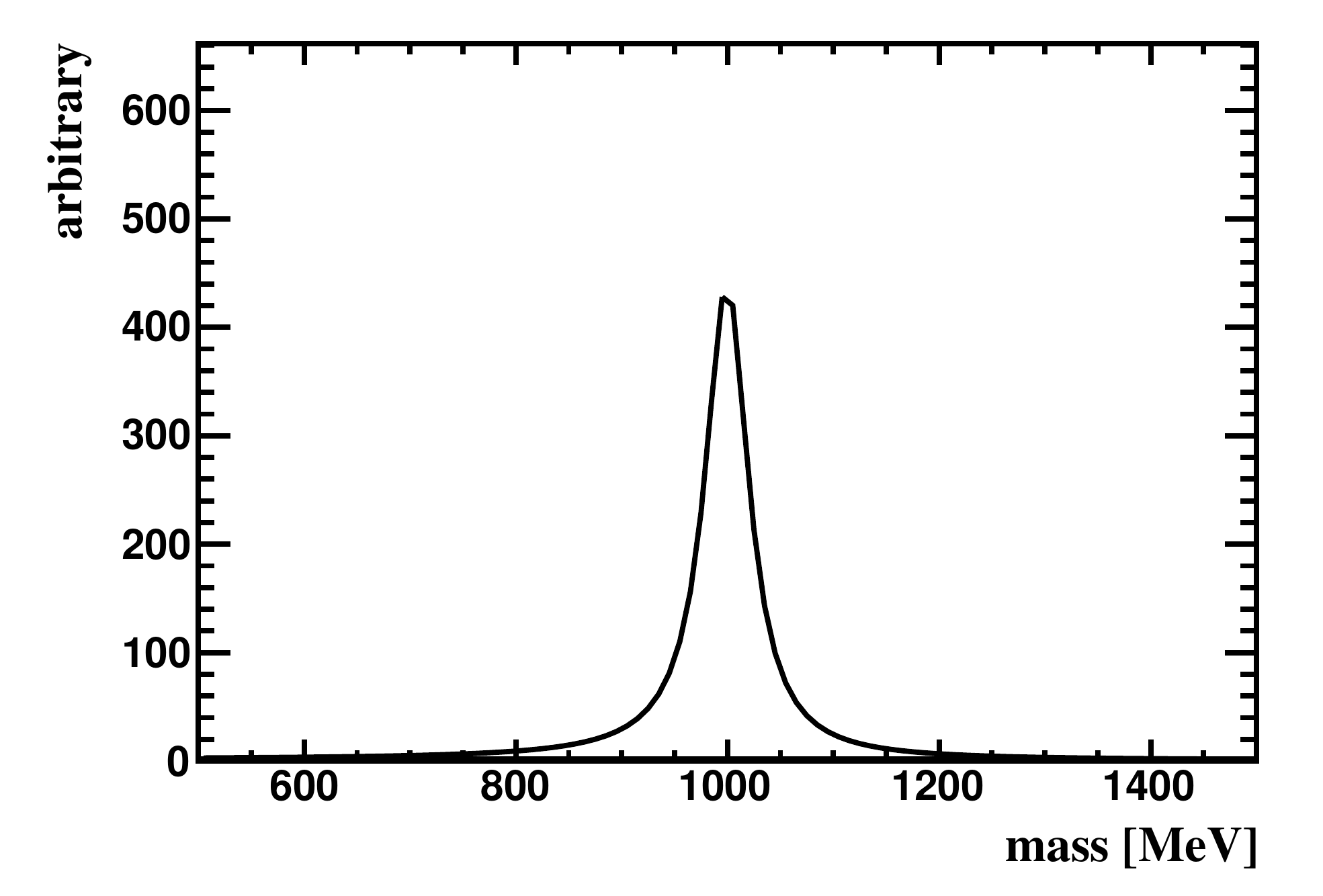}
\includegraphics[width=0.49\textwidth]{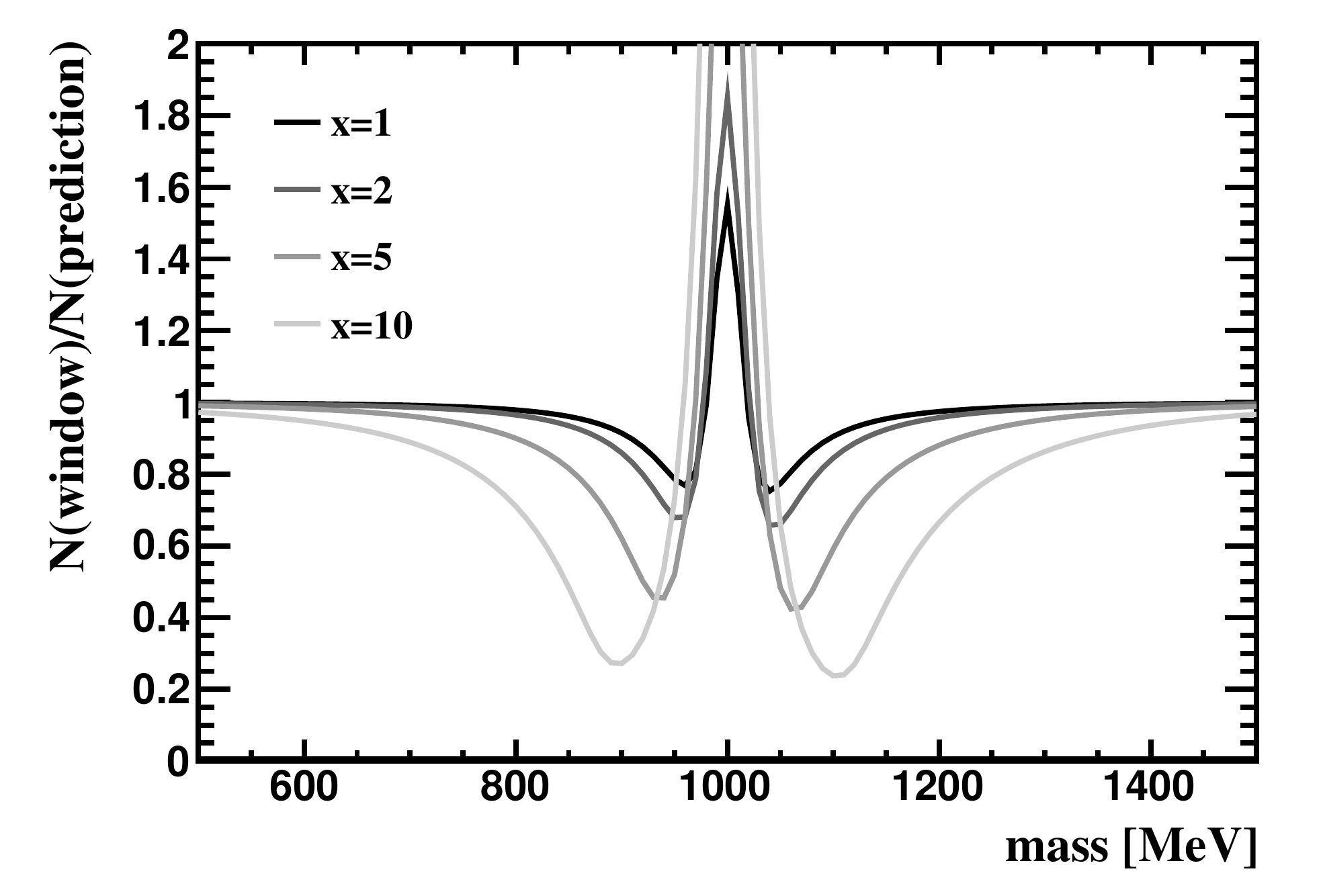}
\caption{\label{fig:res50}Deviations from local-linear for a resonance with $m=1000$~MeV, $\Gamma = 50$~Mev, where $\sigma(m) = 5$~MeV.  The left plots show the yield distribution (in arbitrary units) for various choices of resonance contribution fraction.  The right plots show the ratio of events expected in a signal region to the prediction from the sidebands for various test masses ($x$-axis) and for various sideband to signal region size ratios $x$. }
\end{figure}

\begin{figure}[]
\centering
\includegraphics[width=0.49\textwidth]{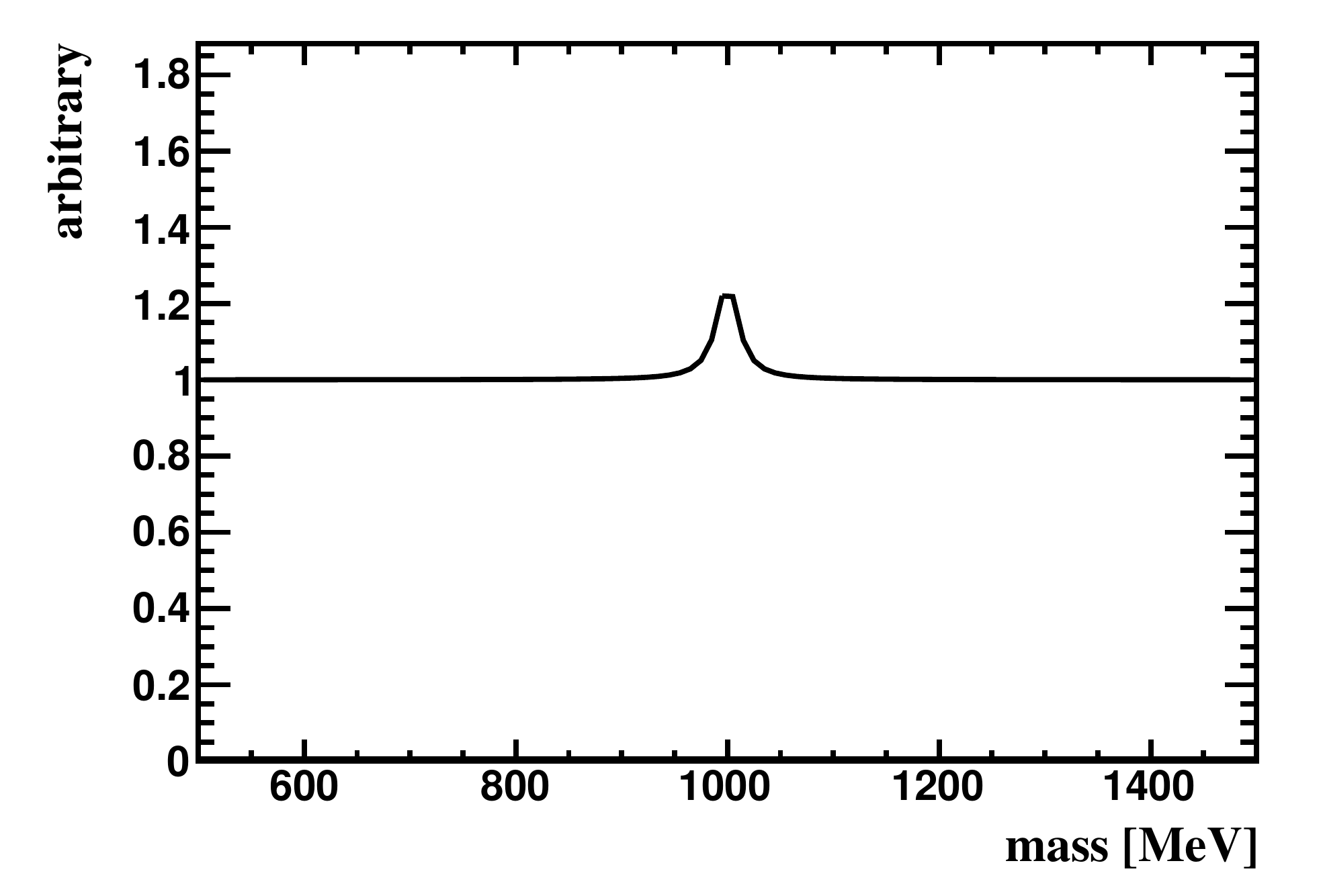}
\includegraphics[width=0.49\textwidth]{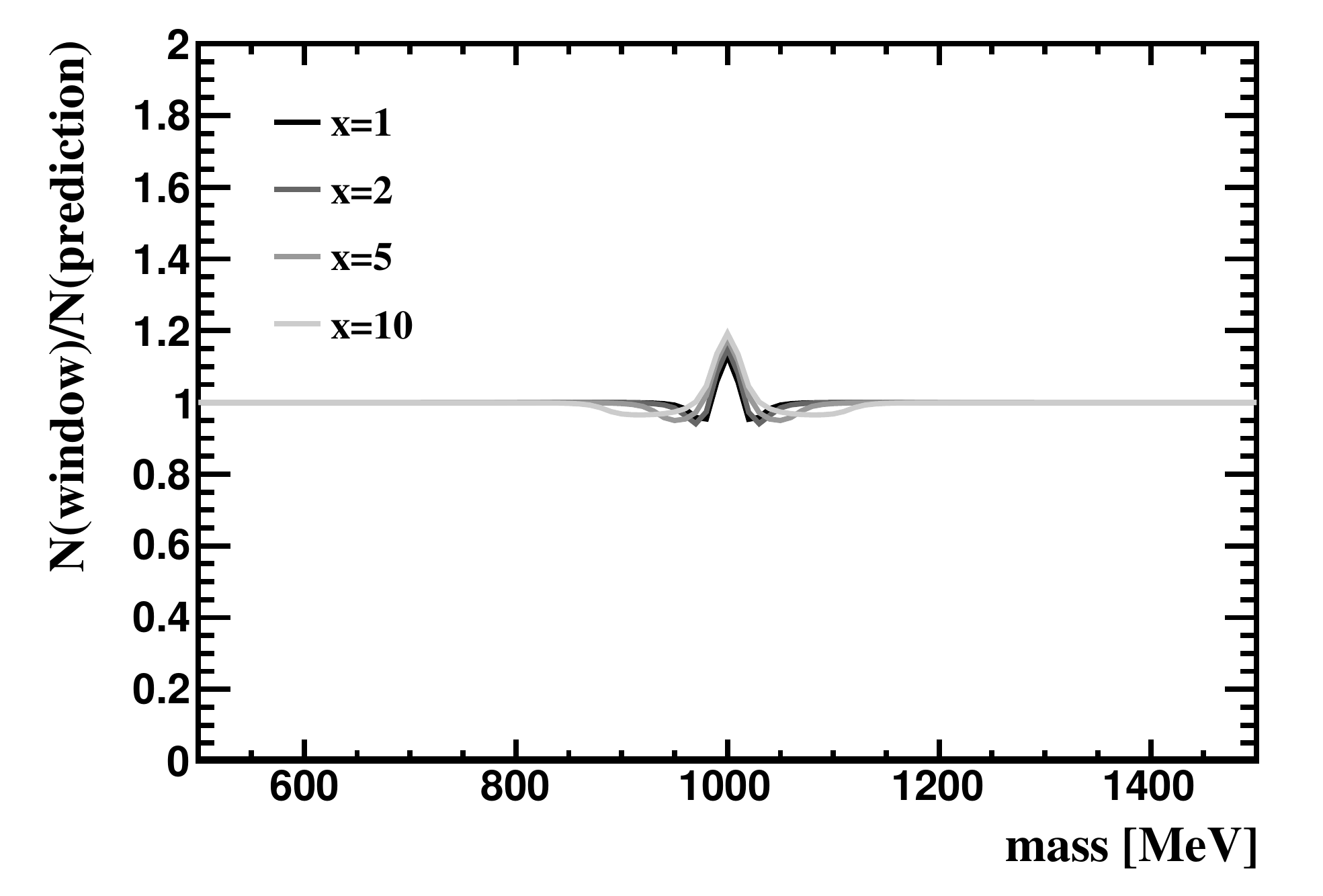}\\
\includegraphics[width=0.49\textwidth]{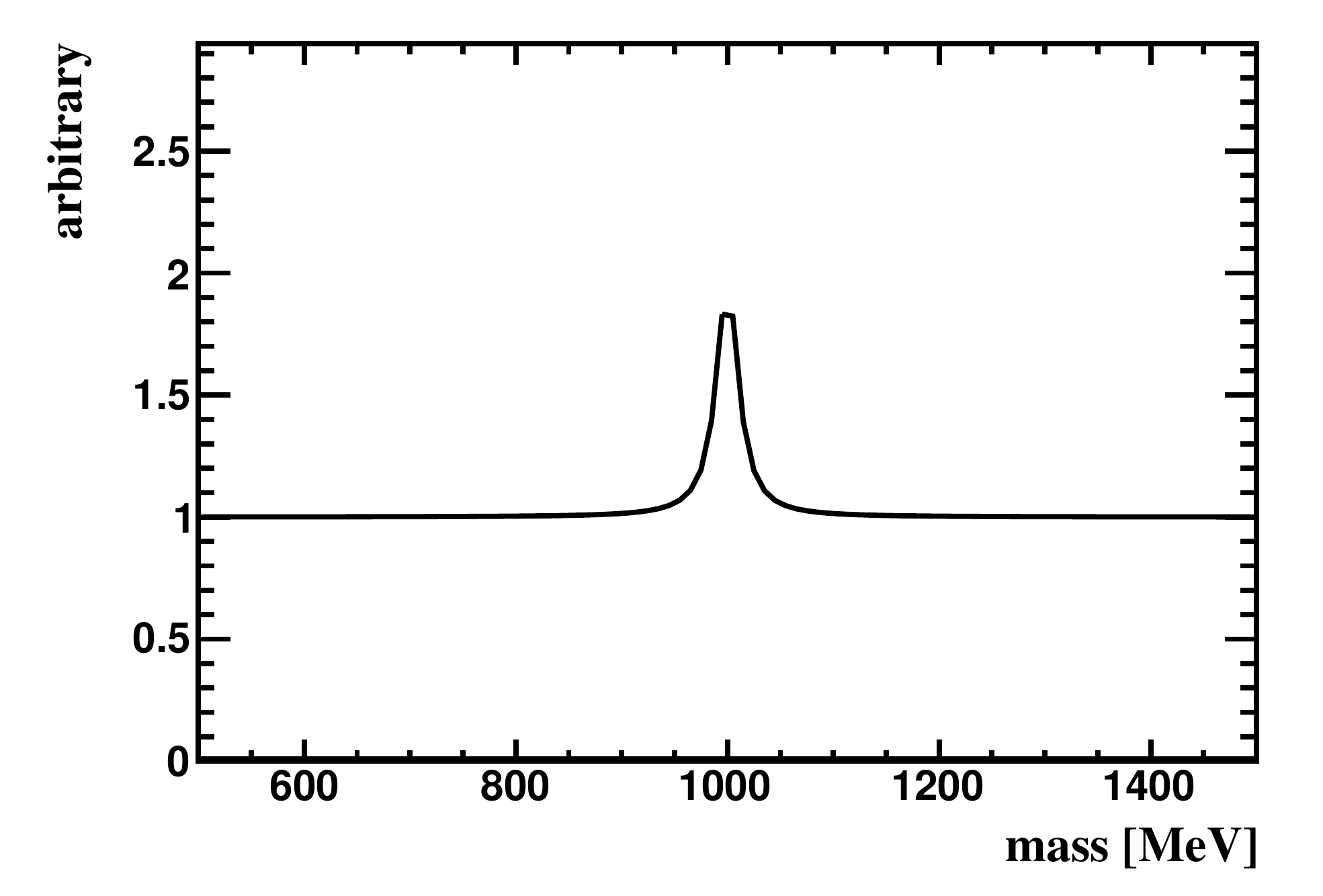}
\includegraphics[width=0.49\textwidth]{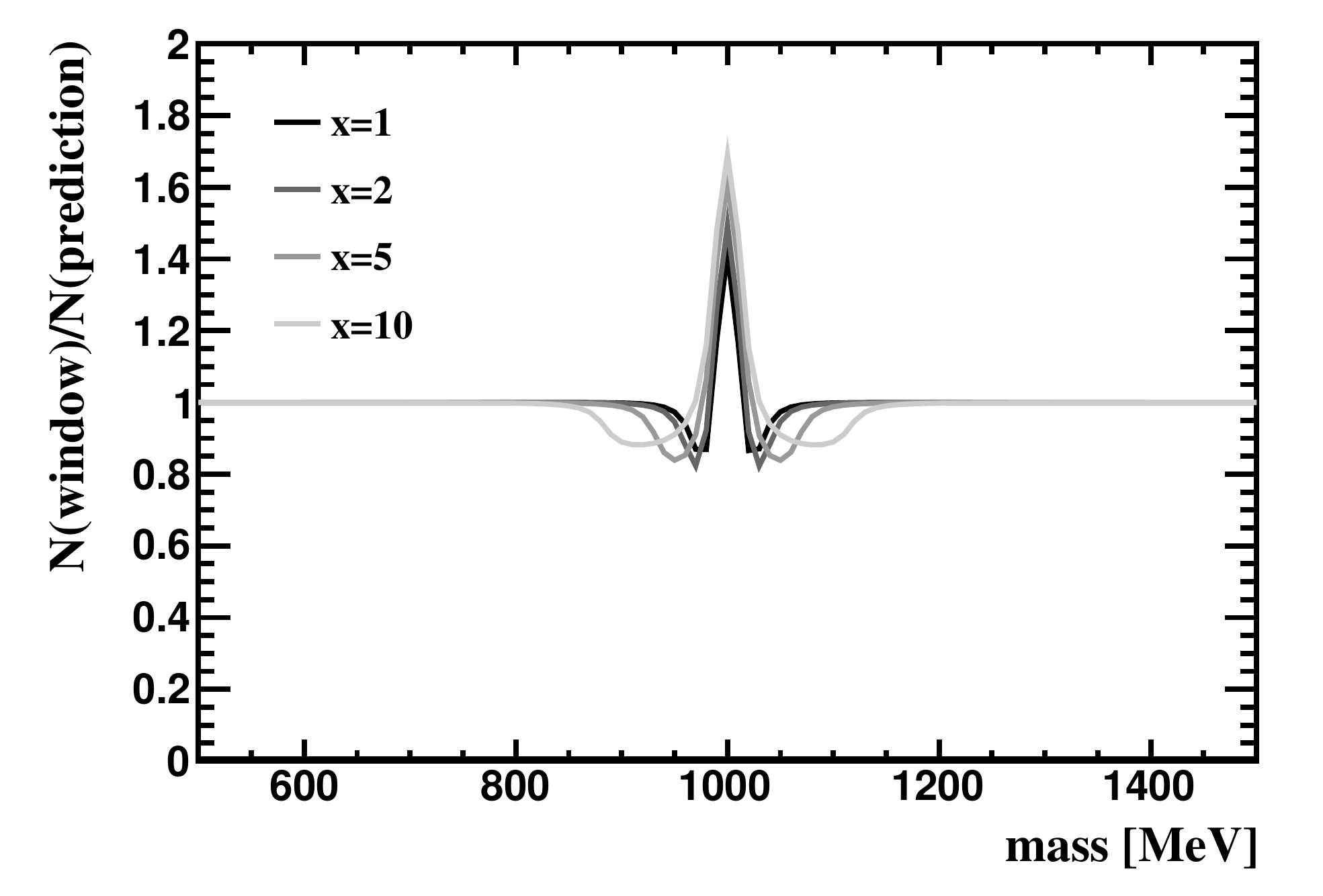}\\
\includegraphics[width=0.49\textwidth]{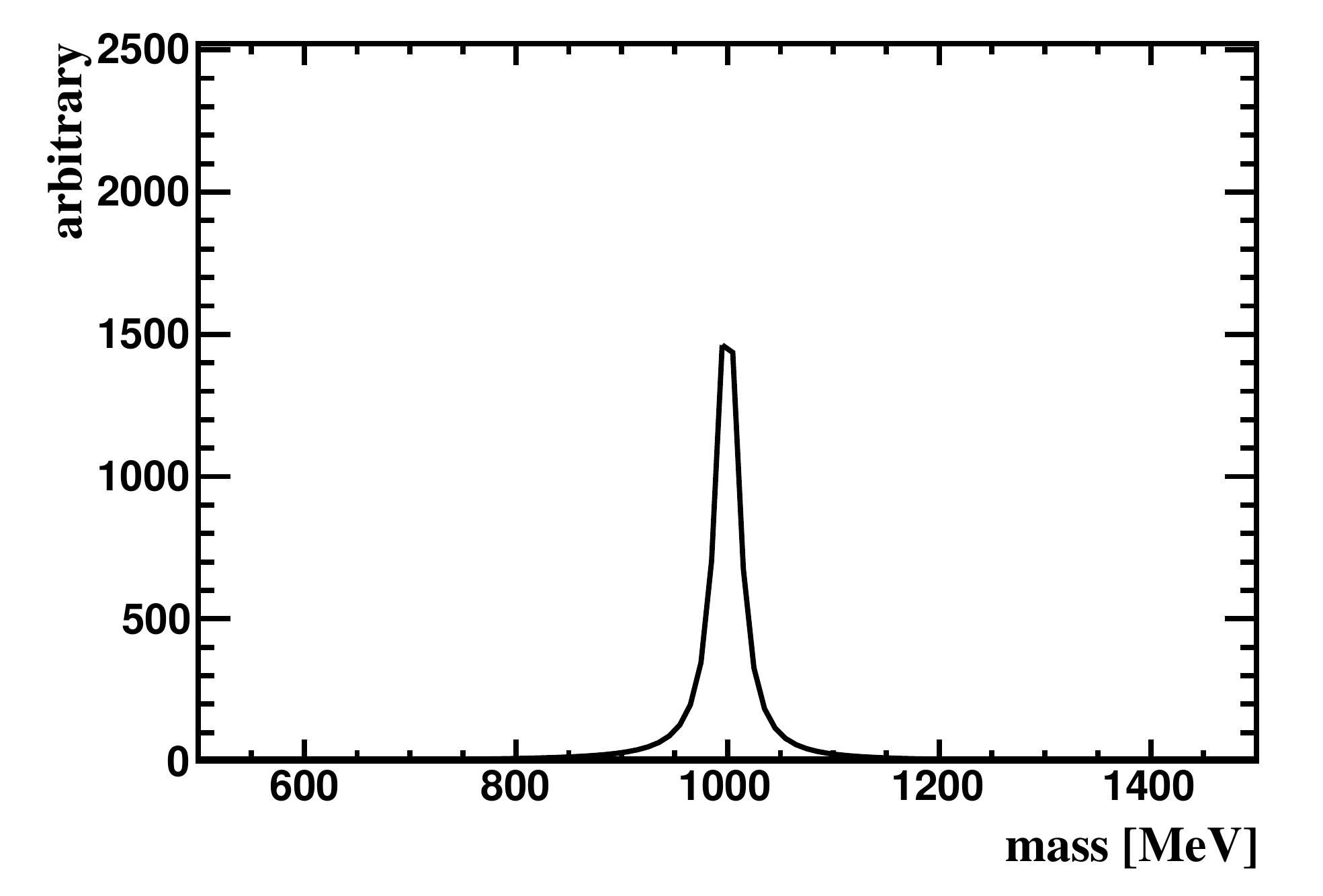}
\includegraphics[width=0.49\textwidth]{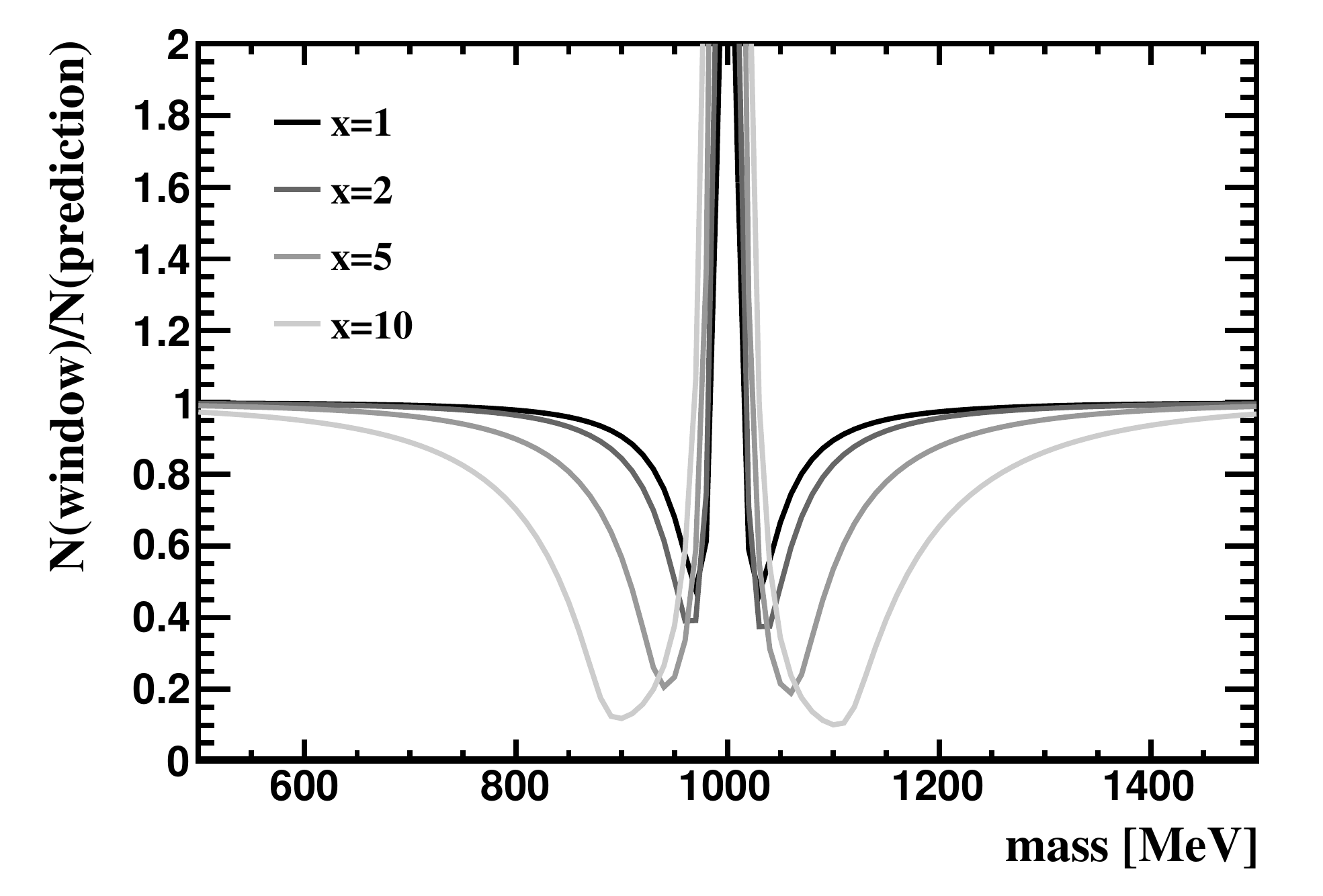}
\caption{\label{fig:res25}Deviations from local-linear for a resonance with $m=1000$~MeV, $\Gamma = 25$~Mev, where $\sigma(m) = 5$~MeV.  The left plots show the yield distribution (in arbitrary units) for various choices of resonance contribution fraction.  The right plots show the ratio of events expected in a signal region to the prediction from the sidebands for various test masses ($x$-axis) and for various sideband to signal region size ratios $x$.}
\end{figure}

This appendix considers the relationship between deviations from the local-linear approximation due to resonance contributions.  Figure~\ref{fig:res100} shows the expected deviations from local-linear for a resonance with $\Gamma/\sigma(m) = 20$.  For this case, even if the resonance makes up close to 100\% of the total PDF the choice $x=1$ is still local-linear to about 10\%.  For smaller resonance contributions larger values of $x$ are local-linear at this level.  Figure~\ref{fig:res50} and \ref{fig:res25} show similar plots for  $\Gamma/\sigma(m) = 10$ and $\Gamma/\sigma(m) = 5$.  For the case $\Gamma/\sigma(m) = 10$, the local-linear approximation is valid at the 10\% level up to resonance contributions of about 50\% of the total PDF, while for $\Gamma/\sigma(m) = 5$ it is only valid at this level for small resonance contributions.  

These results are not surprising.  For the case $\Gamma/\sigma(m) = 5$, the signal region (which is $\pm2\sigma(m)$) contains almost half of the resonance probability.  Any large contribution from such a resonance will need to be vetoed.  Contributions from wide resonances, however, are {\em safe} even if they make up the entire PDF.  For resonances with widths in the range $20 < \Gamma/\sigma(m) < 5$, applying a veto of the region $|m-m({\rm resonance})| < \Gamma$ will typically be safe for any size resonance contribution.  However, if it is known (or can be shown) that the resonance is not dominant, then such a veto may not be required.

\clearpage

\clearpage

\section{Two-Sample Non-Parametric Goodness-of-Fit Tests}

Assuming that the lifetime distributions of all non-signal PDFs are the same in the signal and background regions, a two-sample non-parametric goodness-of-fit test to the hypothesis that the $\tau$ PDF for the signal and background data is the same can be used to test whether a new particle with a resolvable lifetime is contributing to the data.    A well-known test is the Kolmogorov-Smirnov (KS)\cite{ks} test which uses as test statistic
\begin{equation}
  T = {\rm MAX}|F_s(\tau)-F_b(\tau)|,
\end{equation}
where $F$ denotes the cumulative distribution.  The asymptotic formula is used in this study to obtain an approximate $p$-value. 

Despite the KS test's popularity in particle physics, it is known to not be as powerful as the Cramer-von-Mises (CvM)\cite{cvm} and Anderson-Darling (AD)\cite{ad} tests under many conditions.   The CvM and AD tests build statistics
\begin{equation}
  T_{\rm CvM} = \frac{n_sn_b}{n_s+n_b} \int \left(F_s(\tau)-F_b(\tau)\right)^2 dF_{s+b}(\tau),
\end{equation}
and
\begin{equation}
  T_{\rm AD} = \frac{n_sn_b}{n_s+n_b} \int \frac{\left(F_s(\tau)-F_b(\tau)\right)^2}{F_{s+b}(\tau)(1-F_{s+b}(\tau))} dF_{s+b}(\tau),
\end{equation}
respectively.  Both are based on the cumulative distributions like the KS test but instead of using only the maximum discrepancy, they use an integrated discrepancy.   The AD test provides more weight to the ``tails''.  
The approximate $p$-values for these tests are obtained using toy simulated data sets in this study. 

These two-sample tests are for shape only (in how they are used in this study).  To test both size and shape two tests are run: (1) either the KS, CvM or AD tests for a shape comparison of sidebands to search window and (2) Poisson profile likelihood  for yield comparisons.  
As the size and shape tests are uncorrelated (to very good approximation), Fisher's method\cite{fisher} is used to combine the two $p$-values to get a single $p$-value.  
In this way both shape and size anomalies are tested and a single $p$-value is obtained.  

\clearpage

\section{Lifetime Test Results}

\begin{figure}[h!]
\centering
\includegraphics[width=0.49\textwidth]{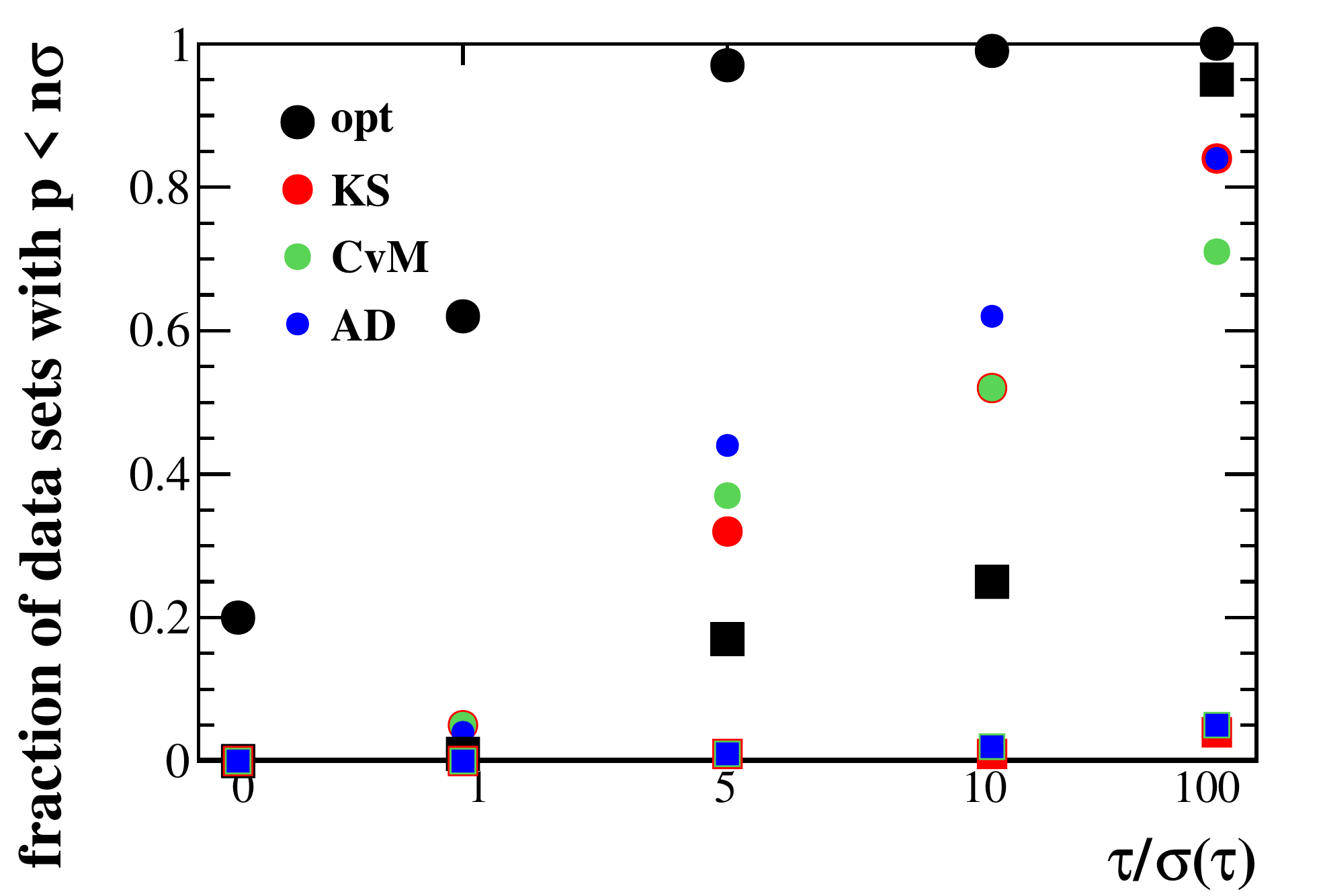}
\includegraphics[width=0.49\textwidth]{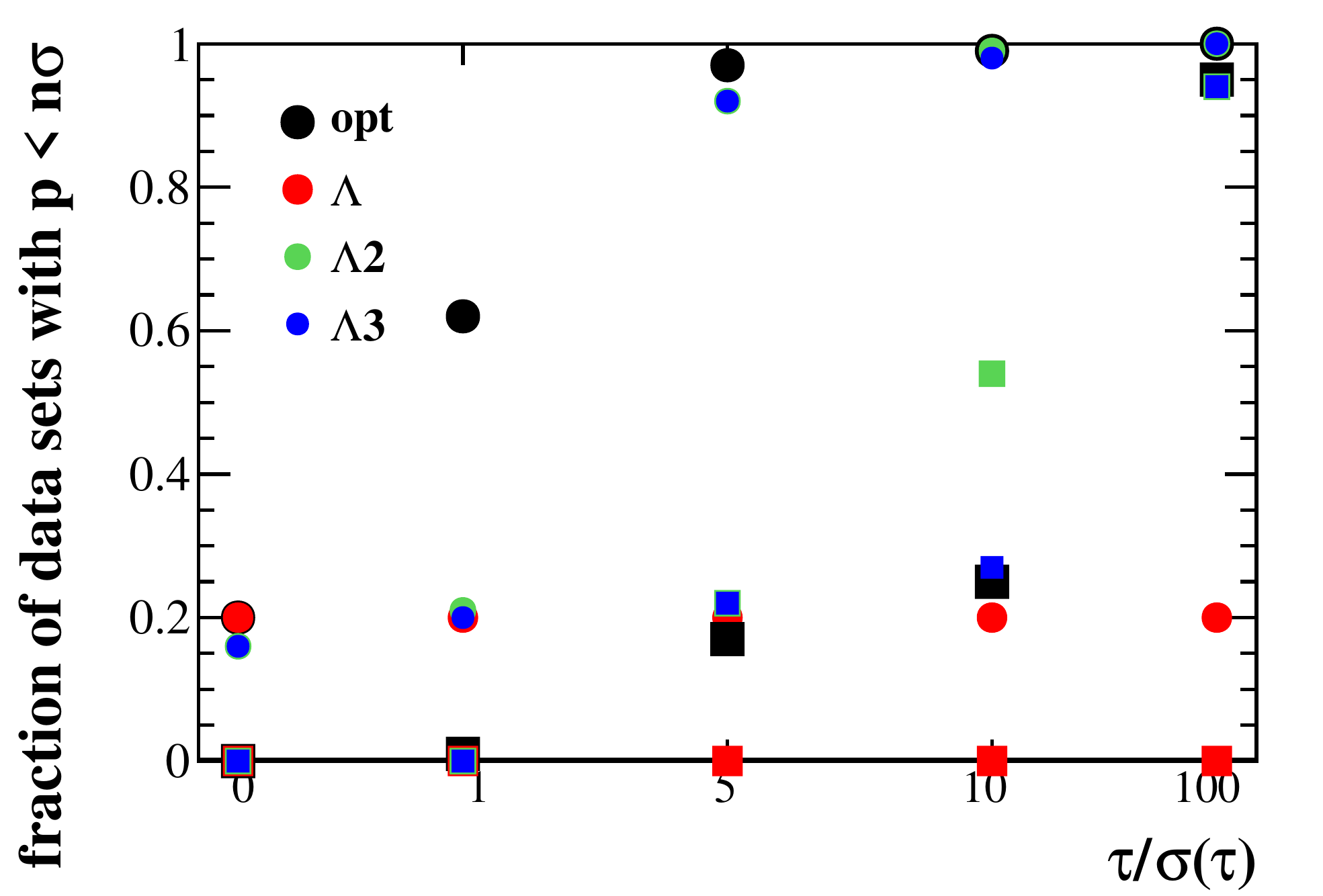}\\
\includegraphics[width=0.49\textwidth]{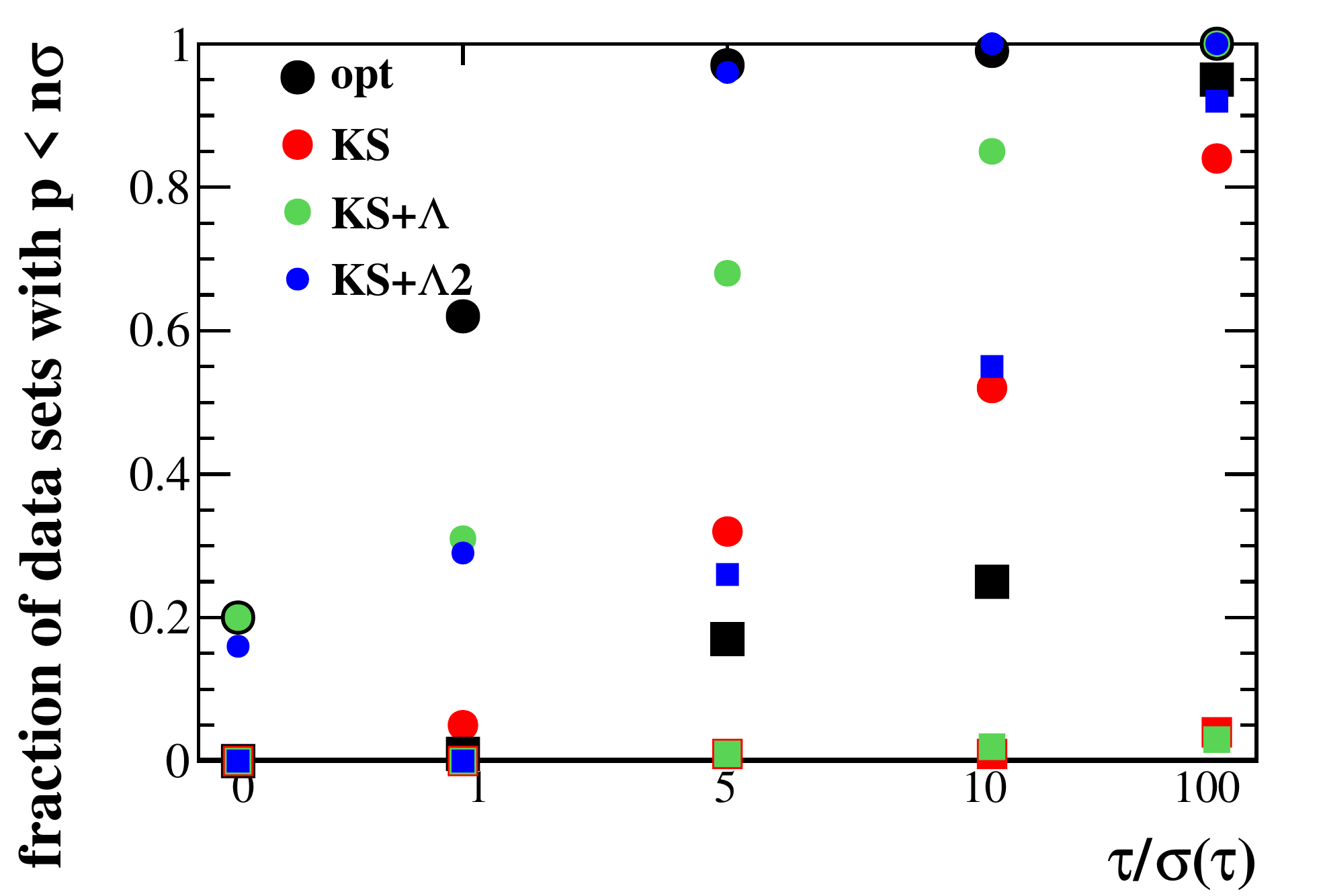}
\includegraphics[width=0.49\textwidth]{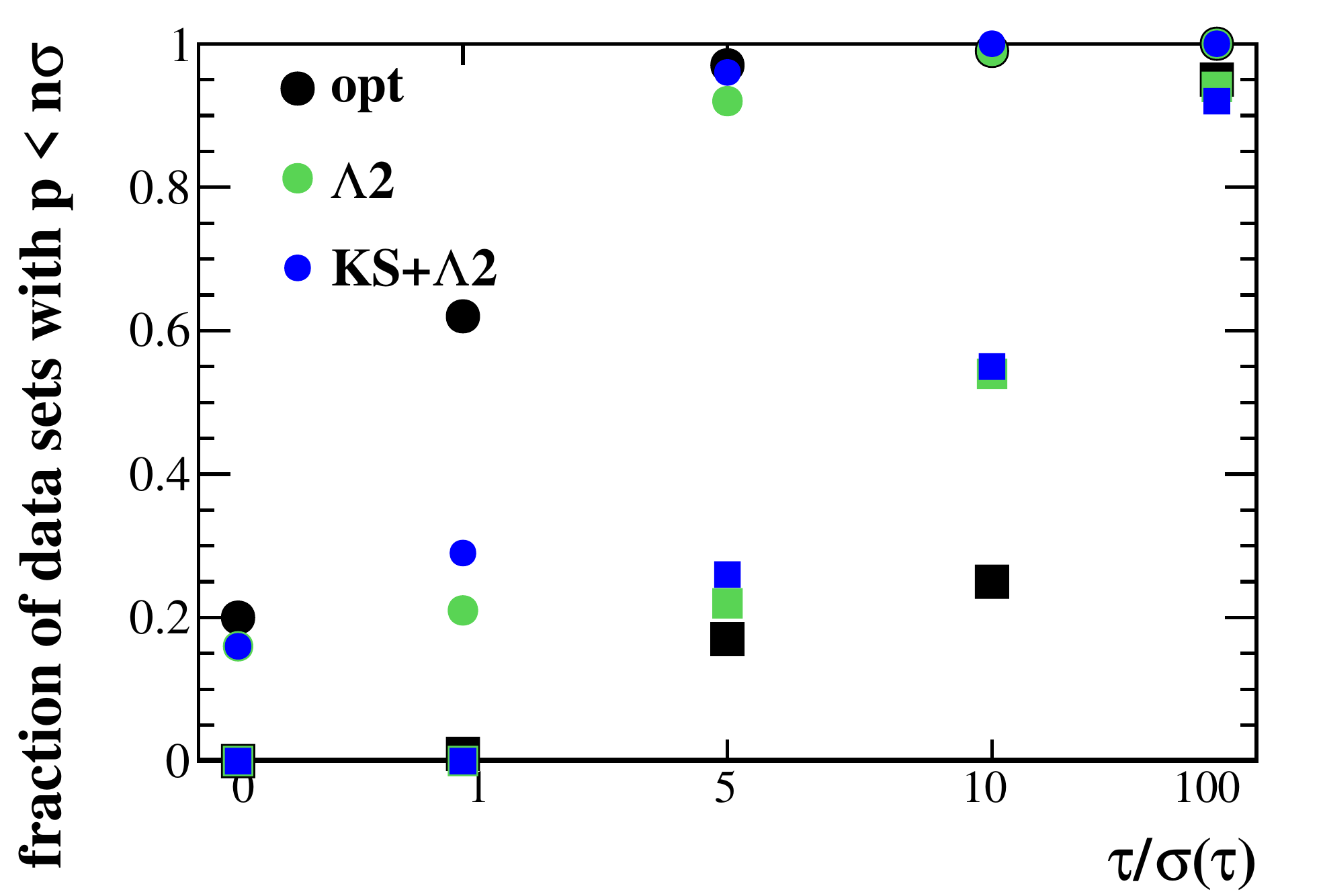}
\caption{\label{fig:tau_tests} 
The (circles,squares) represent the fraction of data sets with $p < (3,5)\sigma$. 
The test labels are defined as follows: (opt) the {\em cheat} PDF-based likelihood fit; (KS,CvM,AD) the 2-sample tests (see text); ($\Lambda(2,3)$) the profile likelihood test, including the $(2,3)$-region versions; ($KS+\Lambda(2)$) combination of both the KS and profile likelihood tests.
{\em N.b.}, the size of the markers has no meaning; this was done to aid in viewing markers for tests with similar power.
}
\end{figure}

Figure~\ref{fig:tau_tests} shows the results of a study of lifetime-only based testing for a given test mass.  Pseudo-datasets are generated using a Gaussian $\tau$ PDF for prompt backgrounds, and an exponential with effective lifetime of $10\sigma(\tau)$ for displaced backgrounds.  Signal events are also generated using an exponential PDF (of varying choices of $\tau$) convolved with a Gaussian to mimic the resolution.  The parameters for the background are taken to be $<n_s^{\rm prompt}> = 10$, $<n_s^{\rm displ}> = 0.1$ and $x=5$.  In each case 10 signal events are added. 
The following tests are studied:
\begin{itemize}
\item For comparison an ``optimal'' test is run where the true background PDF (including for displaced backgrounds) is used.  This is a {\em cheat} since I assume that the displaced background PDF is unknown.  The signal PDF is used but with $\tau$ as a free parameter.
\item Pure shape-based two-sample tests (see Appendix~C) are shown in the top left panel.  There is not much difference between the KS, CvM and AD tests for this particular data set.  As expected these tests provide no power for $\tau \ll \sigma(\tau)$ and increase in power with increasing $\tau$.
\item Profile likelihood tests are shown in the top right panel.  The $\Lambda$ test ignores lifetime information (hence its performance is independent of $\tau$).  As expected this test is optimal for $\tau \ll \sigma(\tau)$.  The $\Lambda2$ test is a two-region (prompt and displaced) counting experiment, where the likelihoods from each region are combined (via a product as usual). This simple test is nearly optimal except when $\tau \sim \sigma(\tau)$.  Adding a third region does not improve the performance for this particular displaced background (it may if the displaced background had some additional structure).  
\item The bottom left panel shows the results of performing the profile likelihood tests along with the shape-based KS test.  This greatly enhances the performance of the KS test.
\item The bottom right panel, however, shows that performing the KS test with the two-region profile likelihood test only provides a small gain in the region $\tau \sim \sigma(\tau)$.   
\end{itemize}
While the combination of KS and $\Lambda2$ in theory improves the performance for $\tau \sim \sigma(\tau)$, it adds some complexity to the analysis. For example, an additional assumption has now been employed: that all lifetime PDFs are the same in the signal and sideband regions.  This is likely to be true for the prompt SM background; however, it may not be true for other types of background.  
Furthermore, the gain in significance obtained by increasing the number of generated signal events from 10 to 11 for $\tau \sim \sigma(\tau)$ is much greater than that obtained by performing the KS test along with the profile likelihood; {\em i.e.}, the maximal gain in sensitivity in this example of using both the KS and $\Lambda2$ tests is < 10\% in signal rate sensitivity.  This gain requires adding an assumption which may not be valid and is expected to be difficult to validate/study.  Thus, my conclusion is that the nominal test should simply be $\Lambda2$.

\clearpage 

\section{Fast Algorithm}

The two-region profile likelihood method can be made extremely fast.  In Appendix~A I obtained analytic expressions that maximize the likelihood (including the case where a Gaussian uncertainty is included on the scaling factor that relates the sideband yields to those in the signal region); thus, no numerical minimization routine, {\em e.g.}, MINUIT, needs to be run.  
Note that the data can be binned in mass (in both the prompt and displaced regions) to perform this test.  This means that time consuming event loops are not required.  
Furthermore, as one scans the mass range, rather than summing up the yields in the signal and background regions, it is faster to simply subtract the bin(s) removed from each region and then add only the bin(s) added.  
Combining all of these optimizations results in a test that can be performed on 10M data sets with $\mathcal{O}(1000)$ test masses in about 2 hours on a single CPU core.  
This process is trivially parallelized to run on multiple cores.
Therefore, it is possible to confirm a significance of $>5\sigma$ without the need to rely on an asymptotic $p$-value.  
This is a desirable feature since for rare decays the asymptotic formulae tend to underestimate the significance for very low $p$-values. 

\end{document}